\documentclass[preprint,manuscript,showpacs,superscriptaddress,preprintnumbers,amsmath,amssymb,prb]{revtex4-1}
\usepackage{graphicx}
\usepackage{dcolumn}
\usepackage{bm}
\usepackage{multirow}
\usepackage{rotating}

\begin{document}

\title{
Electronic Structures of N-doped Graphene with Native Point Defects
}

\author{Zhufeng Hou}
\affiliation{Department of Organic and Polymeric Materials, Graduate School of Science and Engineering,
Tokyo Institute of Technology, 2-12-1 S5-20, Ookayama,Tokyo 152-8552, Japan}
\author{Xianlong Wang}
\affiliation{Department of Organic and Polymeric Materials, Graduate School of Science and Engineering,
Tokyo Institute of Technology, 2-12-1 S5-20, Ookayama,Tokyo 152-8552, Japan}
\author{Takashi Ikeda}
\affiliation{Condensed Matter Science Division, Quantum Beam Science Directorate, Japan Atomic Energy
Agency (JAEA), 1-1-1 Kouto, Sayo, Hyogo 679-5148, Japan}
\author{Kiyoyuki Terakura}
\affiliation{Research Center for Integrated Science, Japan Advanced Institute of Science and Technology
(JAIST), 1-1 Asahidai, Nomi, Ishikawa 923-1292, Japan}
\affiliation{Department of Organic and Polymeric Materials, Graduate School of Science and Engineering,
Tokyo Institute of Technology, 2-12-1 S5-20, Ookayama,Tokyo 152-8552, Japan}
\author{Masaharu Oshima}
\affiliation{Department of Applied Chemistry, The University of Tokyo, 7-3-1 Bunkyo-ku, Tokyo 113-8656, Japan}
\author{Masa-aki Kakimoto}
\affiliation{Department of Organic and Polymeric Materials, Graduate School of Science and Engineering,
Tokyo Institute of Technology, 2-12-1 S5-20, Ookayama,Tokyo 152-8552, Japan}
\begin{abstract}
Nitrogen doping in graphene has important implications in graphene-based devices and catalysts.  We have performed the density functional theory calculations to study the electronic structures of N-doped graphene with vacancies and Stone-Wales defect. Our results show that monovacancies in graphene act as hole dopants and that two substitutional N dopants are needed to compensate for the hole introduced by a monovacancy. On the other hand, divacancy does not produce any free carriers.  Interestingly, a single N dopant at divacancy acts as an acceptor rather than a donor. The interference between native point defect and N dopant strongly modifies the role of N doping regarding the free carrier production in the bulk $\pi$ bands. For some of the defects and N dopant-defect complexes, localized defect $\pi$ states are partially occupied. Discussion on the possibility of spin polarization in such cases is given. We also present qualitative arguments on the electronic structures based on the local bond picture.  We have analyzed the 1\textit{s}-related x-ray photoemission and adsorption spectroscopy spectra of N dopants at vacancies and Stone-Wales defect in connection with the experimental ones. We also discuss characteristic scanning tunneling microscope (STM) images originating from the electronic and structural modifications by the N dopant-defect complexes. STM imaging for small negative bias voltage will provide important information about possible active sites for oxygen reduction reaction.

\end{abstract}
\maketitle
\section{Introduction}
\label{sec:1}
Graphene, because of its unique electronic structure, attracts strong attention in the basic science field as well as in the application-oriented field.  The direct way of controlling its physical and chemical properties is to dope different elements. For example, nitrogen-doped graphene (N-graphene) has shown wide applications in \emph{n}-type graphene-based field-effect transistors,~\cite{Wang2009,Zhang201004110,Guo2010,lin133110} electrochemical biosensors,~\cite{Wang2010acs,shao2008} anodes of lithium-ion batteries,~\cite{Reddy2010,Wang2011jmc} catalysis for various chemical reactions,~\cite{yu2010} and so on. Among them, we pay particular attention to the carbon alloy catalysts (CACs) which are regarded as among the strong candidates of Pt-substitute electro-catalysts for oxygen reduction reaction (ORR) at the cathode of a fuel cell.~\cite{Ozaki2006,Zhang2006,Lefevre2009,Wu22042011,Proietti2011,DHLee2011,biddinger2010,Qu2010,Deng2011cm,Jafri2010}  The basic structural components of CACs are multilayered nanographene (nanographite) including carbon nanotubes.  Although intensive study has been made, the reaction site and reaction process are not yet well understood.  The main motivation of the present study is to clarify the basic aspects related to N doping to graphene so that we can obtain some insights into the possible reaction sites of ORR.  However, as the basic structural and electronic properties of N-graphene form a general background of physics and chemistry of graphene, we present the results of our extensive study of electronic structures for various local structures of N-graphene in a general form so that researchers working on graphene with various aims may utilize the information given below.

Recent experimental studies~\cite{Guo2010,lin133110,Zhang201004110,Deng2011cm,Imamura2011} suggest that some mutual effects between structural defects and N doping are present during the incorporation of N into graphene. In our previous paper,~\cite{Hou2011ar} from the total energy calculations based on density functional theory (DFT) we found that the native point defect (NPD) and N dopant attract each other, i.e., they display cooperative effect, suggesting that N dopants can prompt the creation of point defects and vice versa. Meanwhile, we have analyzed the energetic stability of substitutional N dopants in graphene with vacancies and Stone-Wales (SW) defect. In the controllable N doping of graphene by NH$_3$ annealing after ion irradiation~\cite{Guo2010} or by NH$_3$ plasma exposure,~\cite{lin133110} stepwise carrier doping from \textit{p}-type to \textit{n}-type has been observed in the measured conductivity of N-graphene.

In the present paper, we concentrate our analysis on how the electronic structure of graphene is modified by NPDs and N dopants.  Defect states appear at each of monovacancy (MV), divacancy (DV) and SW defect in the undoped graphene.  The characteristic features of defect states are analyzed in detail for each type of defect.  Then the effect of N doping on the electronic structures is discussed.  We have found several interesting consequences of the coexistence of NPDs and N dopants concerning the carrier doping in the bulk $\pi$ bands.  Although N dopant is generally regarded as electron donor, the N-NPD interaction greatly modifies such a naive picture.  Several examples of non-trivial behaviors of N dopants are presented below. For some of NPDs and N dopant-NPD complexes, defect $\pi$ states are partially occupied. In such cases, the possibility of spin polarization may be an interesting issue. Some detailed discussion is given about it.  We also present discussions on the electronic structures based on the local bond picture and try to shed light on the origin of defect states from a different viewpoint. Recently some theoretical works have been published on the detailed electronic structures of MV and N doping in graphene.~\cite{Palacios2012,Nanda2012,Lambin2012}  We make comments on their results at relevant parts in the later sections.

Scanning tunneling microscopy (STM) is a powerful experimental technique to investigate the electronic features of graphene induced by individual NPD and N dopant~\cite{Zhao12011s,Joucken2012} with an atomic resolution. In particular, the bright spots in the STM image under a small negative bias voltage (e.g., -0.1 or -0.2 V) would directly provide useful information for the possible catalytically active sites in N-graphene for ORR. Although the STM images for the clusters of two or more N dopants were reported, the atomic configurations were not yet solved.~\cite{Zhao12011s} In addition, the STM image simulations for the configurations of an isolated graphitelike N and multiple pyridinelike N dopants at vacancies were also reported.~\cite{Zheng2010acsn,Fujimoto2011} The actual distribution of different N bonding configurations in N-graphene in experiment would depend strongly on the energetic stabilities. In the present study, based on our results for the energetic stabilities of N dopants in graphene,~\cite{Hou2011ar} we investigate the STM images of different N bonding in the most stable configurations including the pyridinelike N at vacancies, the pyridiniumlike N at MV, and the modified-graphitelike (m-graphitelike) N at a five-membered ring of 5-8-5 DV and SW defect.

The spectroscopy measurements of the N 1\textit{s} core state have indicated that N atoms are incorporated into graphene in the form of pyridinelike N, pyridiniumlike
N, pyrrolelike N, and graphitelike N.~\cite{Matter2006,Subramanian200938,Niwa20111006,wei2009,lin133110,Zhang201004110,Niwa200993,Zhao12011s,Usachov2011} Note that the first two configurations of doped N would exist at the edges or the defect sites only. The possible N configurations at the edges of graphene nanoribbons and clusters have been investigated before to understand the effect of edge states on the N doping~\cite{Liyf2009,Huang09} and the electron transport properties.~\cite{Blanca2009} In our previous work~\cite{Wang2011prb} as well as in our recent one,~\cite{Wang2013} we have theoretically analyzed the N 1\textit{s} x-ray photoelectron, absorption and emission spectra (XPS, XAS and XES) of N dopants along edges of graphene nanoribbons. In this work, we extend our analysis for the 1\textit{s} XPS and XAS spectra of N dopants near NPDs in the bulk graphene.

The remainder of this paper is organized as follows. In Sec.~\ref{sec:2}, we introduce the computational methods for electronic structure calculations and for the simulations of STM images and the N 1\textit{s} XPS and XAS spectra. Our computed electronic structures, STM images, and N 1\textit{s} XPS and XAS spectra of N-graphene are presented in Sec. \ref{sec:3}. Finally, we draw conclusions in Sec. \ref{sec:4}.

\section{Method and computational details}
\label{sec:2}
In the present work, the electronic structure calculations and STM image simulations have been performed with the PWSCF code of the Quantum ESPRESSO suite.~\cite{Baroni} The detailed computational setup was given in our previous paper.~\cite{Hou2011ar} We only mention here that a $9\times9$ supercell is used for most of the calculations.  The simulated constant-height STM images are obtained using the Tersoff-Hamann approximation.~\cite{Tersoff1998}
For a positive (negative) bias voltage $V_b$, the STM image provides the information for the unoccupied (occupied) states of the sample. A fixed sample-tip distance of $d=2$ \AA~is used throughout our STM image simulations.

The calculations of the N 1\textit{s} XPS and XAS spectra are performed with the CP2K code.~\cite{cp2kcode,Lippert1999,Iannuzzi2007} Here a $12\times12$ supercell with a vacuum thickness of 20 \AA~is employed in the XAS calculations to avoid the interference between the electrons excited to unoccupied states as much as possible. The atomic structures of N-graphene are taken from those optimized with the PWSCF code.~\cite{Hou2011ar} Two different ways are employed to estimate the N 1\emph{s} binding energy [$E_\mathrm{b}$(N 1\emph{s})] for XPS spectra: (i) $E_\mathrm{b}$(N 1\emph{s}) is obtained by the Kohn-Sham (KS) energy level of ground state for N dopants to take into account the initial state effect and in the present study it is given with respect to the Fermi level; (ii) $E_\mathrm{b}$(N 1\emph{s}) is calculated from the total energy difference ($\Delta\mathrm{SCF}$) between the excited states with a full core-hole  and the ground state and then a correction is taken into account to change the reference from vacuum to the Fermi level. We make some further comments on the method of $\Delta\mathrm{SCF}$ calculation in the present work.  The core-level binding energy with respect to the Fermi level is given by
\begin{equation}\label{eq:2}
E_\mathrm{b}(\mathrm{N}~1s)=E_\mathrm{tot}(n_c-1, n_v)-E_\mathrm{tot}(n_c, n_v)-W,
\end{equation}
where the first term in the right-hand side of Eq.~(\ref{eq:2}) is the total energy of a system with one core-hole, the second term is the total energy of the unperturbed ground state with $n_c$ ($n_v$) denoting the number of core (valence) electrons, and the last term $W$ denotes the workfunction. In our previous work,~\cite{Wang2011prb} we evaluated $W$ as the difference between the Hartree potential at the middle of the vacuum region of the supercell and the Fermi level of the unperturbed system.  If the size of supercell is large enough, this approach does not have ambiguity.  However, we have found that with the current supercell size, the actual value of the core-level binding energy depends on the size of the vacuum region.  One of the reasons for this inconvenient aspect of Eq.~(\ref{eq:2}) may be due to the different numbers of electrons in the first term and the second term, though a uniform negative background is added to compensate the removed core electron.  To avoid this inconvenience, we used the following expression for $W$:
\begin{equation}\label{eq:3}
W=E_\mathrm{tot}(n_c, n_v-1)-E_\mathrm{tot}(n_c, n_v),
\end{equation}
where the first term on the right-hand side of Eq.~(\ref{eq:3}) is the total energy of a system with one electron removed from the Fermi level ($E_\mathrm{F}$).  Then, the core-level binding energy is simply given by
\begin{equation}\label{eq:4}
E_\mathrm{b}(\mathrm{N}~1s)=E_\mathrm{tot}(n_c-1, n_v)-E_\mathrm{tot}(n_c, n_v-1).
\end{equation}
This expression is simple and easy to be calculated.  Both systems appearing on the right-hand side of Eq.~(\ref{eq:4}) have the same number of electrons, i.e., $N-1$ and we have found that the supercell size dependence is virtually removed.
Other computational procedure and numerical algorithms are basically identical to those in our previous publication.~\cite{Hou2011jpcc}

\section{Results and discussion}
\label{sec:3}

\subsection{N doping of defect-free graphene}
\label{sec:3A}
\begin{figure*}[hbtp!]
\begin{center}
\begin{minipage}[t]{6.5cm}
\includegraphics*[width=6.5cm]{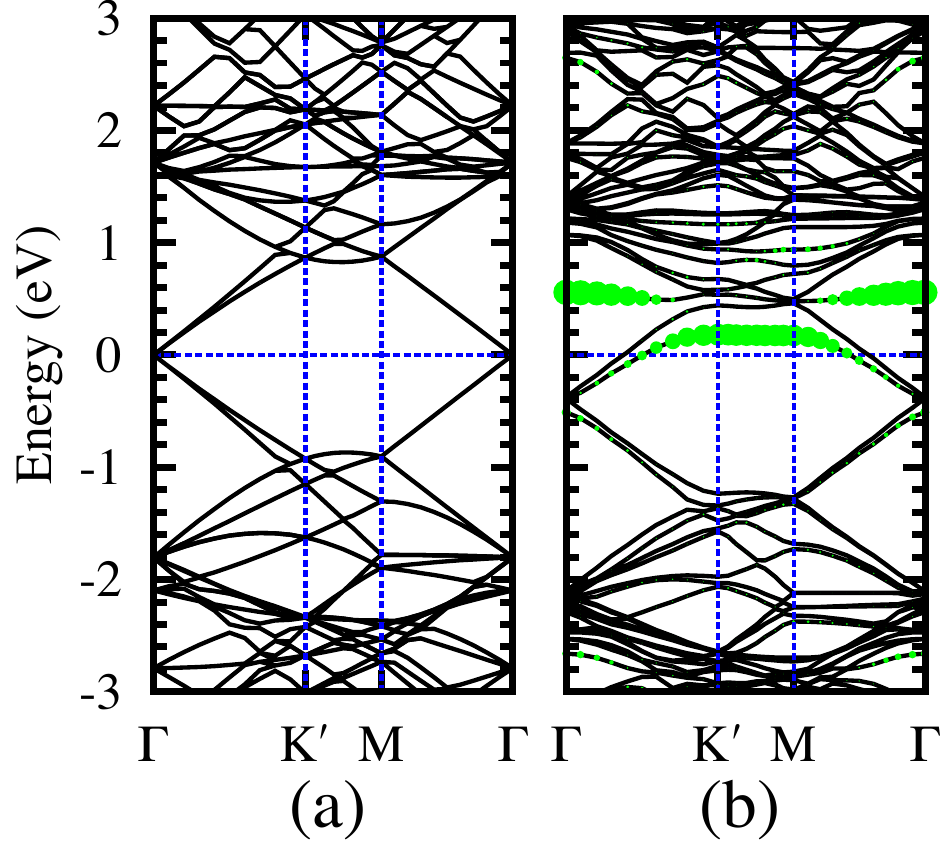}
\end{minipage}
\vspace*{0.2cm}
\begin{minipage}[t]{6.5cm}
\includegraphics*[width=6.5cm]{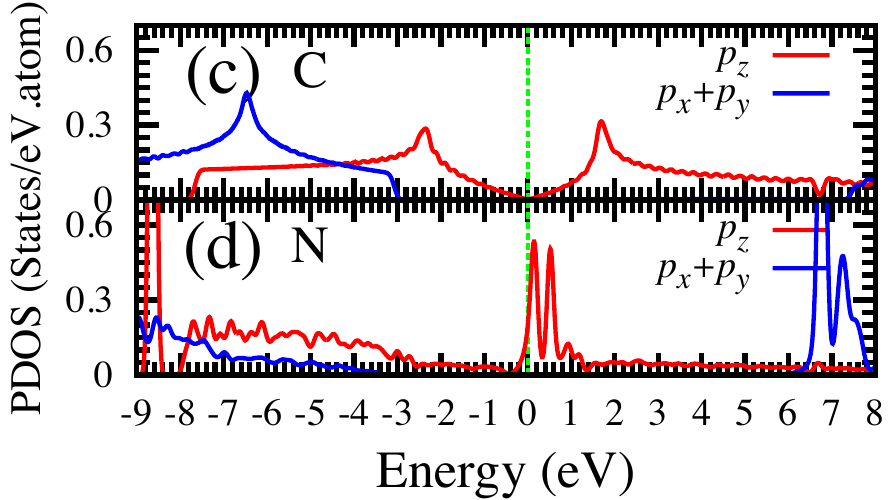}
\end{minipage}
\caption{\label{fig:1} (Color online) The band structures of (a) perfect graphene and (b) N-graphene with single substitutional N atom in a $9\times9$ supercell. Partial density of states (PDOS) for the 2\textit{p} orbitals: (c) C atom in perfect graphene and (d) an isolated substitutional N atom in N-graphene. The zero of energy is set at $E_{\mathrm{F}}$. Panel (b) shows the fat bands (green solid points) derived from N 2$p_z$ state.  }
\end{center}
\end{figure*}

First, we examine as a reference how the atomic and electronic structures of perfect graphene are altered when graphitelike N is introduced. Our calculations predict the N-C bond length of 1.41 \AA\/, which is slightly shorter than the C-C bond length of 1.42 \AA~in perfect graphene partly due to the smaller atomic size of N.
Figure~\ref{fig:1} shows the electronic structures of perfect graphene and N-graphene. For perfect graphene, the original \emph{K} and \emph{K}$^{\prime}$ points in the first Brillouin zone (BZ) of primitive cell are folded into the BZ center (i.e., $\Gamma$ point) of $9\times 9$ super cell, which is used in our calculations. It is known that for perfect graphene $E_{\mathrm{F}}$ coincides with the Dirac point, and thus both the $\pi$ and the $\pi^{\ast}$ bands near $E_\mathrm{F}$ are doubly degenerate [see Fig.~\ref{fig:1}(a)]. For N-graphene, $E_{\mathrm{F}}$ shifts up into the conduction band due to the electron donation from doped N to the $\pi^{\ast}$ bands of graphene.~\cite{Robertson1995} The doped graphitelike N also removes partly the degeneracy of the $\pi$ and $\pi^{\ast}$ bands near the Dirac point. The impurity resonant state appears at about 0.4 eV, hybridizes with one of the bulk $\pi^{\ast}$ bands, and forms two flat bands at about 0.18 and 0.59 eV above $E_\mathrm{F}$, as shown in Figs.~\ref{fig:1}(b) and ~\ref{fig:1}(d).
The supercell size dependences of these flat bands were studied using $3n \times 3n$ supercell with $n$ changing from 2 to 5.  We found that the upper flat band energy $E_2$ and the lower one $E_1$ are very well fitted (with the least-squares fitting) by
\begin{equation}\label{eq:1}
\begin{array}{l}
E_2=2.416/n + 0.115,  \\
E_1=1.264/n + 0.119,
\end{array}
\end{equation}
where both energies are measured in eV with reference to the Dirac point.  These equations suggest that in the limit of large $n$ both energies converge to the same value given by about 0.12 eV.  Therefore, the two split peaks in Fig.~\ref{fig:1}(d) just above the Fermi level will be a single peak at 0.12 eV above the Dirac point in the dilute limit of doped N.  The result is consistent with the recent STS measurement in which the resonant peak is located at about 0.14 eV.~\cite{Kondo2012,Kondo2012suppl}  The strong size dependence of $E_1$ and $E_2$ suggests strong long range interaction among defects.~\cite{Nanda2012,Lambin2012}


\subsection{Monovacancy and N doping}
\label{sec:3B}
\subsubsection{MV in undoped graphene}
\label{sec:3B1}
\begin{figure}[htbp!]
\begin{center}
\begin{minipage}[t]{10cm}
\includegraphics*[width=10cm]{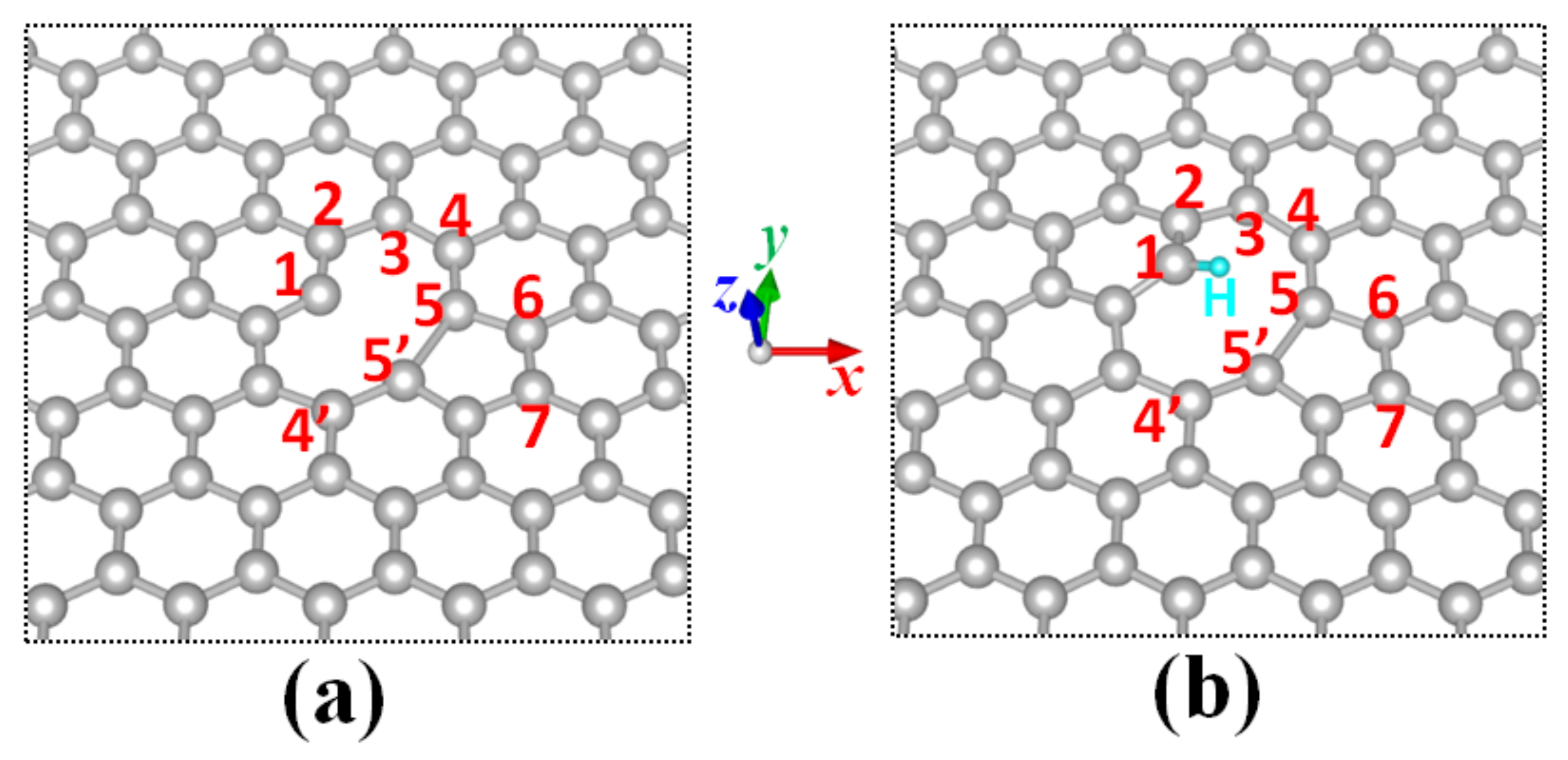}
\end{minipage}
\vspace*{.20cm}
\hspace*{2.50cm}
\begin{minipage}[t]{9cm}
\includegraphics*[width=9cm]{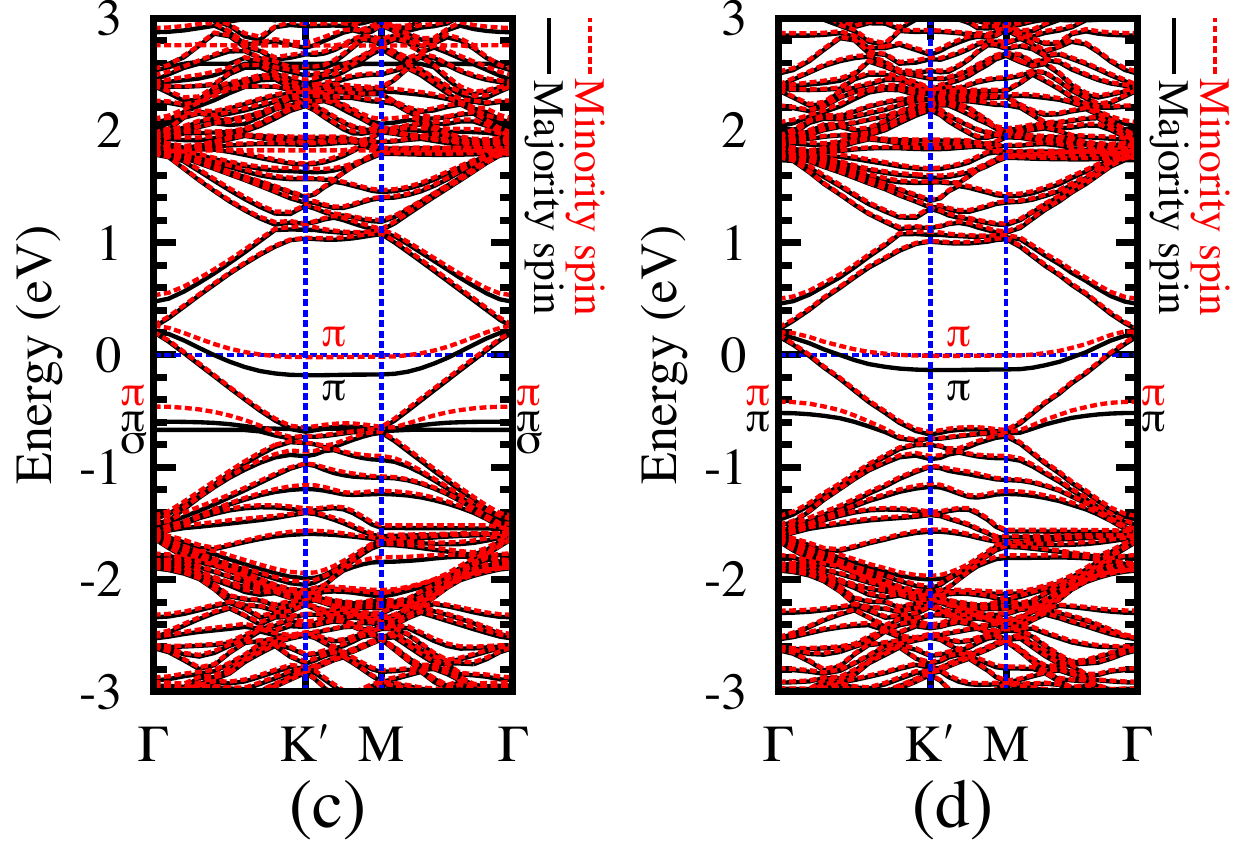}
\end{minipage}
\vspace*{.20cm}
\hspace*{3.00cm}
\begin{minipage}[t]{4.5cm}
\includegraphics*[width=4.5cm]{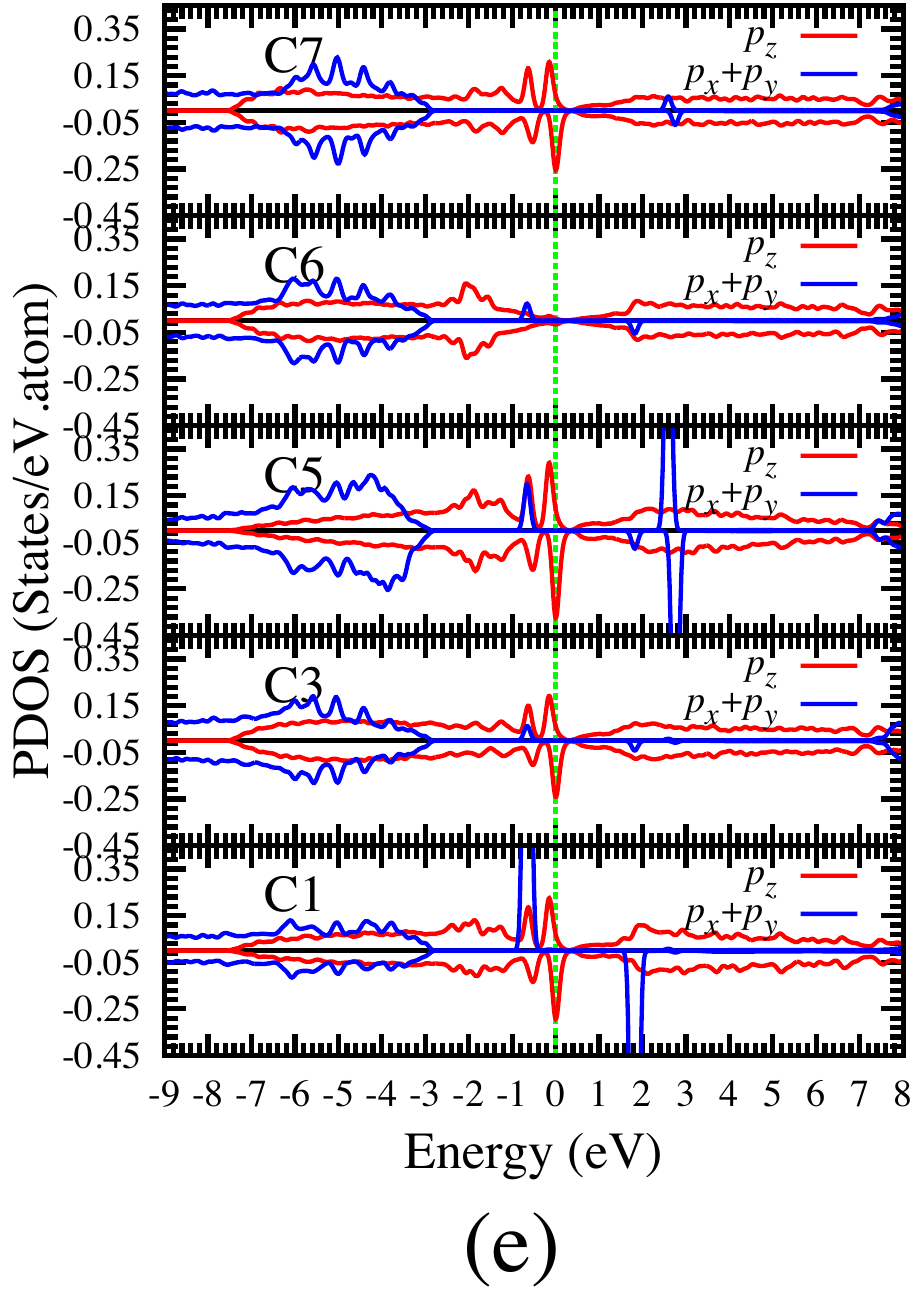}
\end{minipage}
\vspace*{.2cm}
\begin{minipage}[t]{4.5cm}
\includegraphics*[width=4.5cm]{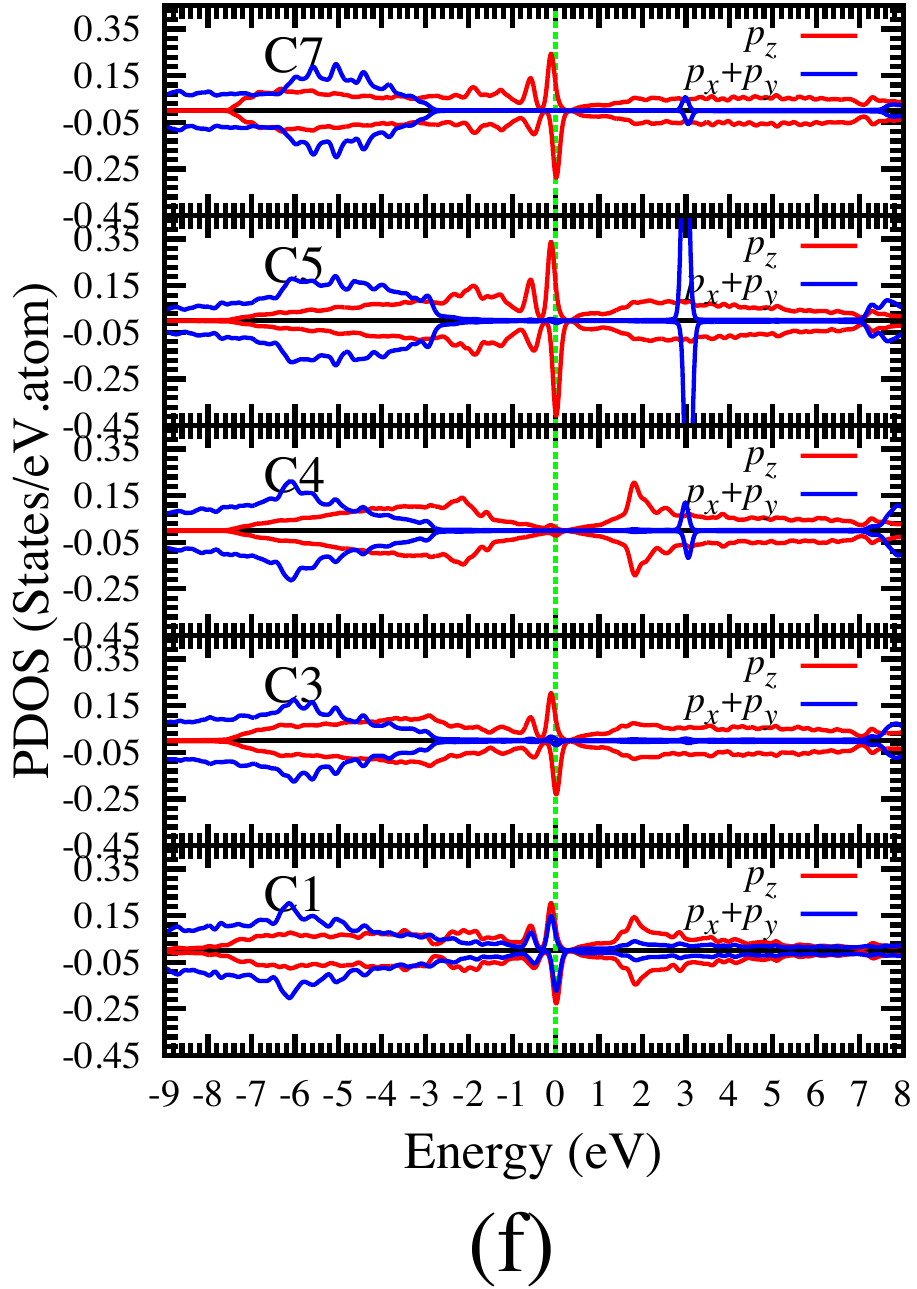}
\end{minipage}
\caption{\label{fig:2} (Color online) The atomic structures of (a) MV and (b) hydrogenated MV [one of three C atoms next to MV terminated by H, denoted as H-MV]. The band structures of graphene with (c) MV and (d) H-MV. The partial density of states (PDOS) for the 2\textit{p} orbitals of C atoms around defect region of (e) MV and (f) H-MV.}
\end{center}
\end{figure}

\paragraph{Band calculations}
For MV in undoped graphene, as three electrons are accommodated to three dangling $\sigma$ orbitals, the three carbon atoms next to the vacancy site would undergo a Jahn-Teller (JT) distortion, i.e., two of them [labeled C5 and C5$^{\prime}$ in Fig.~\ref{fig:2}(a)] will form a new bond.~\cite{Barbary2003,Carlsson2006,Ma2004NJP,Yazyev2007}  As shown in Ref.~\onlinecite{Palacios2012}, the magnitude of JT distortion for MV depends on the employed supercell size. Generally, a smaller supercell predicts weaker JT distortion. In the present study, the spin-polarized GGA calculations on a MV in a $9\times 9$ supercell predict that the C5-C5$^{\prime}$ bond length in the reconstructed MV is 1.96 \AA~and that the C1 atom stays in the graphene plane, which agree with the more recent calculations.~\cite{Palacios2012,Wangb2012prb}  Judging from the results for larger supercells, the converged value of C5-C5$^{\prime}$ bond length may be 1 or 2\% smaller than our value.~\cite{Palacios2012}  Figure~\ref{fig:2}(c) presents the band structure of undoped graphene with MV. It can be seen that $E_{\mathrm{F}}$ is below the Dirac point, indicating that MV acts as a hole dopant. The double degeneracy of the $\pi$ and $\pi^{\ast}$ bands near the Dirac point is also removed by MV in the supercell calculations like the present one.

The electronic structure obtained with the constraint of non spin polarization (for the structure optimized in the spin-polarized state) gives -4.0 and 2.8 eV with reference to $E_{\mathrm{F}}$ as the energy of the bonding state among the three dangling $\sigma$ orbitals and that of an anti-bonding state between C5 and C5$^{\prime}$, respectively.  These states are far away from $E_{\mathrm{F}}$ and do not play any roles in transport and chemical activities.  Only a dangling bond state at the remaining atom labeled C1 has its energy rather near $E_{\mathrm{F}}$ which splits into an occupied majority spin state at -0.6 eV  and the unoccupied counterpart at 1.8 eV with respect to $E_{\mathrm{F}}$. While the exchange splitting is about 2.4 eV for the dangling $\sigma$ bond states, it is only about 0.2 eV in the $\pi$ defect levels, which hybridize strongly with one branch of two $\pi$ bulk bands in each spin state.
The exchange splitting in dangling $\sigma$ bond states produces the magnetic moment of 1.0 $\mu_\mathrm{B}$, while the defect $\pi$ states contribute to a non-integer value of 0.46 $\mu_\mathrm{B}$ due to the hybridization with the bulk $\pi$ bands. The net magnetic moment of 1.46 $\mu_\mathrm{B}$ is in good agreement with the previous studies.~\cite{Lehtinen2004,Ma2004NJP,Yazyev2007}
 (Note, however, the comments on the $\pi$ state spin polarization at the end of this section and the discussion in Sec.\ref{sec:3}E)

The most important point for MV is that the center of gravity of the $\pi$ defect states is located below the Dirac point.  Therefore, the total number of electrons accommodated in the defect related levels is nearly five (three in $\sigma$ and about two in $\pi$) being larger by about one than four needed to charge compensation at MV.  This is the reason why MV acts as a hole dopant.
In the tight-binding model based on the orthonormal basis set for the $\pi$ band of graphene, the $\pi$ orbital level is equal to the Dirac point energy.  Because of this, one may naively think that the defect $\pi$ states will have the center of gravity at the Dirac point. In the model calculations for MV,~\cite{Nanda2012} MV is regarded as an impurity with a strong repulsive potential $U_0$. In this type of treatment, the location of MV defect $\pi$ state is just below the Dirac point for finite strength of $U_0$ simply due to the hybridization repulsion between the impurity orbital and its surrounding orbitals, the latter being the defect $\pi$ states.  This mechanism of locating the defect $\pi$ level below the Dirac point may be correct or may be an artifact of the model.  Below we propose another mechanism in which the orthogonalization effect will lead to such a lowering of the defect $\pi$ level.

In any \emph{ab initio} method with localized basis set, the basis orbitals are nonorthogonal and we solve the following eigenvalue problem.
\begin{equation}\label{eq:5}
\begin{array}{l}
HA=ESA,
\end{array}
\end{equation}
where $H$ is Hamiltonian expressed in terms of nonorthogonal basis, $S$ is the overlap integral between nonorthogonal basis functions and $A$ is the eigenvector.  By performing L\"{o}wdin orthogonalization, we obtain
\begin{equation}\label{eq:6}
\begin{array}{l}
S^{-1/2}HS^{-1/2}S^{1/2}A=ES^{1/2}A,  \\
S=1+\delta S,  \\
S^{-1/2}\approx 1-(1/2)\delta S.
\end{array}
\end{equation}
Therefore, the effective atomic level in the orthogonal tight-binding model which should be equal to the Dirac point energy $E_\mathrm{D}$ is given by
\begin{equation}\label{eq:7}
\begin{array}{l}
E_D=(S^{-1/2}HS^{-1/2})_{ii} \approx H_{ii}-(1/2)\sum_j (\delta S_{ij}H_{ji}+H_{ij} \delta S_{ji})=E_a+\delta E_a,
\end{array}
\end{equation}
where the subscripts $i$, $j$ denote atomic sites, $E_a$ the atomic level given by $H_{ii}$ and $\delta E_a$ the correction term coming from nonorthogonality.  It is important to note that the correction term $\delta E_a$ is positive because the Hamiltonian matrix elements and overlap matrix elements have opposite sign.  In the presence of a vacancy at site 0, the effective atomic level at the surrounding site $n$ is given by
\begin{equation}\label{eq:8}
\begin{array}{l}
E_n ^\mathrm{MV} =E_\mathrm{D} + (1/2)(\delta S_{n0}H_{0n} + H_{0n} \delta S_{n0}).
\end{array}
\end{equation}
As the second term in the above equation must be negative, the effective atomic level at the surrounding sites of MV is lower than $E_\mathrm{D}$.  In this way, we can naturally explain why the center of gravity of defect $\pi$ states is below the Dirac point.

The orbital decomposed partial densities of states (PDOS) of MV are shown in Fig.~\ref{fig:2}(e). The dangling $\sigma$ state near $E_{\mathrm{F}}$ is strongly localized at C1, while the defect $\pi$ states have significant amplitude at C1, C3 (not shown), C5, C5$^{\prime}$, and C7. This is consistent with the tight-binding model analysis results that a single impurity in sublattice A induces an impurity state mostly localized in sublattice B and vice versa due to the existence of two nonequivalent Dirac points.~\cite{Wehling2007,Toyoda2010} The same behavior has also been found in the case of single substitutional N in defect-free graphene~\cite{Hou2011ar} and in the results discussed below for other NPDs even after N doping.

Since C1 next to the vacancy site has dangling bond, it can be terminated by hydrogen~\cite{Hou2011ar,Lehtinen2004} or substituted by nitrogen.~\cite{Hou2011ar,Fujimoto2011} When C1 is terminated by a hydrogen atom [see Fig.~\ref{fig:2}(b)], the atom C1 moves out of the atomic plane by 0.89 \AA, the unoccupied minority spin dangling $\sigma$ state of C1 is pulled down below the Fermi level to accommodate the additional electron from the hydrogen.  The doubly occupied $\sigma$ level with small admixture of $\pi$ states is now merged into the bulk $\sigma$ band and the $\sigma$ spin polarization observed for non-hydrogenated MV disappears.~\cite{Lehtinen2004}
However, the sharp defect $\pi$ states mixed with $\sigma$ states at the Fermi level can spontaneously produce spin polarization, whose magnitude is estimated to be 0.55 $\mu_\mathrm{B}$ in the present calculation.  Nevertheless, as the energy gain from the non-spin-polarized state to the spin-polarized one is only about 5 meV/MV, the spin polarization will disappear at room temperature.  (Note also the comments at the end of this section.)
The band structure of the hydrogenated MV (H-MV) and its PDOS are shown in Figs.~\ref{fig:2}(d) and~\ref{fig:2}(f). Except the position of the $\sigma$ level and exchange splitting, the band structure of H-MV [Fig.~\ref{fig:2}(d)] is very similar to that of MV [Fig.~\ref{fig:2}(c)] and the filling of $\pi$ bands is not modified by hydrogenation.  However, the peaks near $E_{\mathrm{F}}$ in the PDOS of C1 have some components of $\sigma$ states for H-MV because of the out-of-plane displacement of C1.

We make some additional comments on the spin state of MV.  It is claimed in a recent work~\cite{Palacios2012} that the spin polarization of the $\pi$ state will vanish in the dilute limit of vacancy concentration.  As the defect $\pi$ state strongly hybridizes with the bulk $\pi$ band, its tail must be rather extended. This suggests the need of careful treatment of magnetic properties of defect $\pi$ states in the dilute limit.  Moreover, as mentioned above in the context of H-MV, even if the defect $\pi$ states may be spin polarized, the magnetic stabilization energy is only on the order of meV. We also find that when the hydrogen atom and its nearest-neighbor C1 atoms are confined in the atomic plane of graphene, the $\sigma$ states of the C1 atom do not mix with the sharp $\pi$ states near $E_\mathrm{F}$ and the spin polarization of the $\pi$ states of H-MV disappears. Therefore, hereafter up to Sec.~\ref{sec:3}D, we do not take into account the spin polarization in the defect $\pi$ states in similar situations.  The problems of spin polarization are summarized in Sec.~\ref{sec:3}E for all the subtle cases treated in the present work.

\begin{figure*}[htbp!]
\begin{center}
\hspace*{0.50cm}
\begin{minipage}[t]{8.0cm}
\includegraphics*[width=8.0cm]{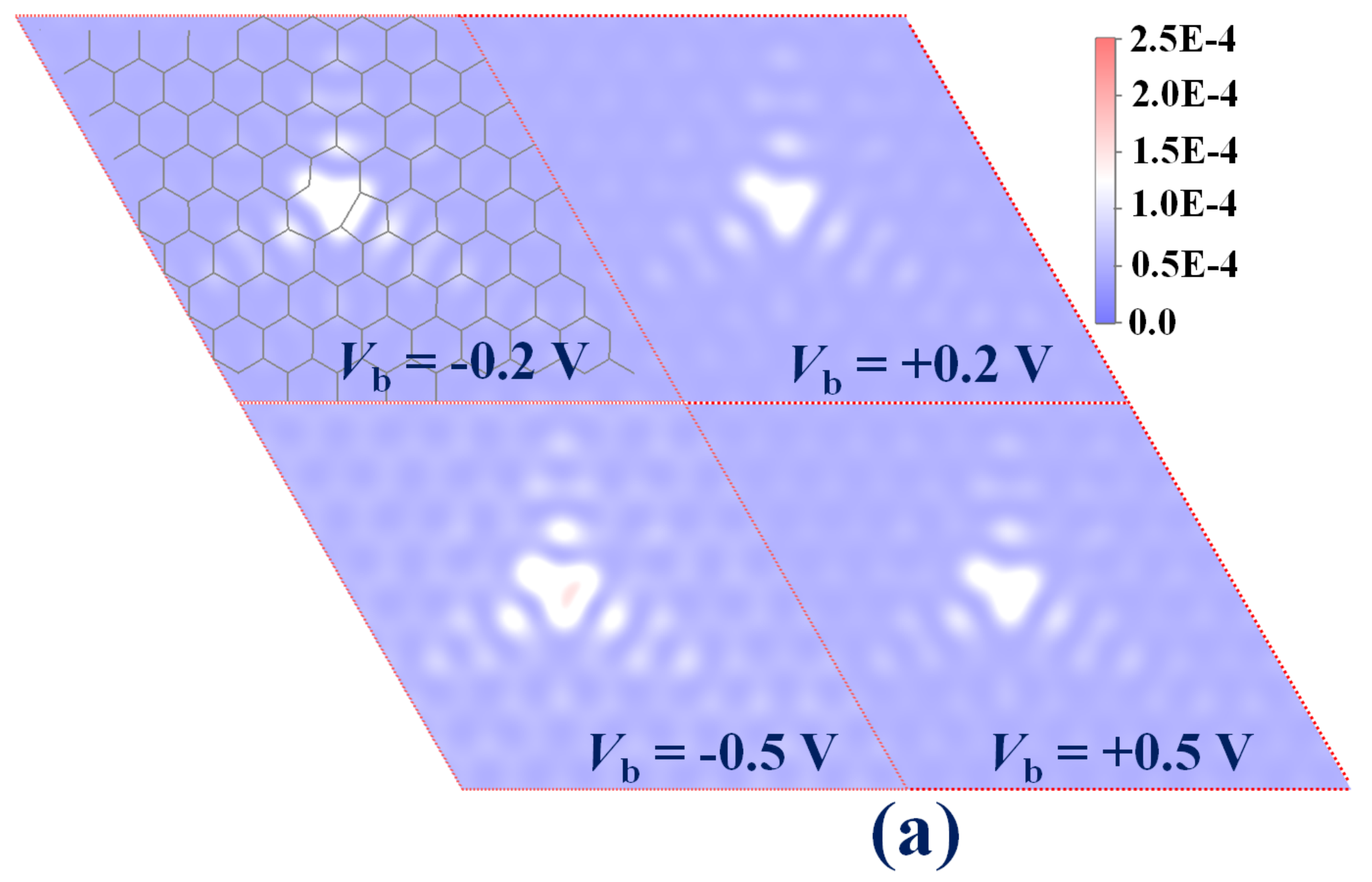}
\end{minipage}
\begin{minipage}[t]{8.0cm}
\includegraphics*[width=8.0cm]{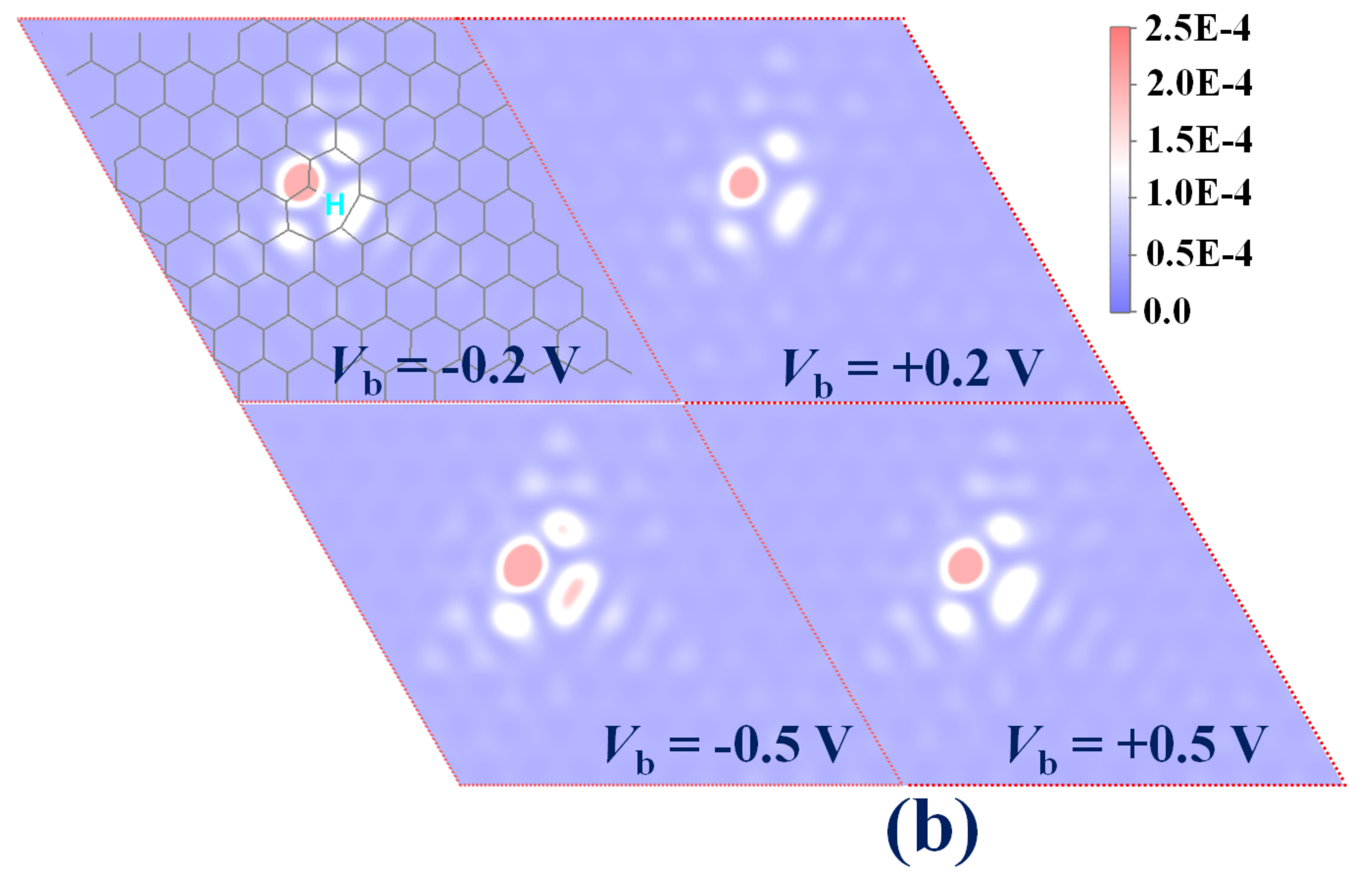}
\end{minipage}
\begin{minipage}[t]{8.0cm}
\includegraphics*[width=8.0cm]{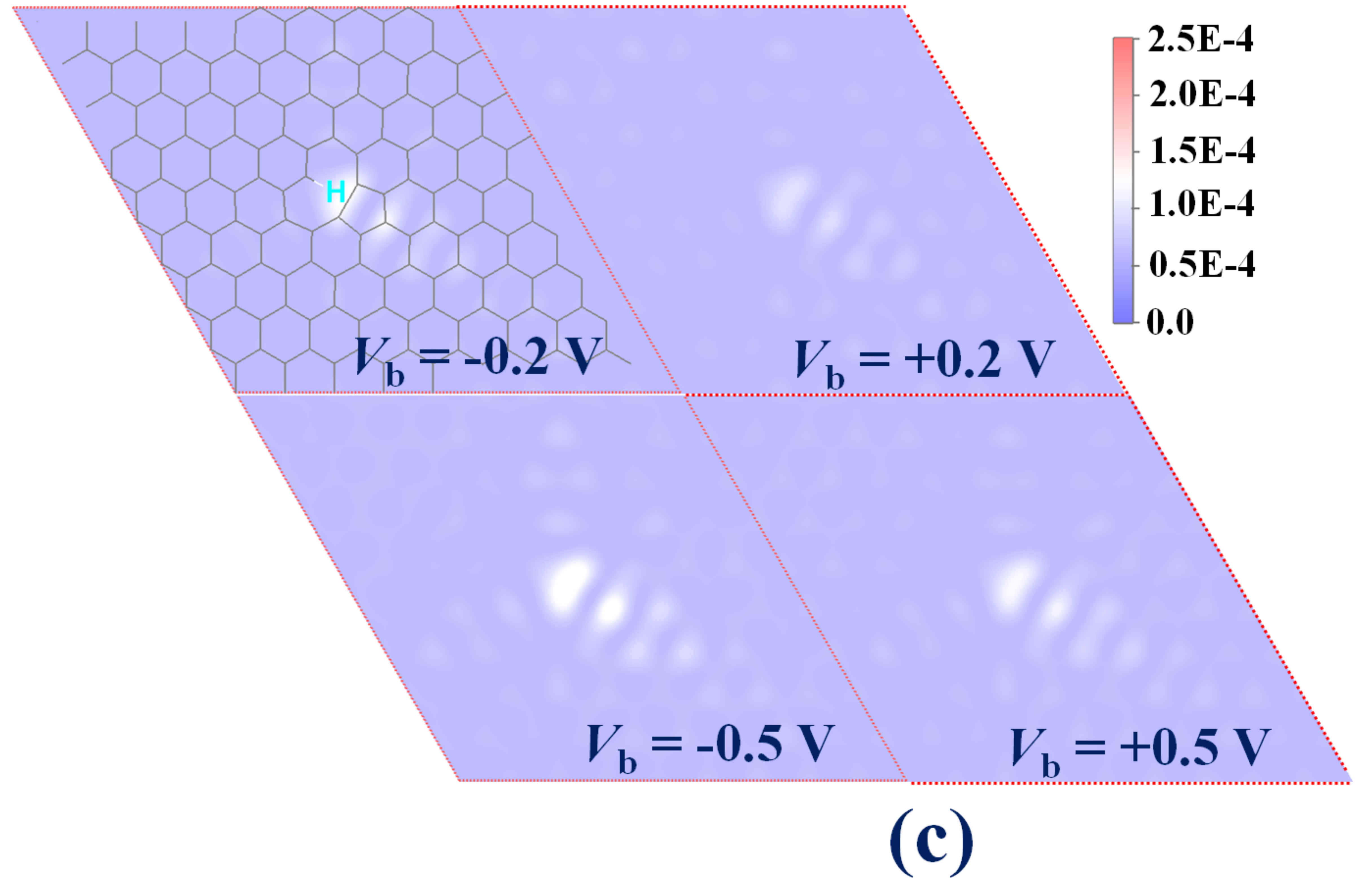}
\end{minipage}
\caption{\label{fig:3} (Color online) Simulated STM images for (a) MV and (b), (c) hydrogenated MV under different bias voltages indicated in each panel and with a sample-tip distance of $d=2$ \AA.  In (b) [(c)], the hydrogenated C1 and STM tip are on the same [opposite] side of graphene plane.}
\end{center}
\end{figure*}

To examine the localized states near $E_\mathrm{F}$, the simulated STM images under different bias voltages are presented in Fig.~\ref{fig:3}.  They serve as a powerful test of the calculated results. Occupied (unoccupied) $\pi$ states are probed by negative (positive) bias voltage. Unless we use a very small tip-sample distance, $\sigma$ states which are mostly confined within a graphene plane are not visible by STM.  Therefore the STM images reflect the $\pi$ state components of PDOS shown in Figs.~\ref{fig:2}(e) and~\ref{fig:2}(f).
Figure~\ref{fig:3}(a) shows the simulated STM images for the undoped MV. The pattern in the STM images under both the positive and the negative bias voltages exhibits a threefold symmetry with respect to the vacancy site and significantly bright spots at C3, C5, and C7 atoms. It is noted that the brightness in the STM images under the negative bias voltages is slightly stronger than that under the positive bias voltages, indicating $p$-type doping behavior for the undoped MV. The simulated STM images for the undoped H-MV are presented in Figs.~\ref{fig:3}(b) and~\ref{fig:3}(c). As the hydrogenated carbon atom C1 is out of the atomic plane, we study two cases with the tip above and below the atomic plane with the same distance of 2 \AA~in the STM simulations. In the case of the tip above the atomic plane in the STM simulations, the C1 atom and the attached hydrogen atom are at the same side with the tip, and thus the C1 atom shows a very strong bright spot, while no bright spot appears at the C1 site in the case of the tip placed blow the atomic plane.
\paragraph{Local bond picture}
Although $\pi$ defect states are fairly extended, we demonstrate that the local bond picture which gives us insight to the electronic structures from a different viewpoint is quite useful.

\begin{figure*}[htbp!]
\begin{center}
\begin{minipage}[t]{5.0cm}
\includegraphics*[width=5.0cm]{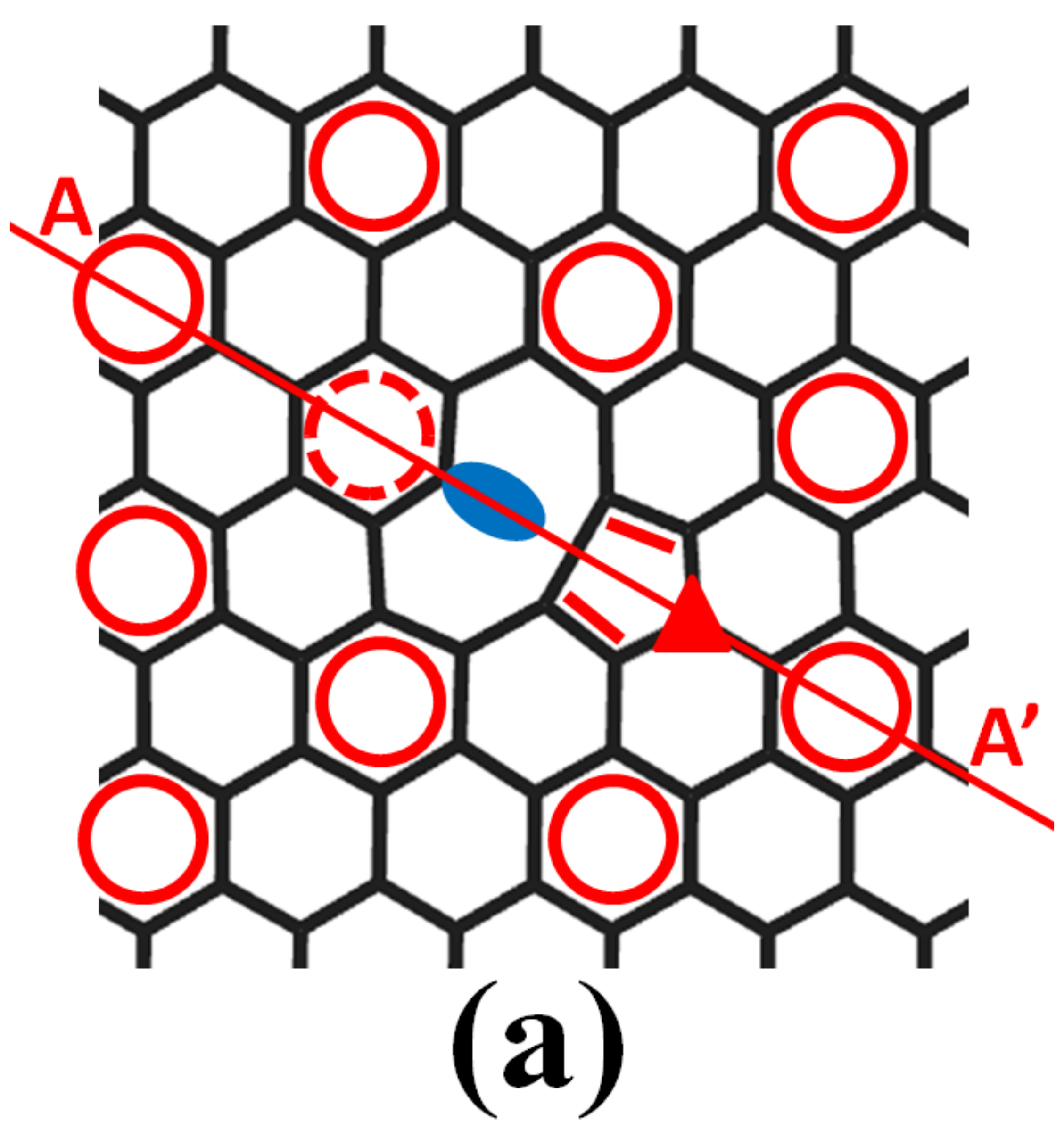}
\end{minipage}
\hspace*{0.2cm}
\begin{minipage}[t]{5.0cm}
\includegraphics*[width=5.0cm]{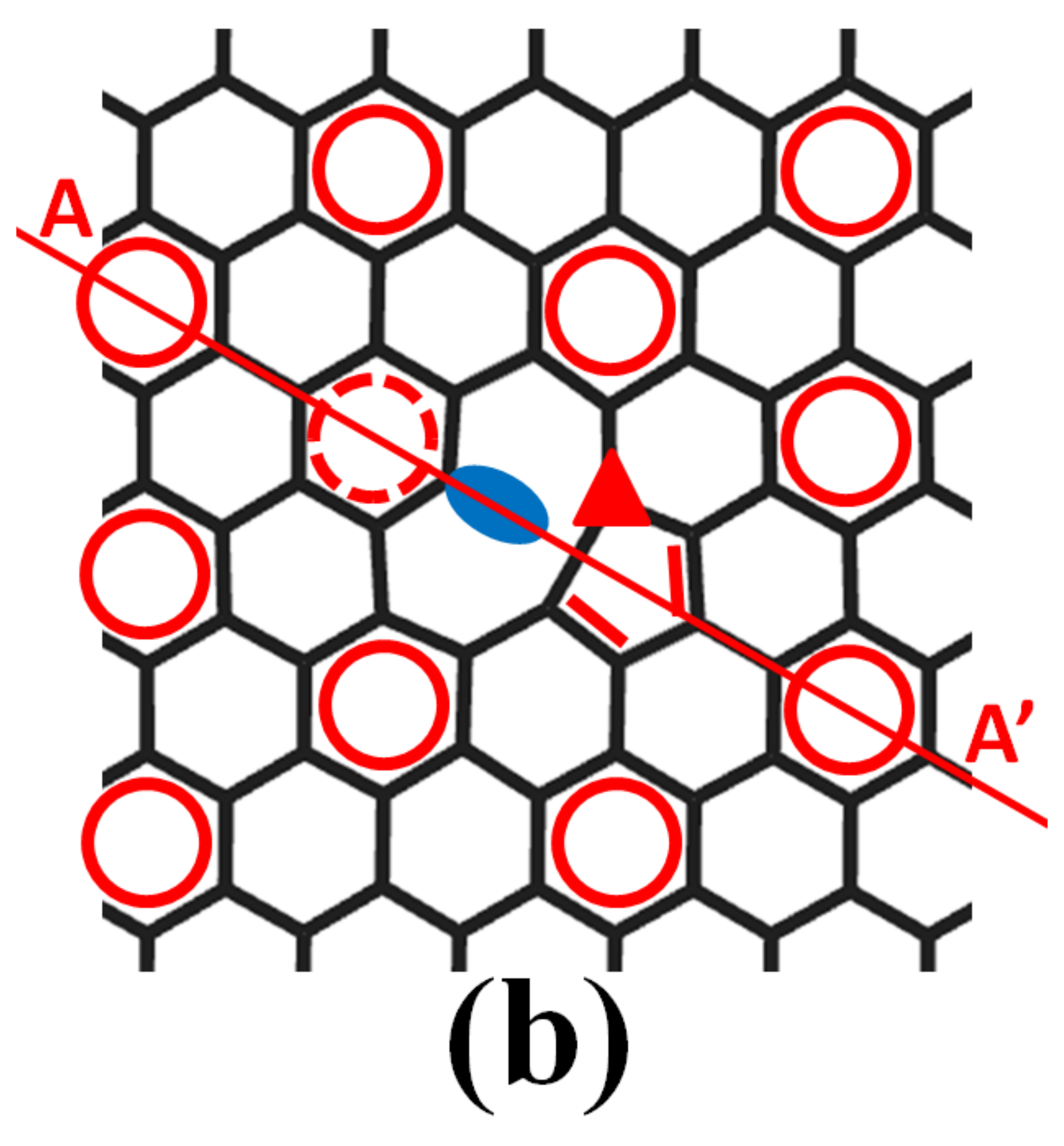}
\end{minipage}
\vspace*{0.2cm}
\hspace*{0.2cm}
\begin{minipage}[t]{5.0cm}
\includegraphics*[width=5.0cm]{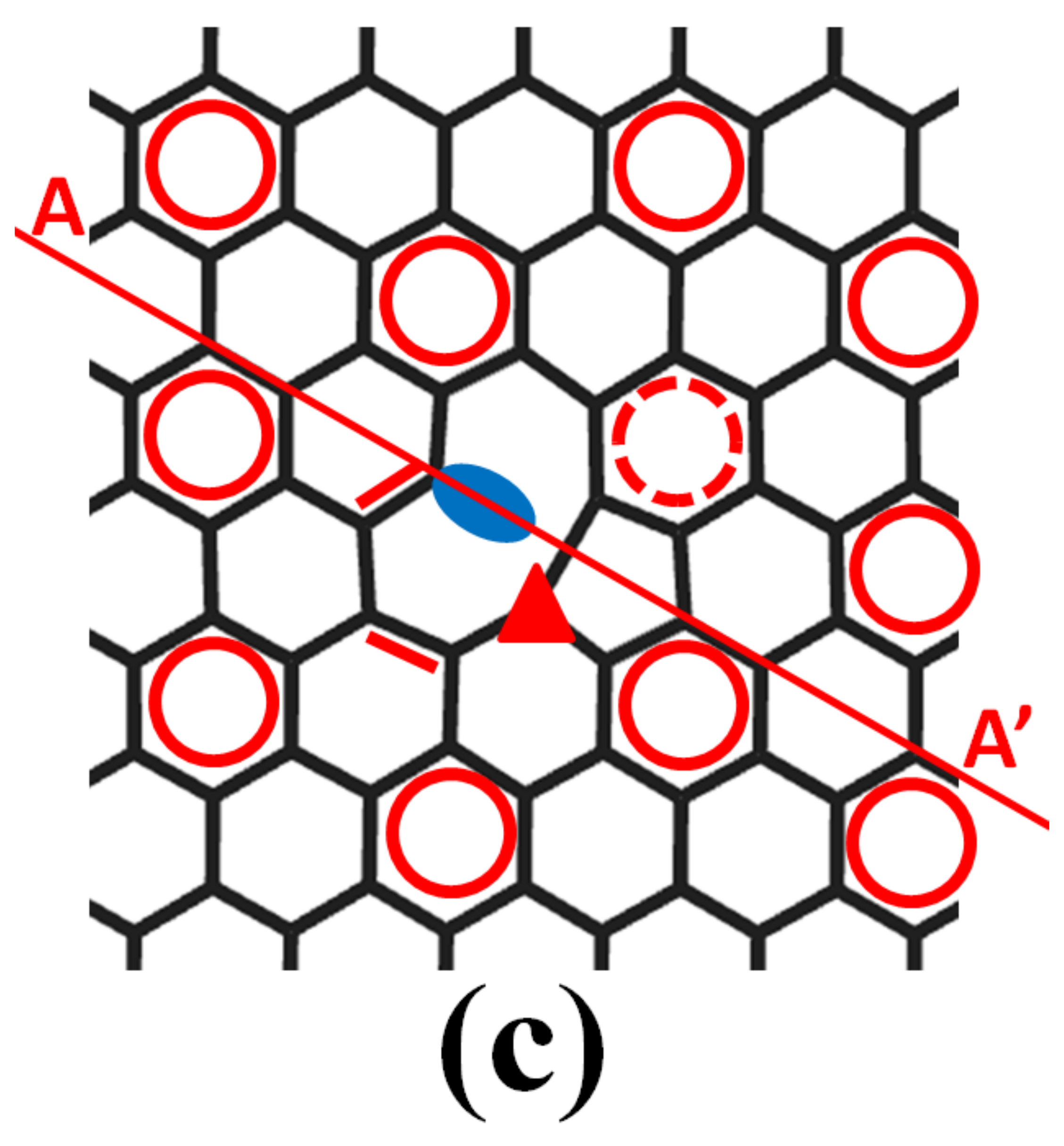}
\end{minipage}
\vspace*{0.2cm}
\hspace*{0.2cm}
\begin{minipage}[t]{5.0cm}
\includegraphics*[width=5.0cm]{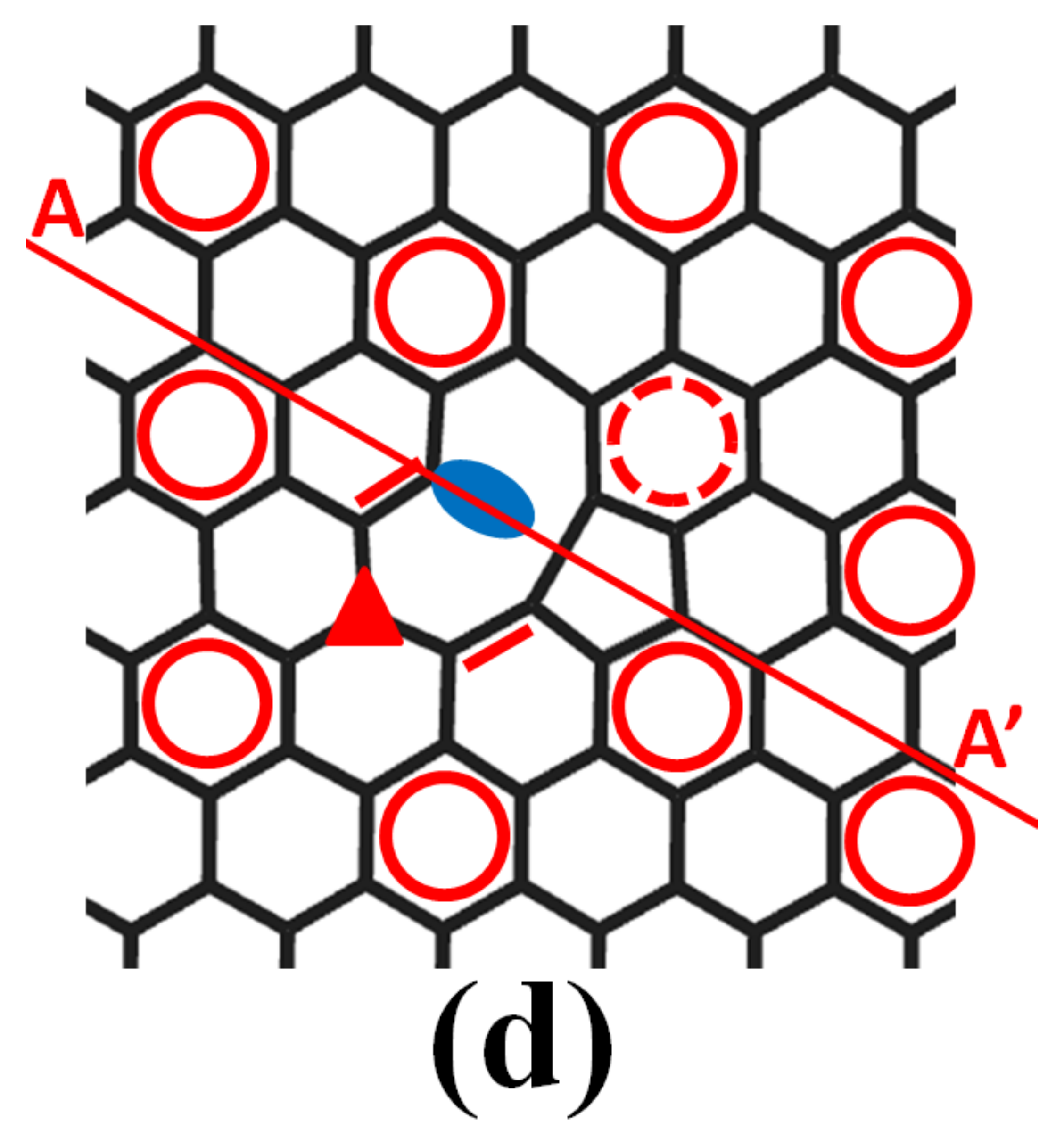}
\end{minipage}
\vspace*{0.2cm}
\hspace*{0.2cm}
\begin{minipage}[t]{5.0cm}
\includegraphics*[width=5.0cm]{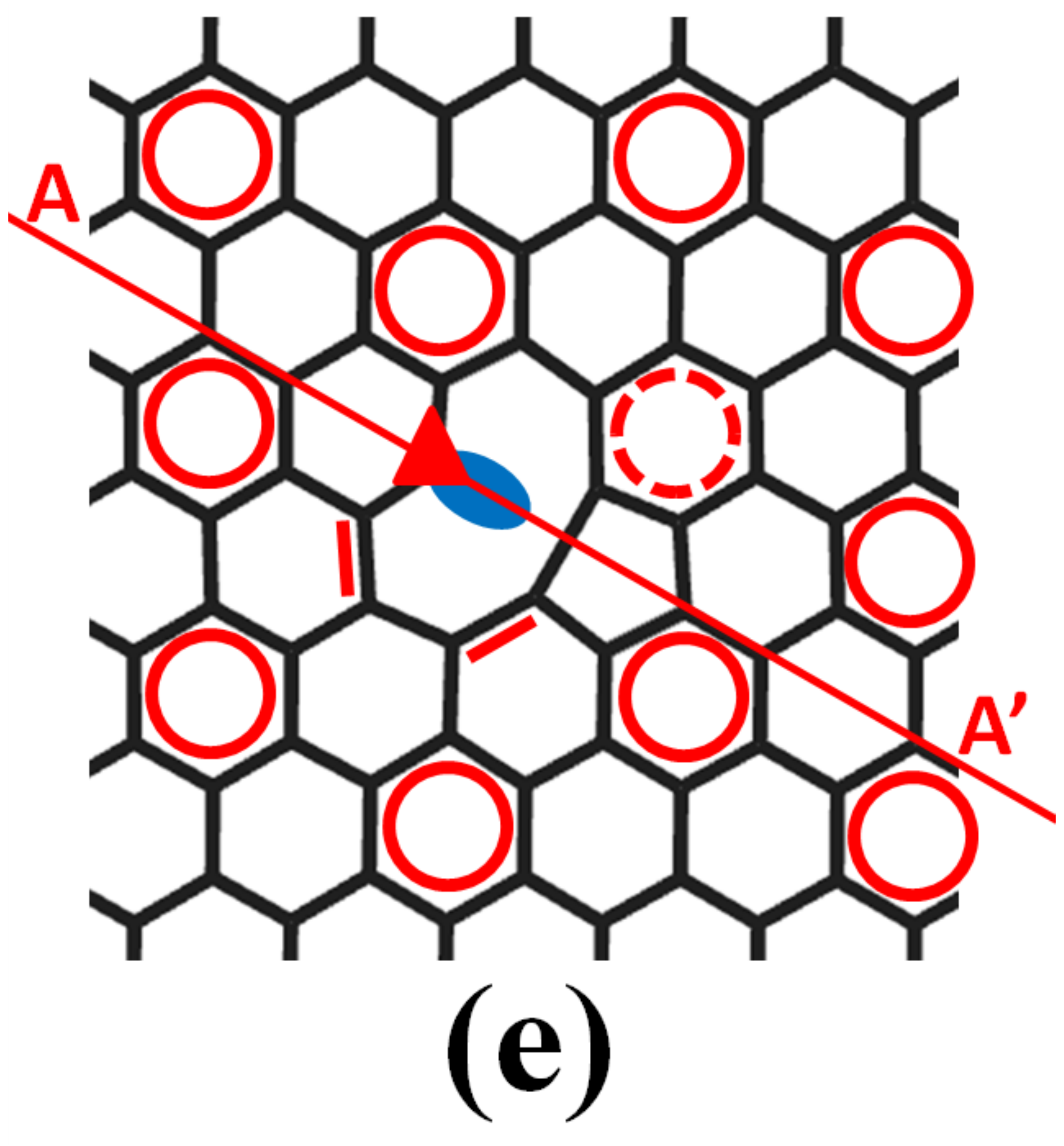}
\end{minipage}
\caption{\label{fig:4} (Color online) Five possible ways of arranging Clar sextets and double bonds around MV. The solid line AA$^{\prime}$ shows the mirror symmetry axis. The dashed circle stands for a pseudo-Clar sextet. The lobe and triangle symbols stand for dangling $\sigma$ and $\pi$ orbitals, respectively. }
\end{center}
\end{figure*}

Figure~\ref{fig:4} shows five possible ways of arranging Clar sextets~\cite{Clar72,Wassmann2010jacs,Fujii2012} and double bonds in the graphene network with MV. As the system has a mirror symmetry with respect to the line AA$^{\prime}$, there is another counter part connected by the mirror symmetry for each pattern shown in Figs.~\ref{fig:4}(b),~\ref{fig:4}(c),~\ref{fig:4}(d), and~\ref{fig:4}(e).  In these figures, a blue lobe denotes a singly occupied dangling $\sigma$ orbital attached to C1.  The dashed circle means that the Clar sextet is somewhat distorted (called pseudo Clar sextet hereafter) and the triangle shows the atom with a dangling $\pi$ orbital.  We can see that the sites with the triangle correspond to those with partially filled sharp $\pi$ DOS which are bright in the STM image with a bias voltage of -0.2 V.  Moreover, the triangles exist at C5 and C5$^{\prime}$ in both Figs.~\ref{fig:4}(b) and \ref{fig:4}(c) and the weight of the configuration, Figs.~\ref{fig:4}(d) and \ref{fig:4}(e), will be small because of the lower probability for the double-bond formation at the elongated C4$^{\prime}$-C5$^{\prime}$ bond (1.49 \AA).  These observations are consistent with the strongly highlighted spots in the STM image.

The dangling bonds discussed here are the consequence of a given atomic structure.  Therefore, once a new chemical bond with another atom or molecule is formed at a given dangling bond, both the electronic and the atomic structures are significantly modified.  However, we expect that the presence of a dangling bond will contribute to the lowering of the activation barrier (if any) when a foreign atom or molecule approaches the atom with a dangling bond.  In this sense, the distribution of dangling bonds predicted by the local bond picture gives us useful information about the chemical reactivity.
The Clar sextet patterns of Fig.~\ref{fig:4} imply that although a dangling bond may exist at different sites, there is no possibility of simultaneous existence of more than one dangling bond.  This means that although both C5 and C5$^{\prime}$ sites are bright in the STM image for a bias voltage of -0.2 V, they are not simultaneously chemically active.  Because of this, a diatomic molecule like O$_2$ will approach the MV with one oxygen atom heading for either C5 or C5$^{\prime}$.  More detailed analysis on the oxygen molecule adsorption at MV will be presented in a future publication.

If a C1 atom is hydrogenated, simply the blue lobe is doubly occupied to form a lone pair and can be removed from Fig.~\ref{fig:4}.  Nothing happens in the $\pi$ states.

\subsubsection{N doping at MV}
\label{sec:3B2}
\paragraph{Band calculations}
Now, we discuss the electronic structures of N-graphene with MV. The results presented below correspond to the most stable configurations of substitutional N dopants near MV.

\begin{figure}[htbp!]
\begin{center}
\hspace*{3.5cm}
\begin{minipage}[t]{10cm}
\includegraphics*[width=10cm]{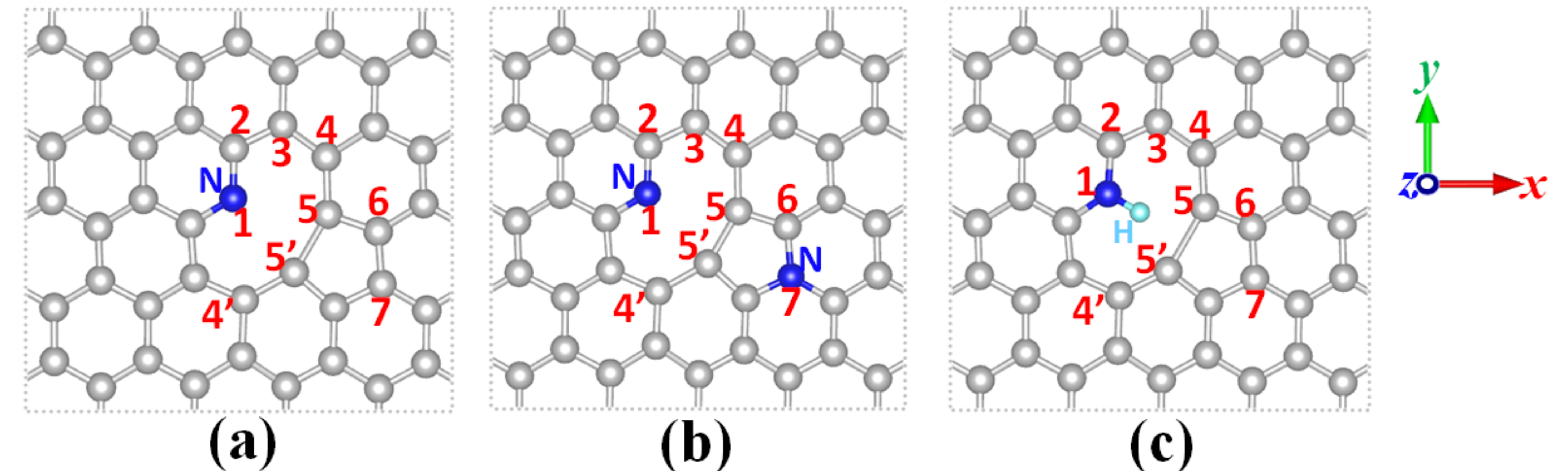}
\end{minipage}
\vspace*{0.2cm}
\hspace*{2.5cm}
\begin{minipage}[t]{3cm}
\includegraphics*[width=3cm]{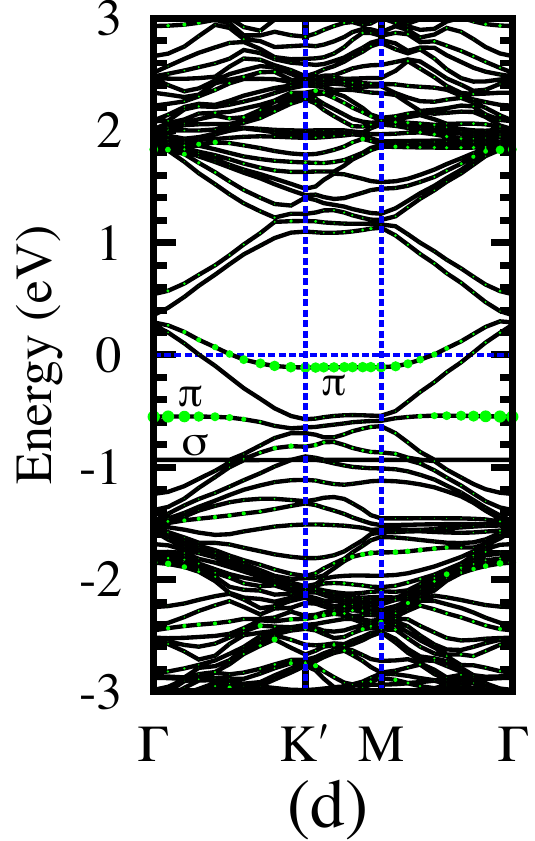}
\end{minipage}
\vspace*{0.2cm}
\begin{minipage}[t]{3.0cm}
\includegraphics*[width=3.0cm]{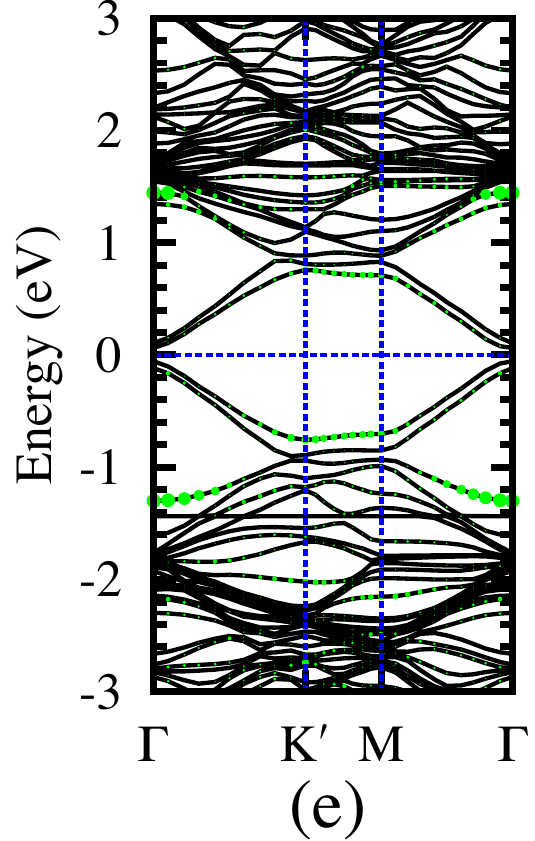}
\end{minipage}
\vspace*{0.2cm}
\begin{minipage}[t]{3.0cm}
\includegraphics*[width=3cm]{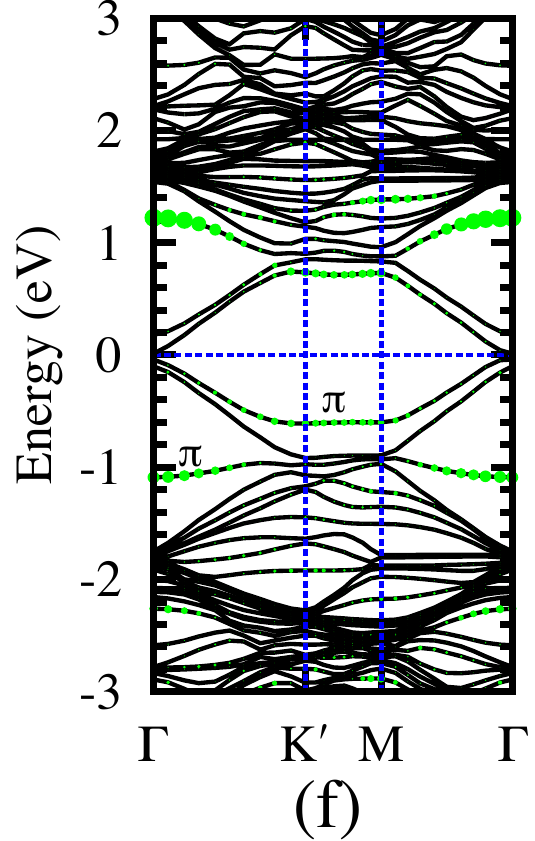}
\end{minipage}
\vspace*{0.2cm}
\caption{\label{fig:5} (Color online) The atomic structures of (a) a pyridinelike N (i.e., N$_\mathrm{C1}$) at MV, (b) a pyridinelike N (i.e., N$_\mathrm{C1}$) plus a m-graphitelike N (i.e., N$_\mathrm{C7}$) at MV, and (c) a pyridiniumlike N at H-MV.  The band structures of N-graphene with MV: (d)  N$_\mathrm{C1}$ at MV, (e)  N$_\mathrm{C1}$ + N$_\mathrm{C7}$ at MV, and (f) N$_\mathrm{C1}$ at H-MV. Panels (d), (e), and (f) show the fat bands (green solid points) derived from N$_\mathrm{C1}$ 2$p_z$ state.}
\end{center}
\end{figure}

The band structures of graphene doped with one and two substitutional N atoms at MV are presented in Figs.~\ref{fig:5}(d) and~\ref{fig:5}(e), respectively.
When one of  three C atoms next to the vacancy site (e.g., C1) is substituted by N to form a pyridinelike N [see Fig.~\ref{fig:5}(a)], the unoccupied minority spin dangling $\sigma$ bond state is pulled down below  $E_{\mathrm{F}}$ to make the $\sigma$ state nonmagnetic similarly to the case of H-MV.  The C5-C5$^{\prime}$ bond length is reduced by 0.18 \AA.  However unlike H-MV, N stays in the plane and the $\sigma$ dangling bond state is located in the bulk $\pi$ band region to form a dispersionless sharp level rather close to $E_{\mathrm{F}}$. Except these differences in the $\sigma$ states, the hydrogenated C1 and the pyridinelike N behave quite similarly for $\pi$ states and the filling of $\pi$ bands is basically the same as that of MV. Interestingly, the bonding $\pi$ bands are completely filled by two N dopants (i.e., N$_\mathrm{C1}$+N$_\mathrm{C7}$).
For H-MV, N doping [pyridiniumlike N, N$_\mathrm{C1}$, as shown in Fig.~\ref{fig:5}(c)] fills the bonding $\pi$ bands similarly to the two N dopants at MV.

\begin{figure*}[htbp!]
\begin{center}
\begin{minipage}[t]{8.0cm}
\includegraphics*[width=8.0cm]{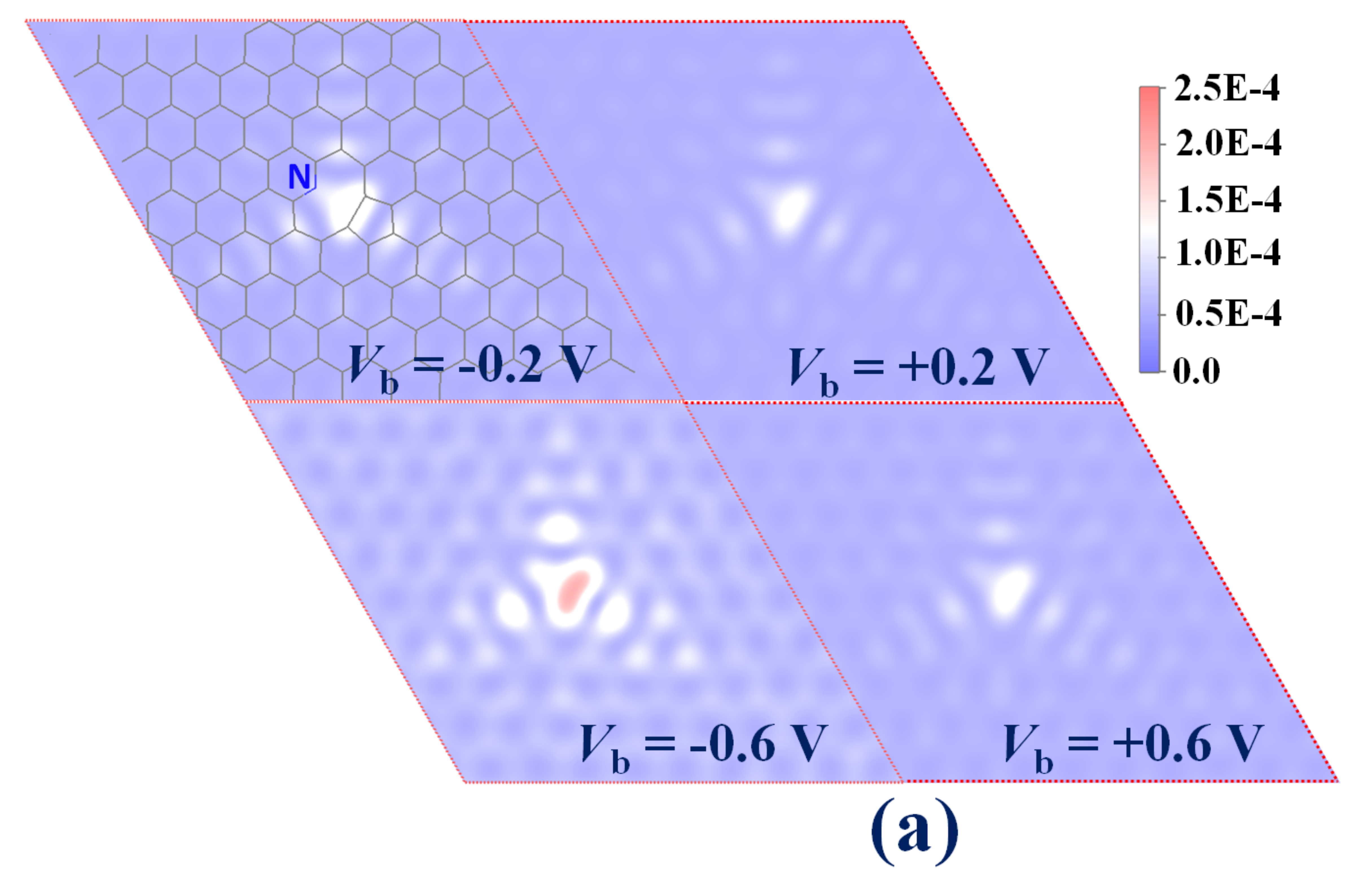}
\end{minipage}
\vspace*{0.2cm}
\begin{minipage}[t]{8.0cm}
\includegraphics*[width=8.0cm]{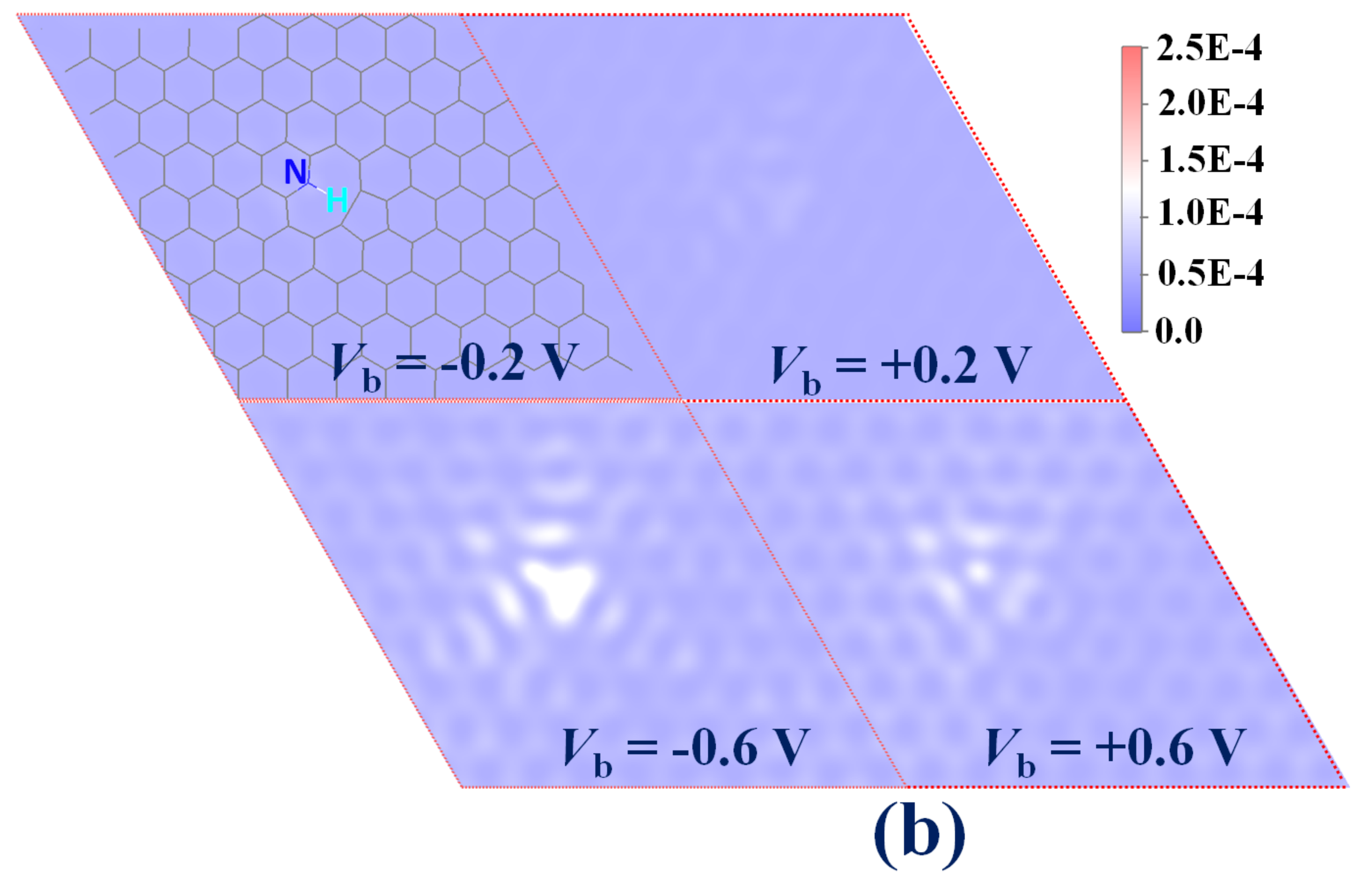}
\end{minipage}
\caption{\label{fig:6} (Color online) Simulated STM images for (a) pyridinelike N and (b) pyridiniumlike N at MV under different bias voltages indicated in each panel and with a sample-tip distance of $d=2$ \AA.}
\end{center}
\end{figure*}

While the PDOSs corresponding to the above three band structures are given in Fig.S1 in the Supplemental Material,~\cite{suppm} we show the simulated STM images for N-doped MV and its hydrogen terminated case in Figs.~\ref{fig:6}(a) and~\ref{fig:6}(b). For the N-doped MV, C5 and C7 atoms are bright for $V_b=-0.2$ V. The problems of $\pi$-state spin polarization in this case are discussed in Sec. \ref{sec:3}E.  On the other hand, the STM images for pyridiniumlike N at MV under  $V_b=-0.2$ eV and  $V_b=+0.2$ eV show much lower contrast reflecting the band structure of Fig.~\ref{fig:5}(f).  The situation is the same for the two N dopant case of Figs.~\ref{fig:5}(b) and \ref{fig:5}(e).

\begin{figure}[htbp!]
\begin{center}
\begin{minipage}[t]{5.0cm}
\includegraphics*[width=5.0cm]{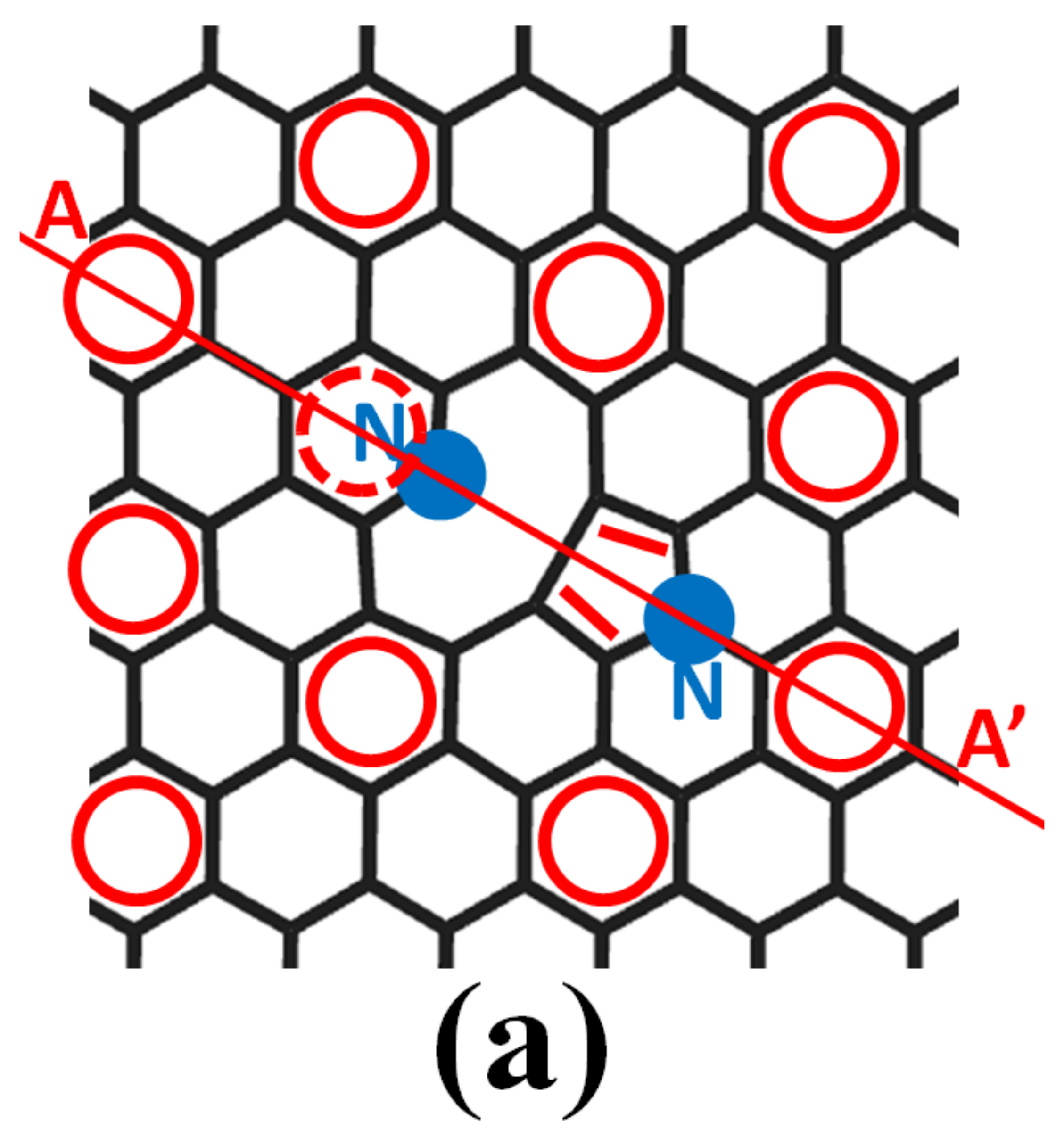}
\end{minipage}
\hspace*{0.5cm}
\begin{minipage}[t]{5.0cm}
\includegraphics*[width=5.0cm]{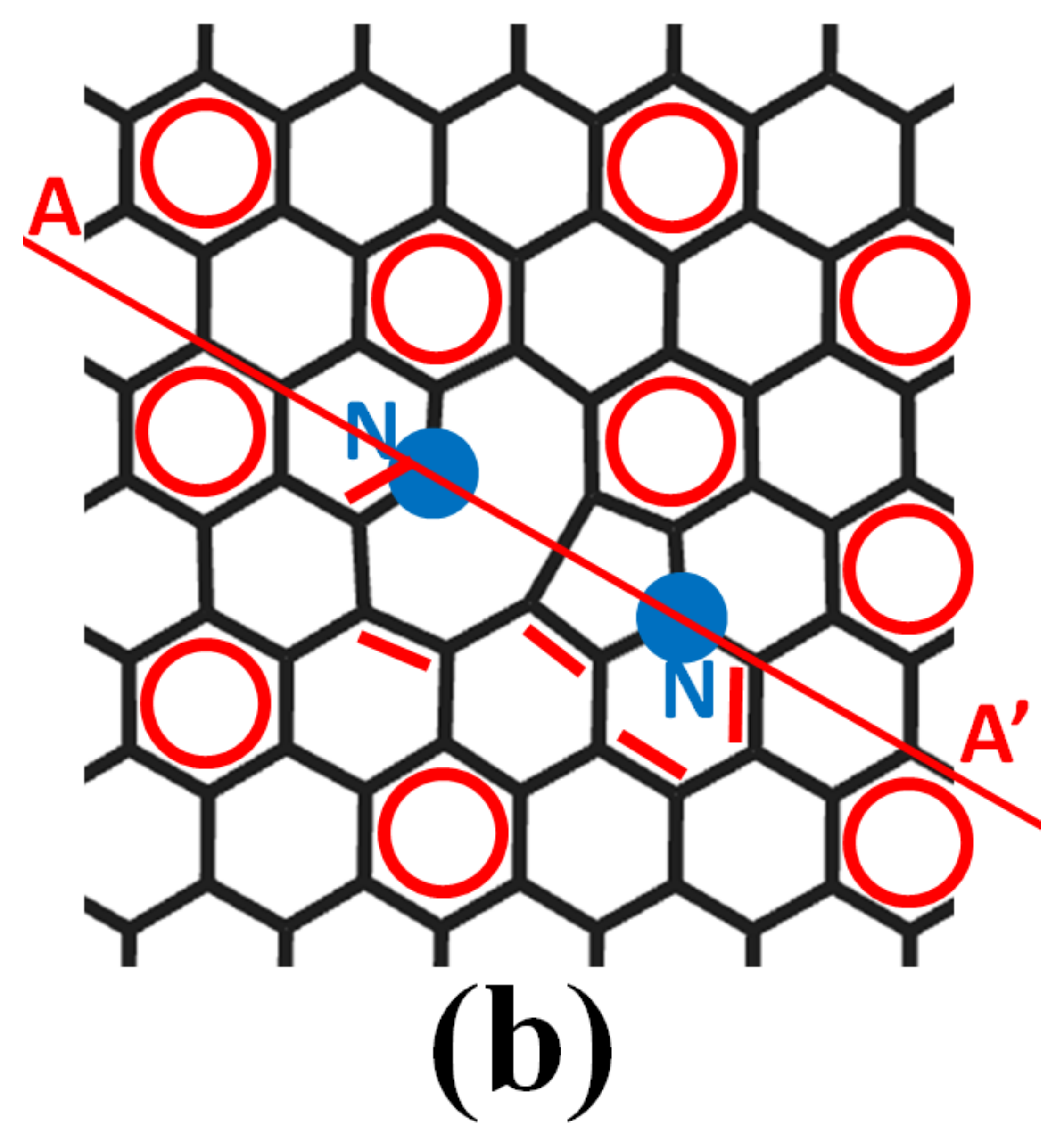}
\end{minipage}
\caption{\label{fig:7} (Color online) Two possible ways of arranging Clar sextets and double bonds around two N substitutions (a pyridinelike N plus a m-graphitlike N) at MV. The solid line AA$^{\prime}$ shows the mirror symmetry axis. The dashed circle and triangle symbol stand for a pseudo-Clar sextet and a dangling $\pi$ orbital, respectively.}
\end{center}
\end{figure}

\begin{figure*}[htbp!]
\begin{center}
\begin{minipage}[t]{5.0cm}
\includegraphics*[width=5.0cm]{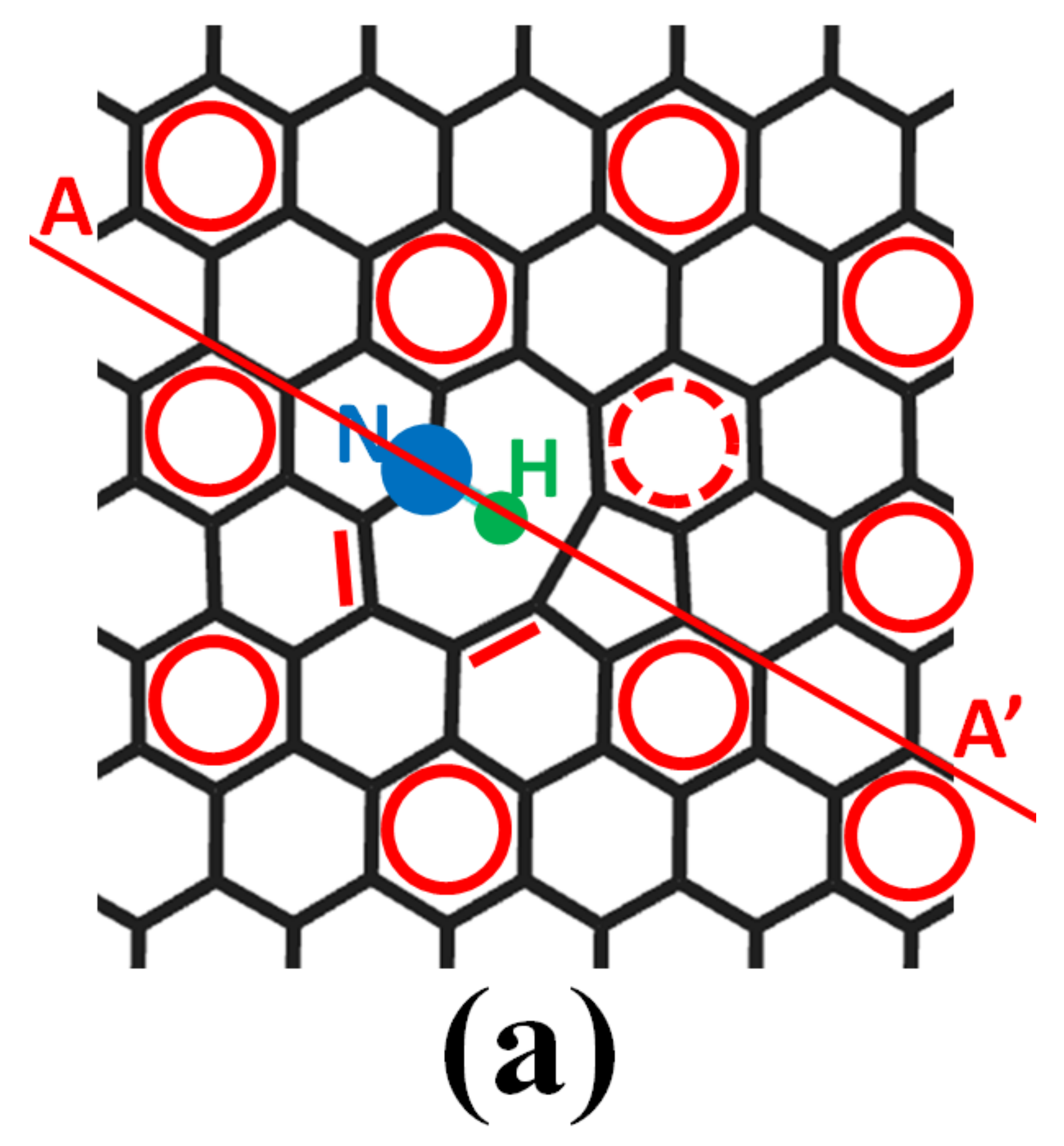}
\end{minipage}
\hspace*{0.3cm}
\begin{minipage}[t]{5.0cm}
\includegraphics*[width=5.0cm]{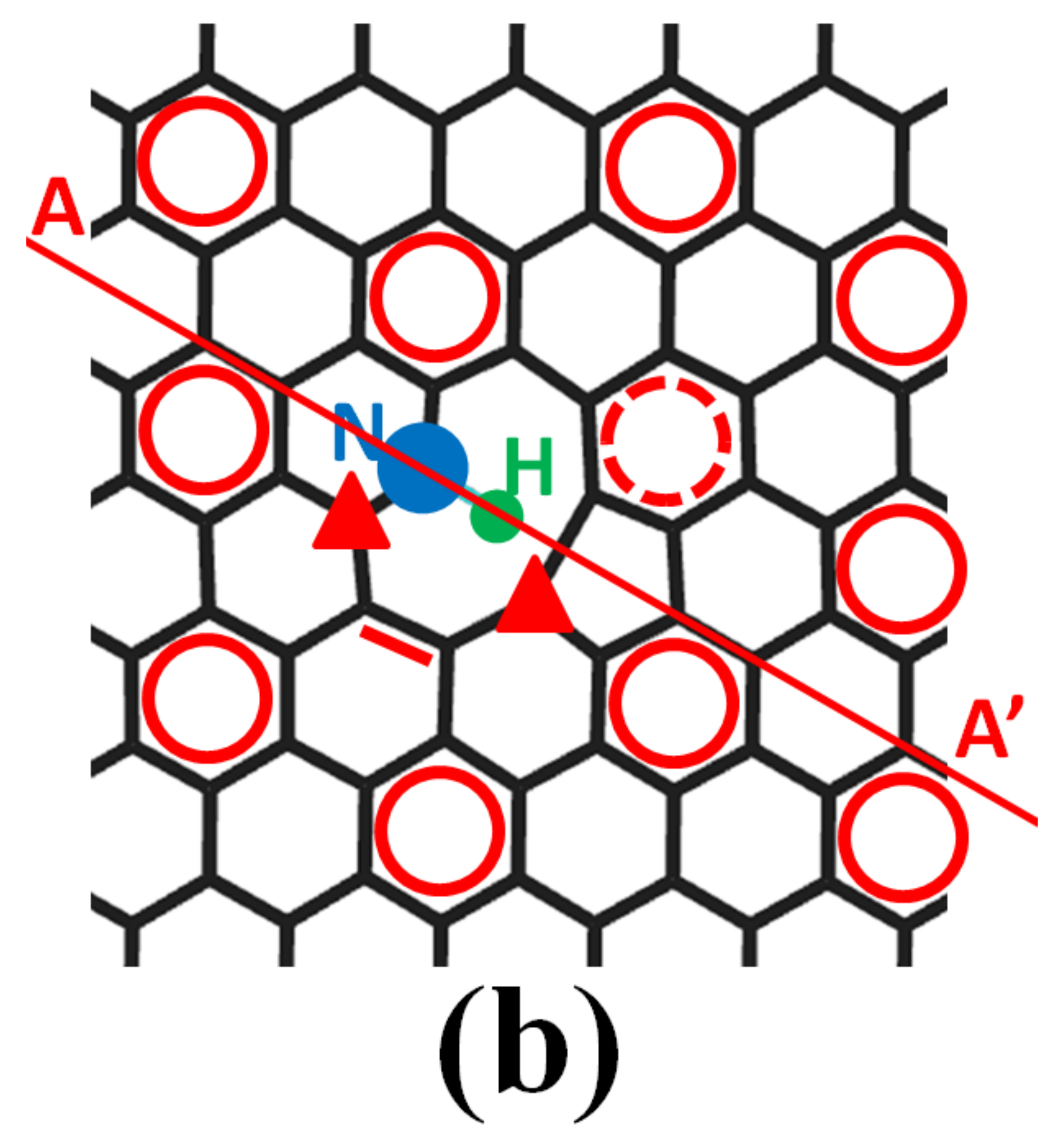}
\end{minipage}
\hspace*{0.3cm}
\begin{minipage}[t]{5.0cm}
\includegraphics*[width=5.0cm]{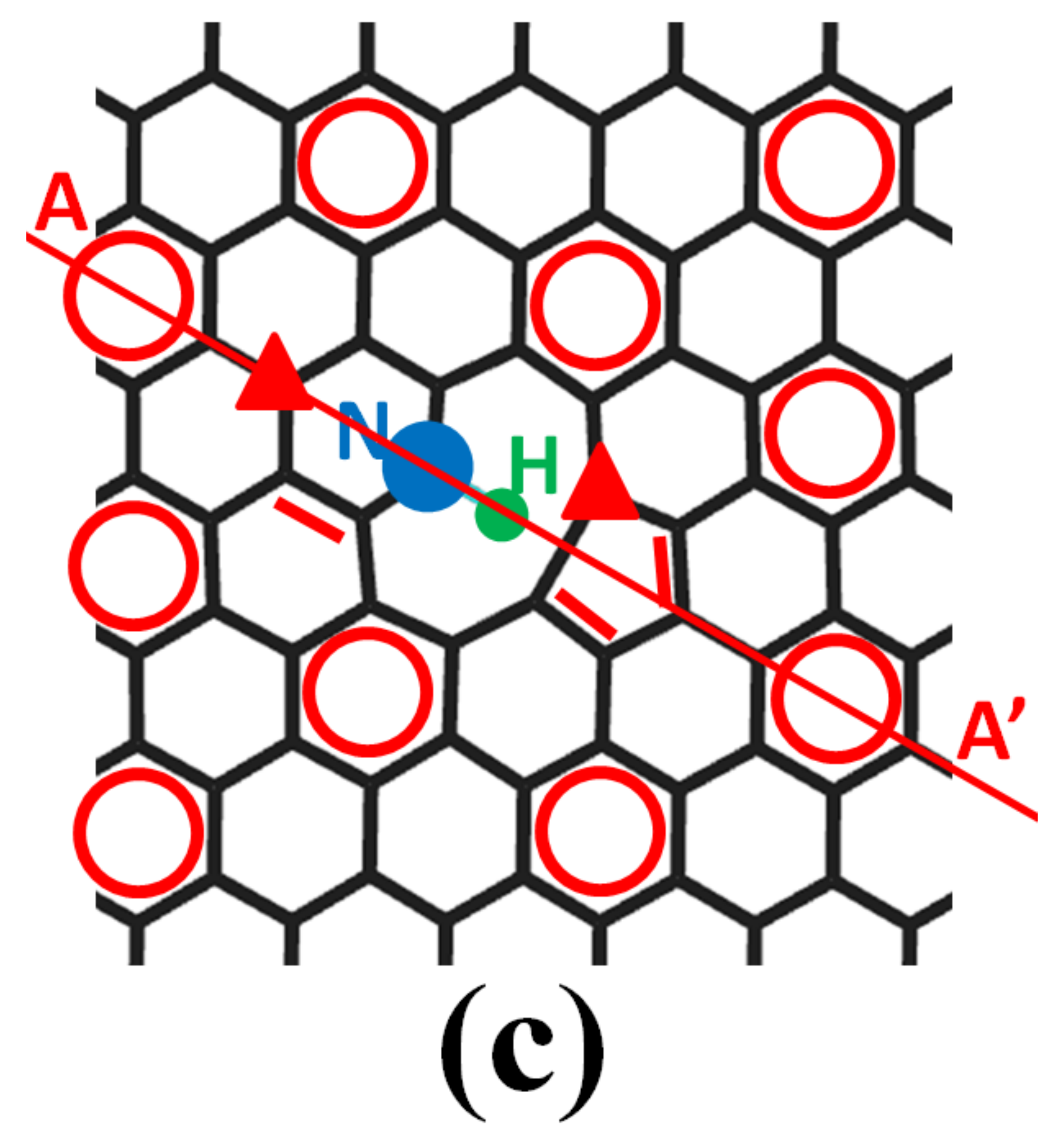}
\end{minipage}
\caption{\label{fig:8} (Color online) Three possible ways of arranging Clar sextets and double bonds around a pyridiniumlike N at MV. The solid line AA$^{\prime}$ shows the mirror symmetry axis. The dashed circle and triangle symbol stand for a pseudo-Clar sextet and a dangling $\pi$ orbital, respectively. }
\end{center}
\end{figure*}

\paragraph{Local bond picture}
How can the situations described above be explained with the local bond picture?  For N-doped MV, we simply remove the blue lobe in Fig.~\ref{fig:4} like for H-MV and there are no changes in the distribution of Clar sextet circles and double bonds. For the case of two N dopants shown in Fig.~\ref{fig:5}(b), we show two distributions of Clar sextets and double bonds in Fig.~\ref{fig:7}.  In both configurations, there are no dangling bonds being consistent with the band structure of Fig.~\ref{fig:5}(e).  However, compared with Fig.~\ref{fig:7}(a), one pseudo-Clar sextet circle is missing and several double bonds are fixed in Fig.~\ref{fig:7}(b).  Therefore, we expect that the configuration of Fig.~\ref{fig:7}(a) may have a larger weight in the ground state. The situation is slightly different for pyridiniumlike N at MV [Fig.~\ref{fig:5}(c)]. The configuration shown in Fig.~\ref{fig:8}(a) and its counterpart obtained by mirror symmetry do not have any dangling bonds being consistent with the band structure of Fig.~\ref{fig:5}(f).  However, C4$^{\prime}$-C5$^{\prime}$ distance is elongated (1.48 \AA) and the probability of double bond formation at this bond will be reduced.  On the other hand, the configurations of Figs.~\ref{fig:8}(b) and \ref{fig:8}(c) have two dangling bonds and one less (pseudo) Clar sextet. These arguments suggest that pyridiniumlike N at MV is less stable than the case of two N dopants. In fact, the defect $\pi$ states are located closer to $E_\mathrm{F}$ for pyridiniumlike N than for two N dopants.

\subsection{Divacancy and N doping}
\label{sec:3C}
\subsubsection{DV in undoped graphene}
\label{sec:3C1}
\begin{figure}[htbp!]
\begin{center}
\hspace*{0.20cm}
\begin{minipage}[t]{5.5cm}
\includegraphics*[width=5.5cm]{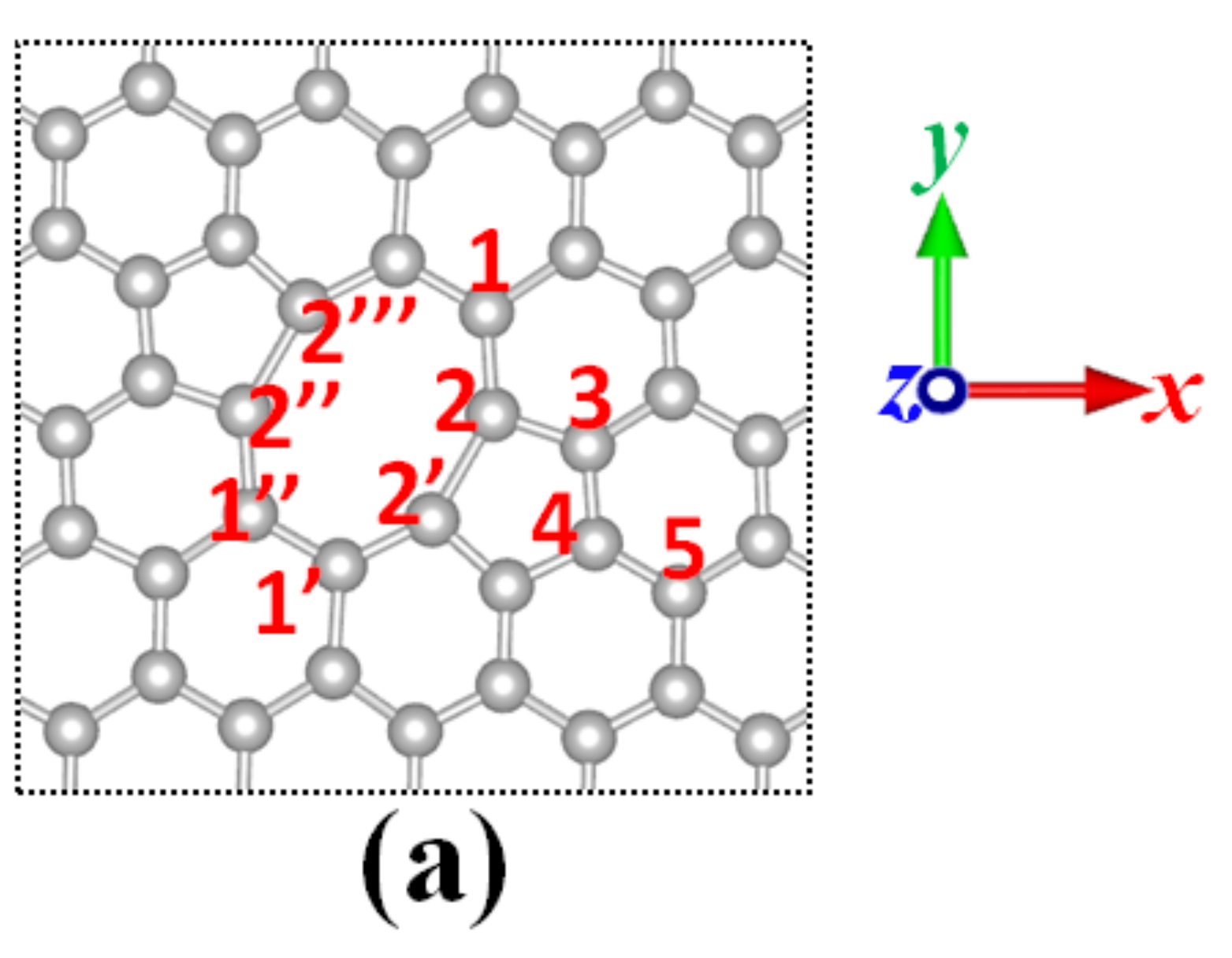}
\end{minipage}
\vspace*{.200cm}
\begin{minipage}[t]{3.4cm}
\includegraphics*[width=3.4cm]{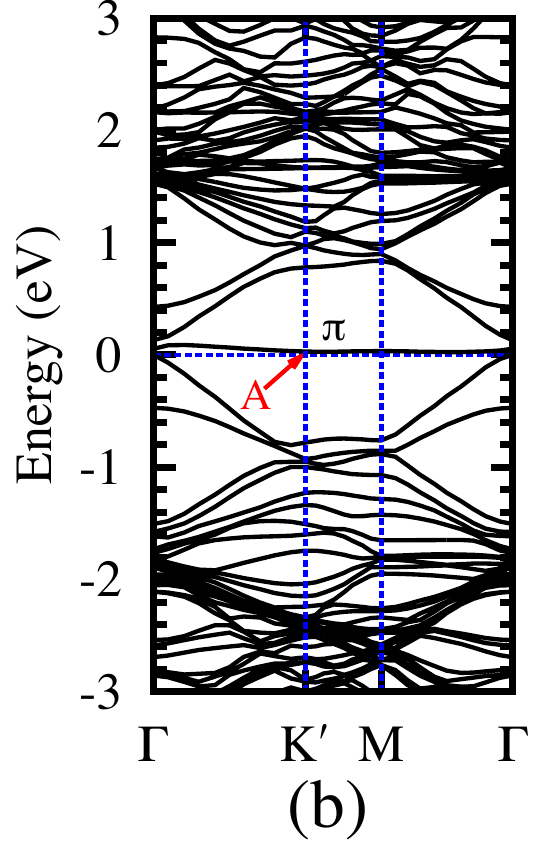}
\end{minipage}
\vspace*{0.20cm}
\begin{minipage}[t]{4.5cm}
\includegraphics*[width=4.5cm]{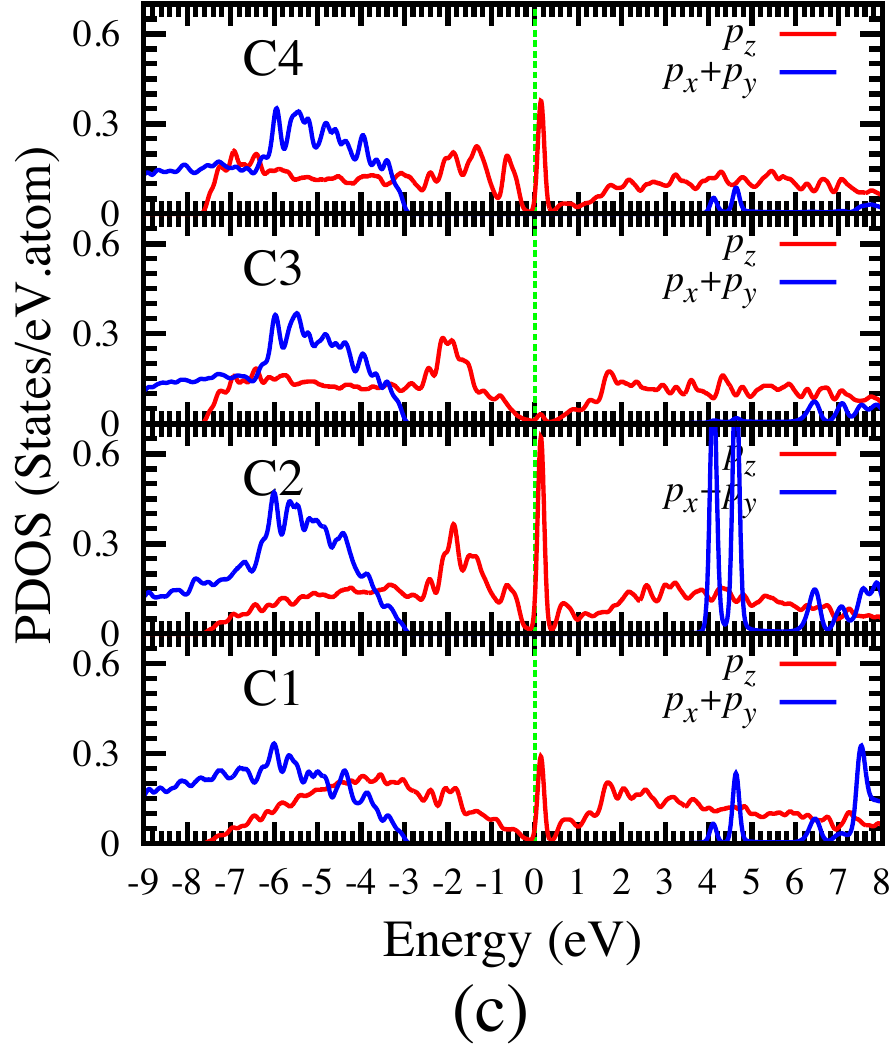}
\end{minipage}
\vspace*{.200cm}
\begin{minipage}[t]{8.0cm}
\includegraphics*[width=8.0cm]{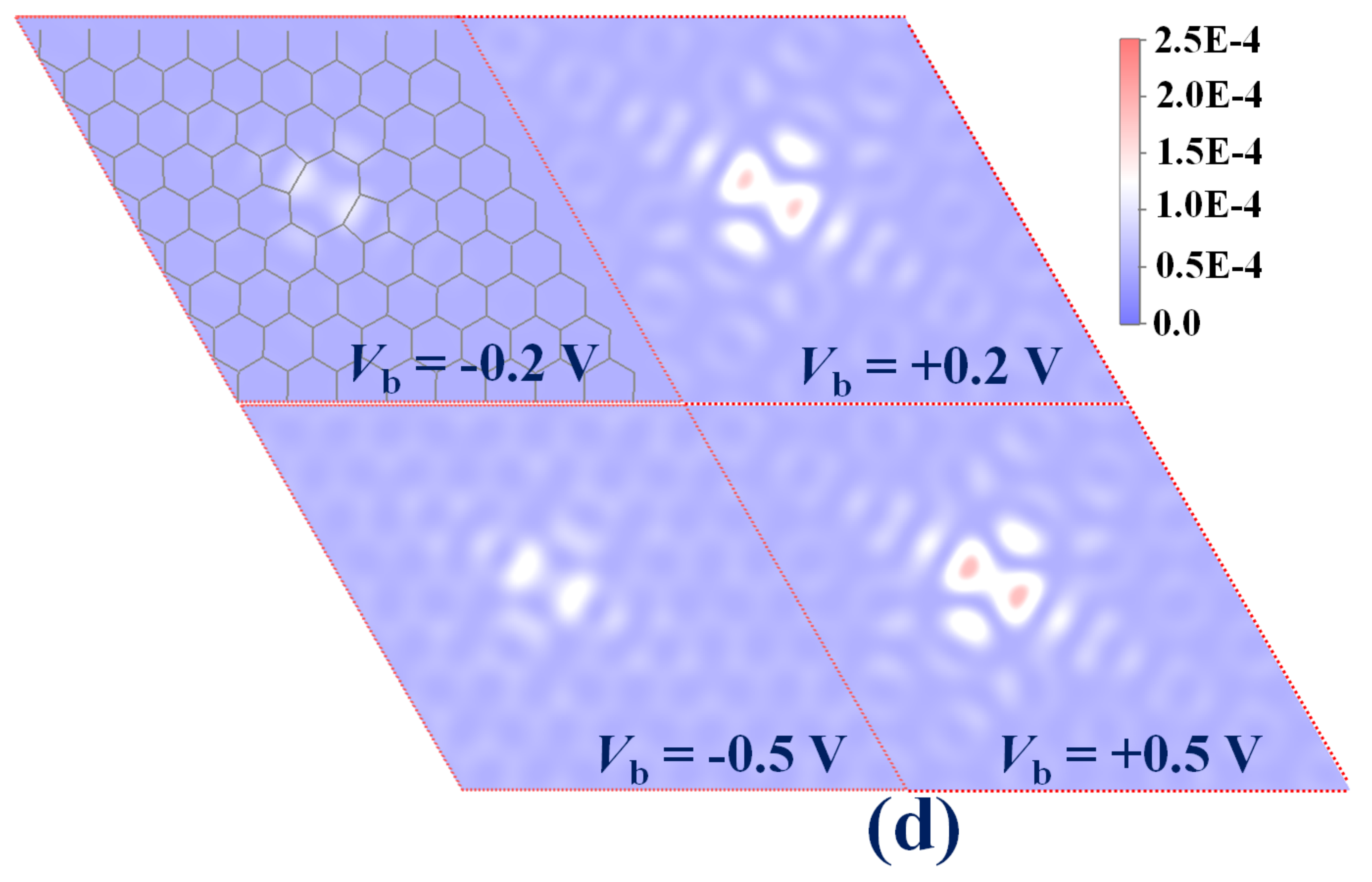}
\end{minipage}
\caption{\label{fig:9} (Color online) (a) The atomic structure of 5-8-5 DV. (b) The band structure of graphene with 5-8-5 DV. The zero of energy is set at  $E_{\mathrm{F}}$.
(c) The partial density of states (PDOS) for the 2\textit{p} orbitals of C atoms.  (d) Simulated STM images under different bias voltages indicated in each panel and with a sample-tip distance of $d=2$ \AA.}
\end{center}
\end{figure}

The ordinary configuration of DVs in graphene which contains two pentagons and one octagon [i.e., 5-8-5 pattern as shown in Fig.~\ref{fig:9}(a)] can reconstruct into more complex ones such as the 555-777 and 5555-6-7777 patterns.~\cite{Banhart2010,Hou2011ar} The stabilities of these patterns of DVs before and after N doping have been discussed in our previous paper.~\cite{Hou2011ar} Although the 555-777 DV is the most stable one among the different patterns of DVs, N doping can enhance the stability of 5-8-5 DV against the 555-777 one. Therefore, we treat only 5-8-5 DV in the present work.

\paragraph{Band calculations}
C2 and C2$^{\prime}$ atoms next to DV form a new bond with a length of 1.73 \AA~[see Fig.~\ref{fig:9}(a)].
The band structure and PDOS for graphene with 5-8-5 DV are presented in Figs.~\ref{fig:9}(b) and \ref{fig:9}(c). As for the $\sigma$ states, there are two sharp peaks in the PDOS of C2 at about 4 to 5 eV above $E_{\mathrm{F}}$.  These states have anti-bonding character between C2 and C2$^{\prime}$ of Fig.~\ref{fig:9}(a) and their bonding counterparts are located in the energy range of the occupied bulk $\sigma$ band to become resonance states.
Therefore four electrons are accommodated to the lower two defect resonance $\sigma$ states, so that the charge consistency is satisfied within $\sigma$ states. A rather subtle feature exists for $\pi$ states.  It is interesting to find that a quite flat defect $\pi$ band exists just above $E_{\mathrm{F}}$ and that no carriers are created in the bulk $\pi$ bands.
As C1, C2, and C4 belong to the same sublattice, the $p_z$ states of C1, C2, C4, and their equivalent ones are the dominant contribution to the flat defect $\pi$ band. In Ref.~\onlinecite{Dubois2011}, however, the authors found that $E_\mathrm{F}$ is below the Dirac point and crosses the $\pi$ bands of graphene. However such results were obtained by a smaller (i.e., $7\times 7$) supercell and a localized orbital basis.  The use of a larger supercell and a plane-wave basis makes our result more reliable.

We make two comments on this flat defect $\pi$ band.  First, the flatness of the band does not necessarily mean the defect state is very localized.  This defect state strongly hybridizes with one of the bulk $\pi$ bands, whose modification is a measure of the strength of hybridization.  Therefore, like in the case of MV, the tail of the defect $\pi$ state may be rather extended.  Second, the meaning of the presence of this flat defect $\pi$ state just above $E_\mathrm{F}$ can be understood in the following way.
The defect $\pi$ state just above $E_{\mathrm{F}}$ has the same symmetry as that of the bonding $\pi$ molecular orbital of the dimer removed from DV as suggested by the STM image with $V_b=+0.2$ V [see Fig.~\ref{fig:9}(d)].
The implication of the flat defect $\pi$ state being unoccupied is that the two $\pi$ electrons originally belonging to the dimer are also removed.
Therefore, the charge consistency is satisfied also within $\pi$ states without shifting the Fermi level with regard to the  bulk $\pi$ bands leading to no carrier formation. This is distinctly different from the case of MV where the  defect $\pi$ level is mostly below $E_{\mathrm{F}}$.

\begin{figure*}[htbp!]
\begin{center}
\begin{minipage}[t]{4.0cm}
\includegraphics*[width=4.0cm]{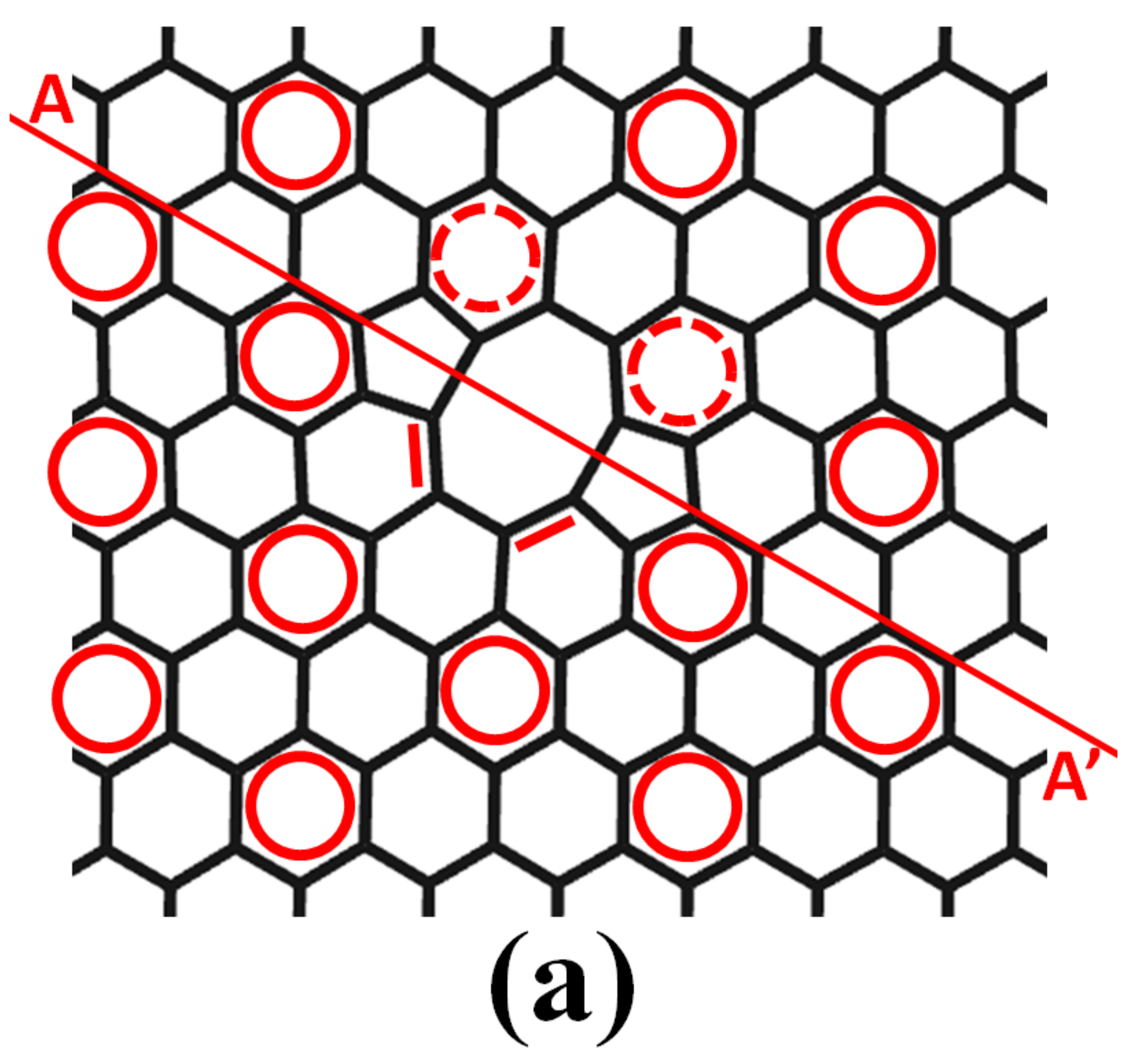}
\end{minipage}
\begin{minipage}[t]{4.0cm}
\includegraphics*[width=4.0cm]{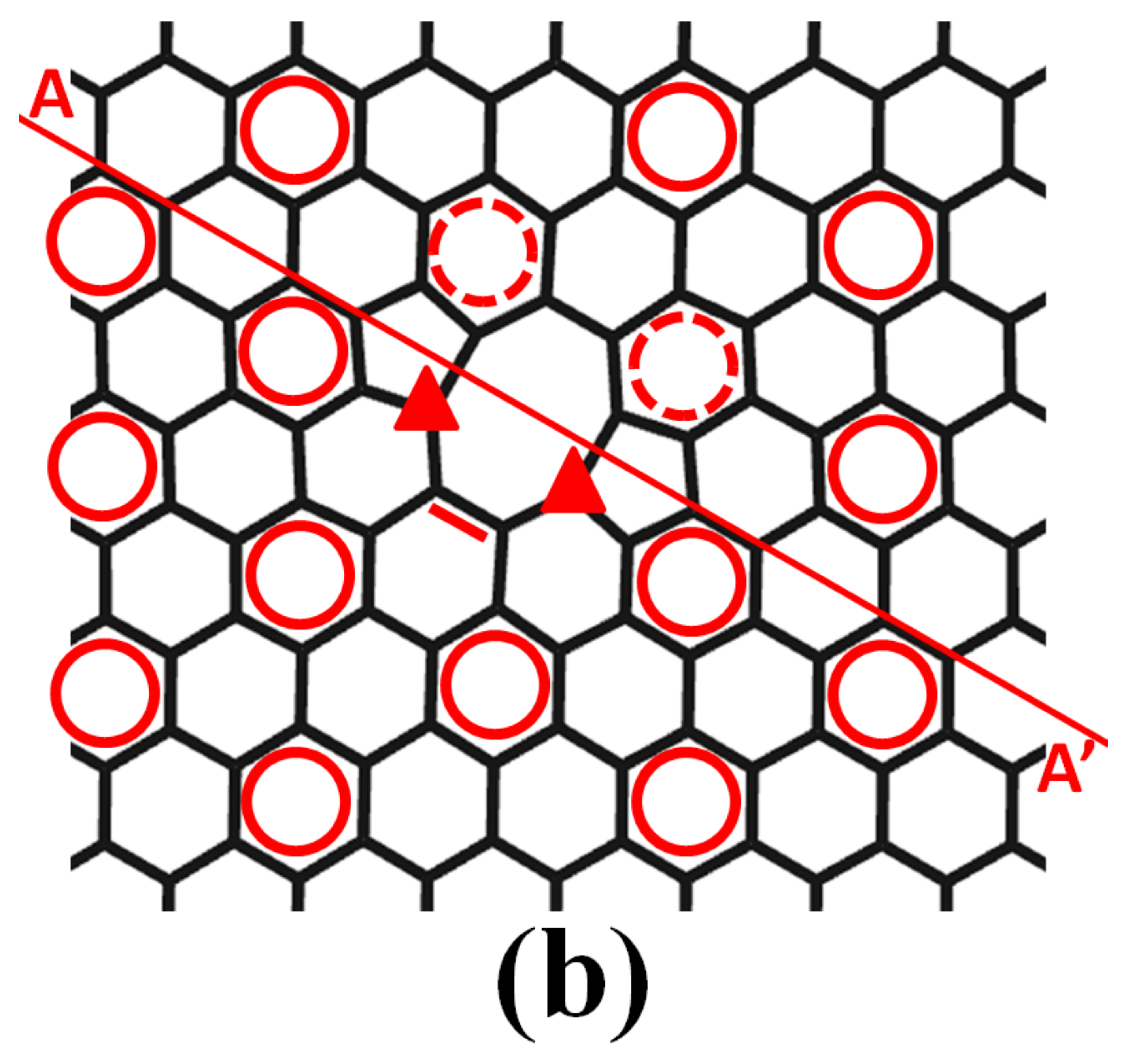}
\end{minipage}
\begin{minipage}[t]{4.0cm}
\includegraphics*[width=4.0cm]{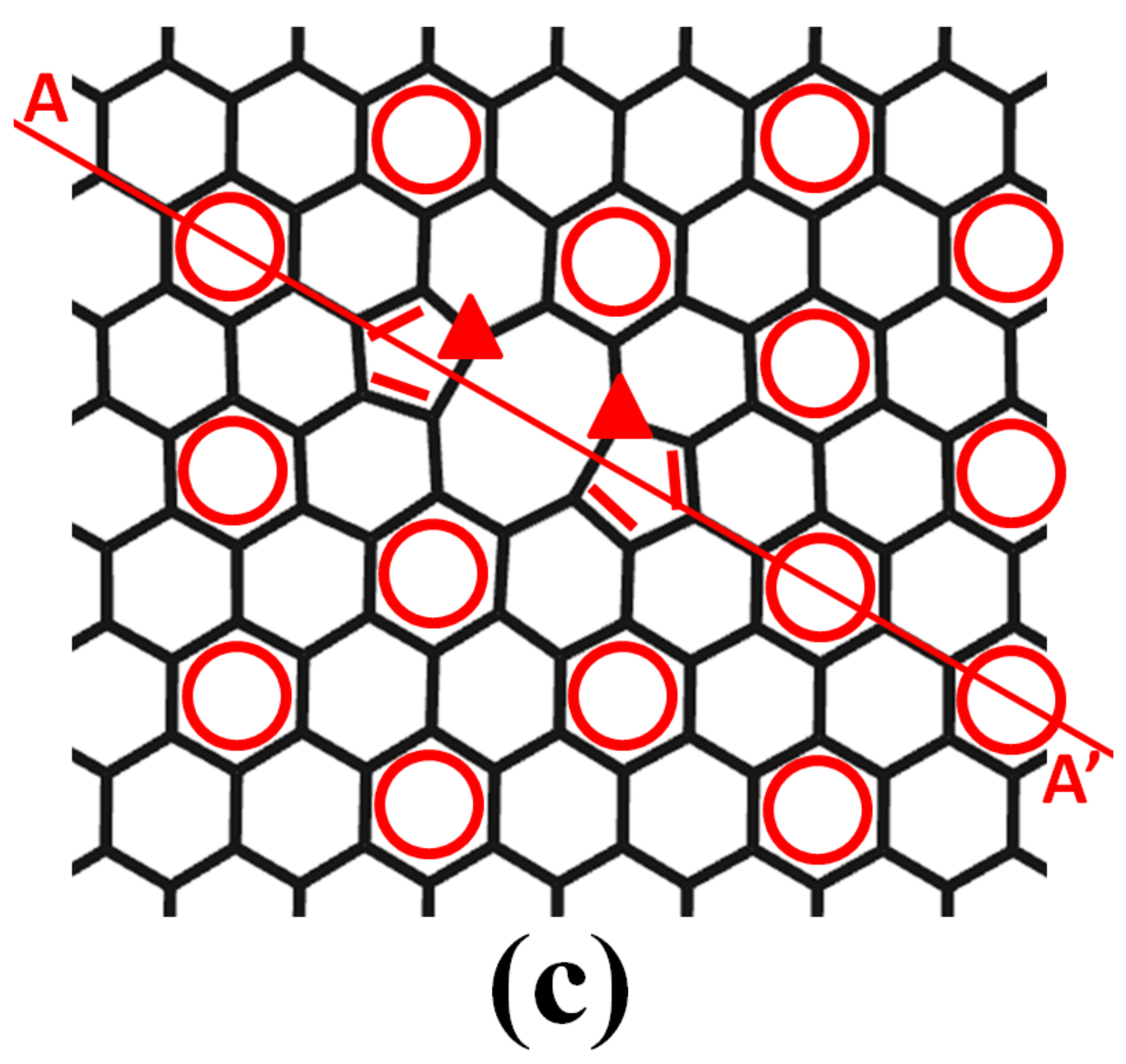}
\end{minipage}
\begin{minipage}[t]{4.0cm}
\includegraphics*[width=4.0cm]{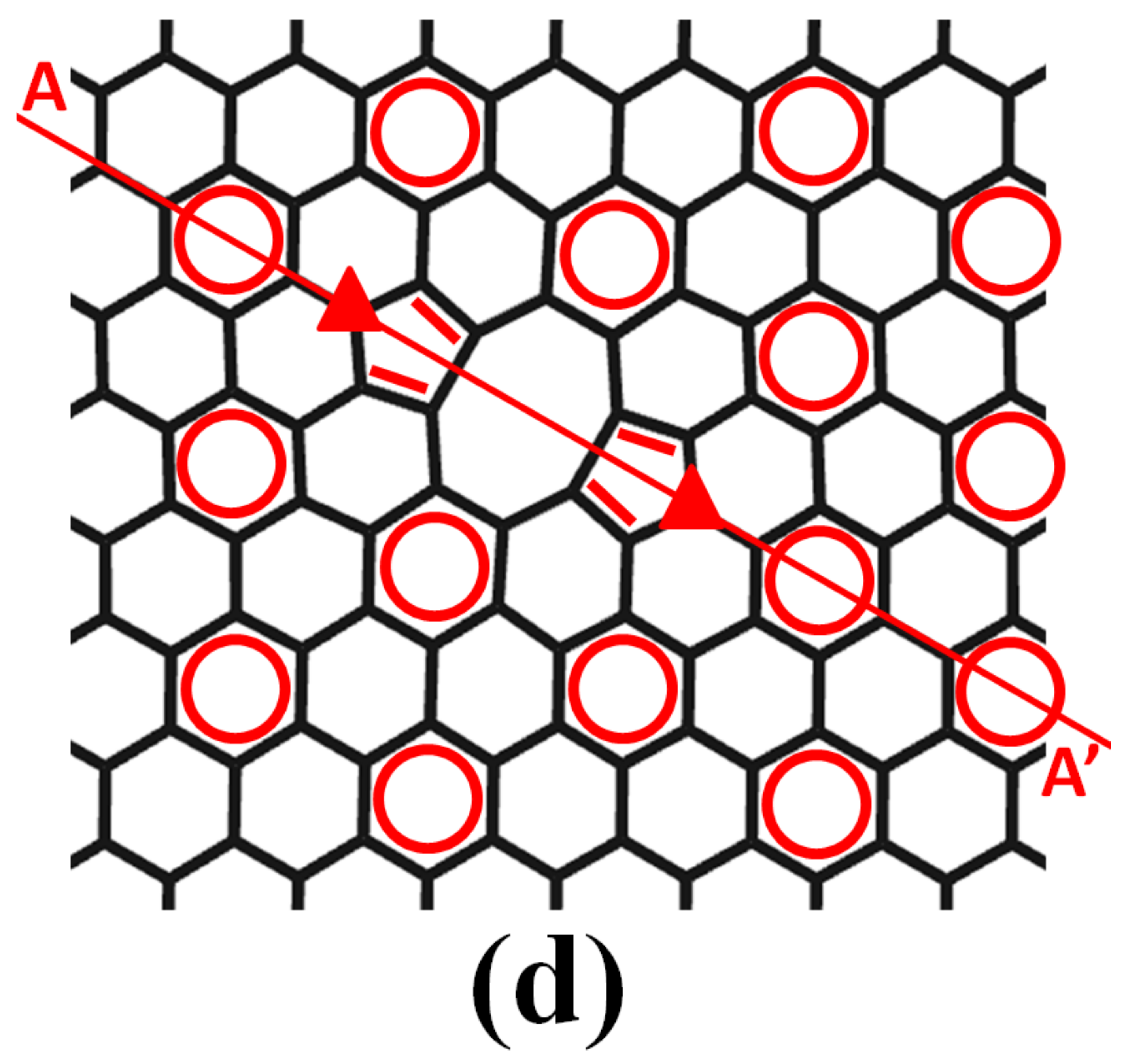}
\end{minipage}
\caption{\label{fig:10} (Color online) Four possible ways of arranging Clar sextets and double bonds around 5-8-5 DV. The solid line AA$^{\prime}$ shows the mirror symmetry axis. The dashed circle and triangle symbol stand for a pseudo-Clar sextet and a dangling $\pi$ orbital, respectively.}
\end{center}
\end{figure*}

Figure~\ref{fig:9}(d) shows the simulated STM images for some different bias voltages as indicated in each panel.  As the defect level is not occupied, the STM images with negative bias voltages have much weaker contrast than those with positive bias voltages.
\paragraph{Local bond picture}
Figure~\ref{fig:10} shows the four ways of Clar sextets and double bonds around 5-8-5 DV. For the pattern in Fig.~\ref{fig:10}(a), there is no dangling $\pi$ orbital by considering a double bonds for C1$^{\prime}$-C2$^{\prime}$ and C1$^{\prime\prime}$-C2$^{\prime\prime}$. However, as these bonds (about 1.47 \AA~in length) are elongated, the pattern may not be strongly stabilized.  On the other hand, configurations shown in Figs.~\ref{fig:10}(b)-(d) do not have these double bonds but have two sites with dangling $\pi$ states near the vacancy. Therefore we expect that these three patterns may also mix with the one in Fig.~\ref{fig:10}(a) and they may correspond to the presence of defect states just above the Fermi level.

\subsubsection{N doping at DV}
\label{sec:3C2}
\begin{figure}[htbp!]
\begin{center}
\hspace*{3.20cm}
\begin{minipage}[t]{12.0cm}
\includegraphics*[width=12.0cm]{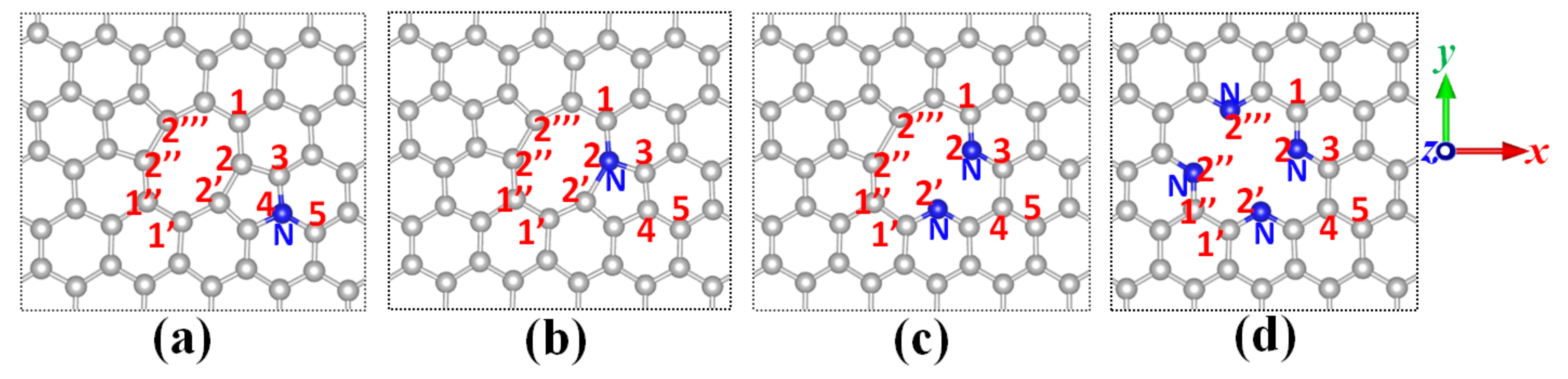}
\end{minipage}
\vspace*{.200cm}
\begin{minipage}[t]{12.0cm}
\includegraphics*[width=12.0cm]{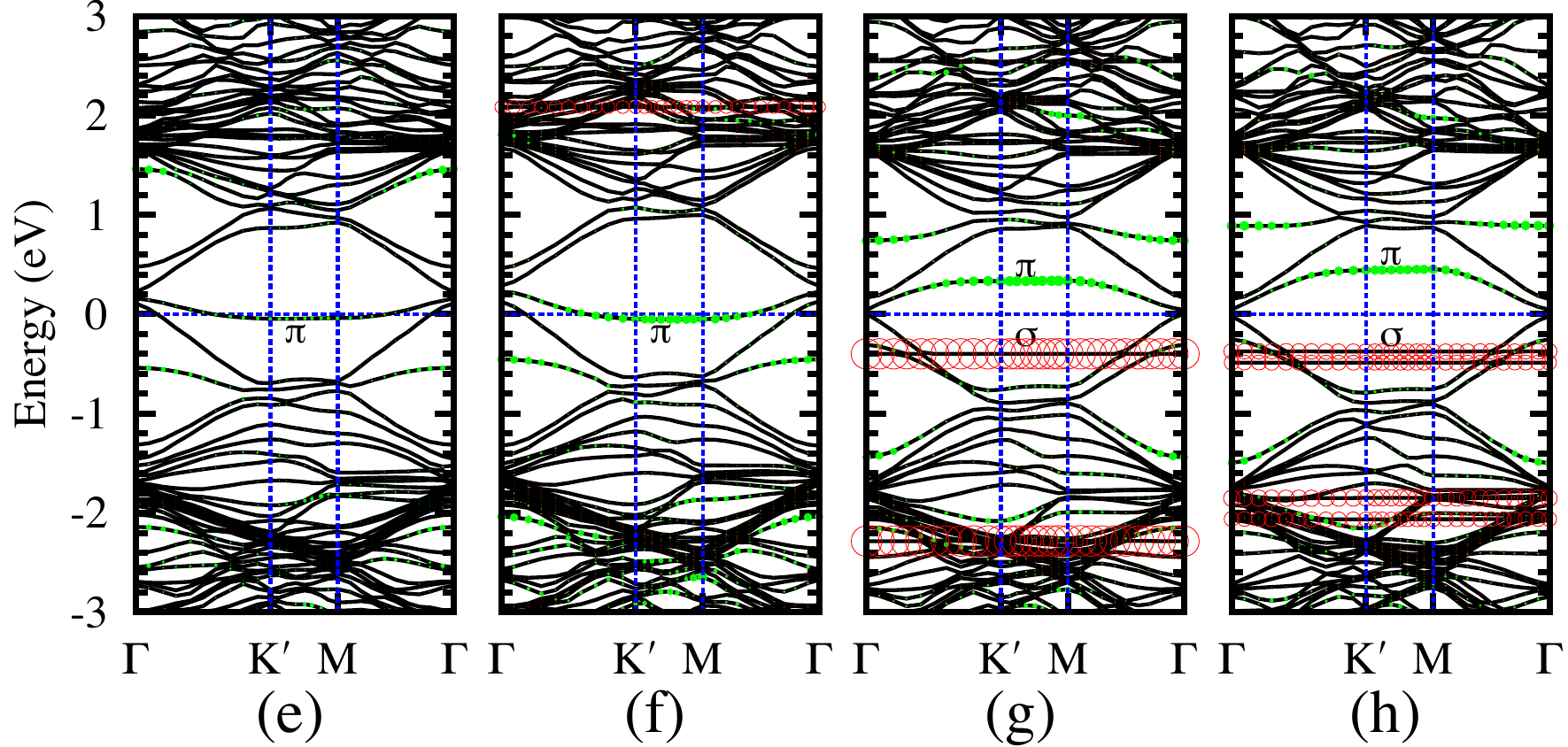}
\end{minipage}
\vspace*{0.20cm}
\begin{minipage}[t]{8.0cm}
\includegraphics*[width=8.0cm]{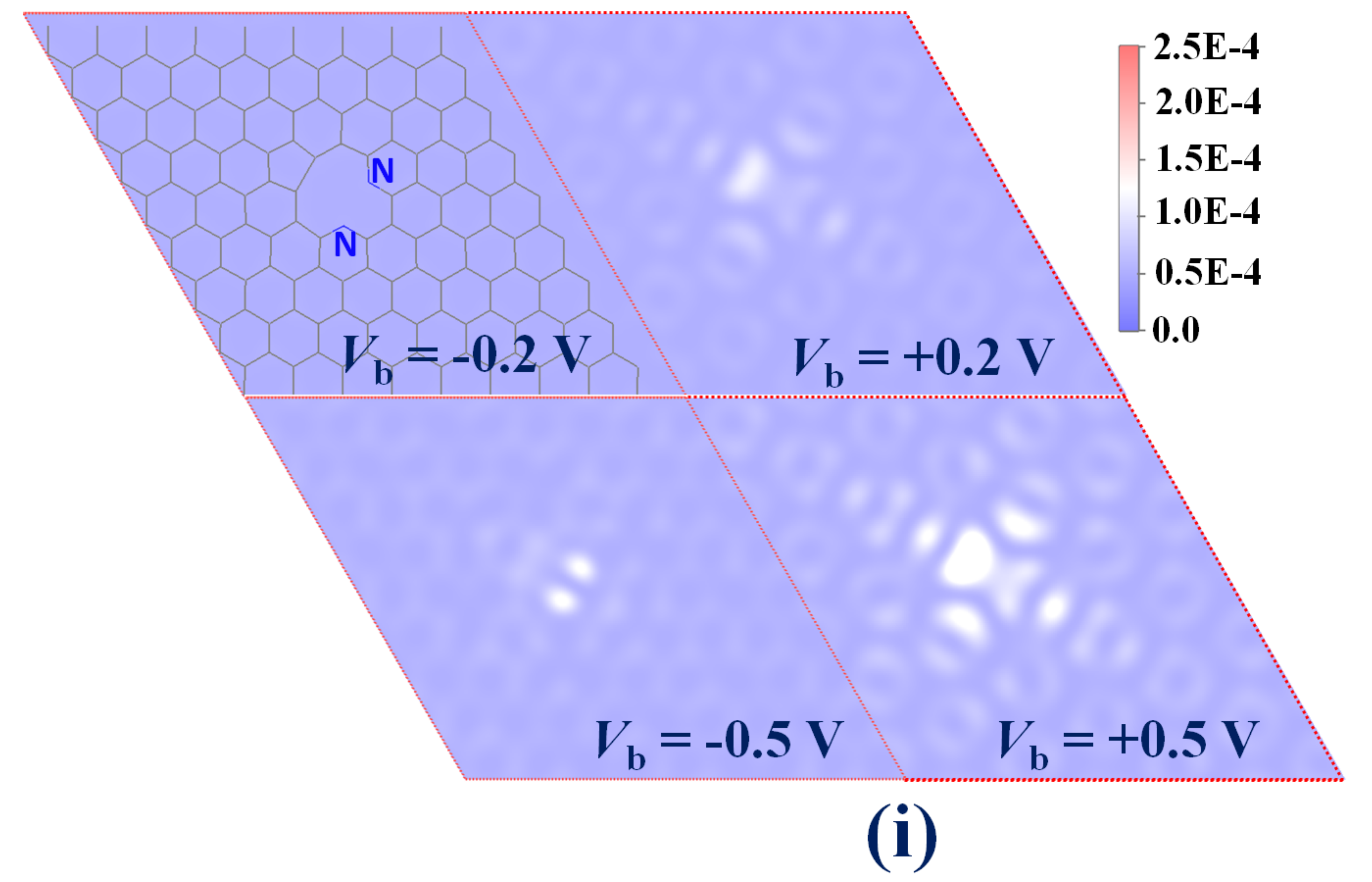}
\end{minipage}
\vspace*{.20cm}
\begin{minipage}[t]{8.0cm}
\includegraphics*[width=8.0cm]{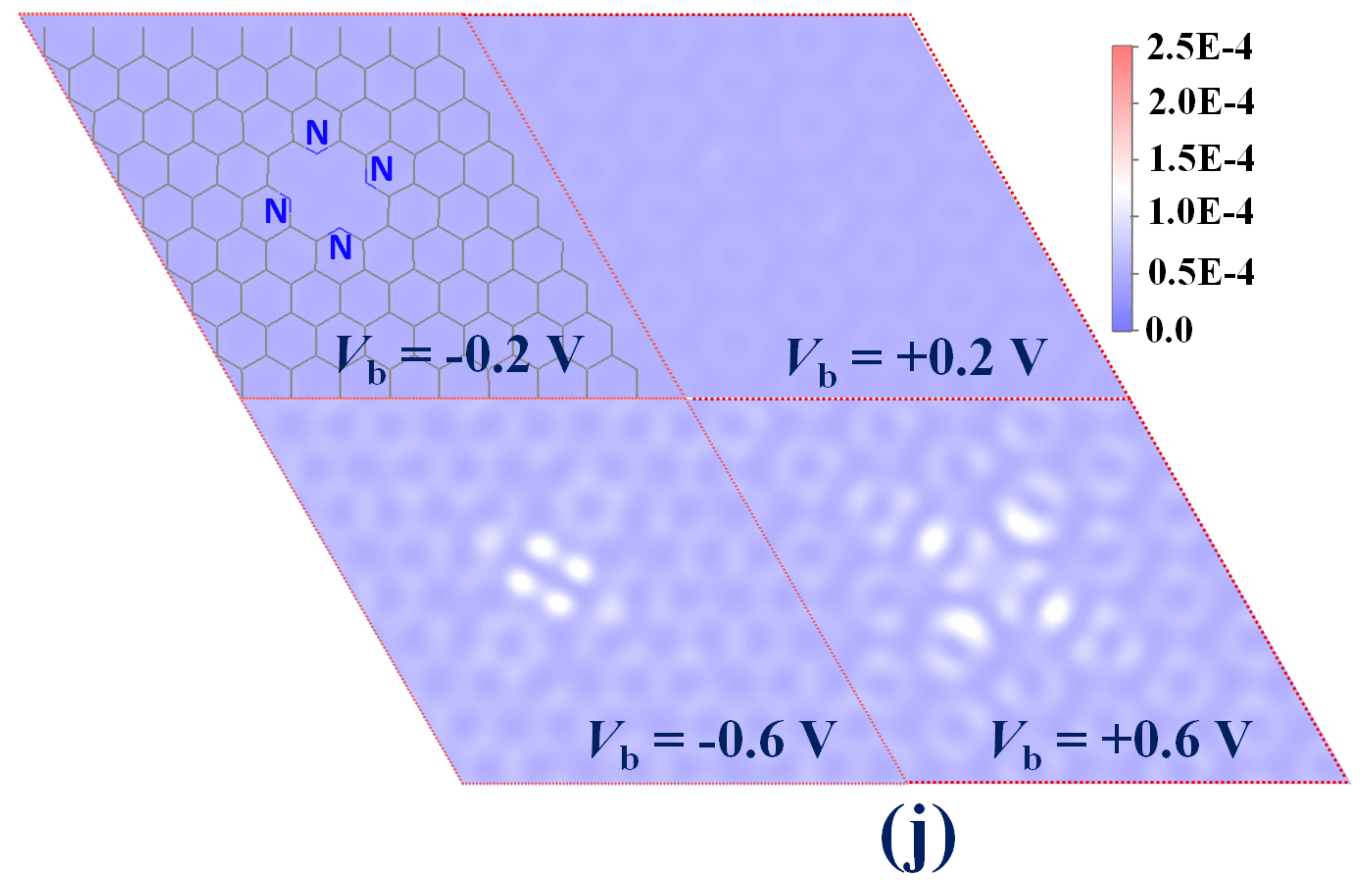}
\end{minipage}
\caption{\label{fig:11} (Color online) The atomic structures of 5-8-5 DV after N doping: (a) a m-graphitelike N (i.e., N$_\mathrm{C4}$), (b) single pyridinelike N (i.e., N$_\mathrm{C2}$), (c) dimerized pyridinelike N (i.e., N$_\mathrm{C2}$ + N$_\mathrm{C2^{\prime}}$), (d) tetramized pyridinelike N.  The band structures of graphene with 5-8-5 DV: (e) N$_\mathrm{C4}$, (f) N$_\mathrm{C2}$, (g) N$_\mathrm{C2}$ + N$_\mathrm{C2^{\prime}}$, and (h) tetramized pyridinelike N. The zero of energy is set at  $E_{\mathrm{F}}$. Panels (e)-(h) show the fat-bands derived from N 2$p_z$ and $2p_x+2p_y$ (green solid and red open points, respectively) states. Simulated STM images under different bias voltages indicated in each panel and with a sample-tip distance of $d=2$ \AA: (i) dimerized pyridinelike N and (j) tetramized pyridinelike N. }
\end{center}
\end{figure}

\paragraph{Band calculations}
Single substitutional N dopant at these DVs prefers the vertex site of a pentagonal ring to form a m-graphitelike N bonding configuration as shown in Fig.~\ref{fig:11}(a). We consider also a pyridinelike N of Fig.~\ref{fig:11}(b) to analyze the site dependence. In particular, two and four pyridinelike N atoms at the 5-8-5 DV are favorable to aggregate.~\cite{Hou2011ar}  Therefore, it is interesting to see how the electronic structures of the 5-8-5 DV changes with N doping.
The band structures for one, two, and four substitutional N atoms at the 5-8-5 DV are presented in Figs.~\ref{fig:11}(e) - (h).

For a single substitutional N dopant at C4 [i.e., N$_\mathrm{C4}$, Fig.~\ref{fig:11}(a)], the defect band of 5-8-5 DV is pulled down and becomes more than half filled [see Fig.~\ref{fig:11}(e)], producing holes in the bulk $\pi$ bands. Therefore, interestingly a single m-graphitelike N at 5-8-5 DV acts as an acceptor rather than a donor. For N$_\mathrm{C4}$ at 5-8-5 DV, the C2-C2$^{\prime}$ bond length is shortened by 0.11 \AA, and thus the corresponding unoccupied $\sigma$ states of this C-C pair are pushed up toward higher energy [see the PDOSs in Fig.~S2(a) in the Supplemental Material~\cite{suppm}]. For a single pyridinelike N at 5-8-5 DV [i.e., N$_\mathrm{C2}$, Fig.~\ref{fig:11}(b)], it is less stable than the configuration of N$_\mathrm{C4}$ by about 0.97 eV.~\cite{Hou2011ar} A single pyridinelike N at 5-8-5 DV does not possess a lone-pair of $\sigma$ electrons, in contrast to the one at MV. N$_\mathrm{C2}$ still forms a $\sigma$ bond with C2$^{\prime}$. However the unoccupied $\sigma$ states of this N$_\mathrm{C2}$-C2$^{\prime}$ pair shift toward lower energy (see the PDOSs in the Fig.~S2(b) in the Supplemental Material~\cite{suppm}) due mostly to a deeper potential of N. The two band structures [Figs.~\ref{fig:11}(e) and \ref{fig:11}(f)] look similar near $E_{\mathrm{F}}$. A single pyridinelike N at 5-8-5 DV also acts as an acceptor.
As the defect $\pi$ states are partially occupied for both N$_\mathrm{C2}$ and N$_\mathrm{C4}$, the problem of $\pi$-state spin polarization is discussed in Sec.~\ref{sec:3}E.

For two N dopants at 5-8-5 DV, they prefer to form a dimerized pyridinelike N configuration as shown in Fig.~\ref{fig:11}(c).
We first note in Fig.~\ref{fig:11}(g) that two flat $\sigma$ bands exist at -0.4 eV and -2.3 eV, which correspond to antibonding and bonding states between dangling $\sigma$ orbitals at N$_\mathrm{C2}$ and N$_\mathrm{C2^{\prime}}$. Because both bonding and antibonding states are occupied for two N dopants, the $\sigma$ bond between them does not contribute to their binding and the N$_\mathrm{C2}$-N$_\mathrm{C2^{\prime}}$ distance (2.51 \AA) is longer than the original C2-C2$^{\prime}$ distance by 0.78 \AA~and even longer than the unrelaxed C2-C2$^{\prime}$ bond of 2.46 \AA.
From the configuration of a single pyridinelike N [Fig.~\ref{fig:11}(b)] to that of a dimerized pyridinelike N [Fig.~\ref{fig:11}(c)], the antibonding $\sigma$ band is pulled down below $E_{\mathrm{F}}$, so that two $\sigma$ electrons are added.  To compensate for the over-screening, the defect $\pi$ band is pushed up to remove one $\pi$ electron and the Fermi level moves back to the Dirac point. A nearly flat defect band appears at 0.2 eV above $E_\mathrm{F}$ and it comes mainly from the contribution of the $p_z$ states of the dimerized pyridinelike N atoms and the dimerized C atoms next to DV. This can be seen from the simulated STM images under different bias voltages. As shown in Fig.~\ref{fig:11}(i), only the C2$^{\prime\prime}$-C2$^{\prime\prime\prime}$ pair at the five-membered ring looks bright by applying positive bias voltage of 0.2 V.

The electronic structures for tetramized pyridinelike N at 5-8-5 DV [Fig.~\ref{fig:11}(d)] are similar to those of dimerized pyridinelike N at 5-8-5 DV. The N-N bond becomes even longer to be 2.64 \AA. The four flat defect $\sigma$ states in Fig.~\ref{fig:11}(h) correspond to the four bonding and anti-bonding combinations among four dangling $\sigma$ orbitals. From Fig.~\ref{fig:11}(g) to Fig.~\ref{fig:11}(h), one more antibonding $\sigma$ state is occupied and two additional electrons introduced by two additional N dopants are accommodated there. In both cases of dimerized and tetramized pyridinelike N at 5-8-5 DV, no carriers are created by the N doping and the presence of the Fermi level at the Dirac point guarantees the stability of these configurations. The PDOSs for N doped DV are given in Fig.~S2(c) in the Supplemental Material~\cite{suppm}.

The simulated STM images for the tetramized pyridinelike N at 5-8-5 DV are presented in Figs.~\ref{fig:11}(j). Under the applied bias voltages of $V_\mathrm{b} =- 0.2$ V and $V_\mathrm{b}=+0.2$ V, the simulated STM images show only the background because of no localized states in this energy range. When a more negative bias voltage (e.g., -0.6 V) is applied and a sample-tip distance of $d =2$ \AA~is employed, the pyridinelike N sites would have very strong brightness due to the unpaired $\sigma$ states. When the sample-tip distance increases to 2.5 \AA, those $\sigma$ states will not be visible in the STM image, because their spatial distribution is confined mostly within the graphene plane.
\paragraph{Local bond picture}
The analysis of stability of N-doped DV and STM images based on local bond picture are given in the Supplemental Material (Fig.~S3 for the simulated STM images of a m-graphitelike N at 5-8-5 DV; Fig.~S4 and S5 for the local bond pictures of a single m-graphitelike N and dimerized pyridinelike N at 5-8-5 DV, respectively.)~\cite{suppm} for Figs.~\ref{fig:11}(a) and \ref{fig:11}(c).  We present the local bond analysis only for the tetramized pyridinlike N of Fig.~\ref{fig:11}(d).

\begin{figure*}[htbp!]
\begin{center}
\begin{minipage}[t]{5.0cm}
\includegraphics*[width=5.0cm]{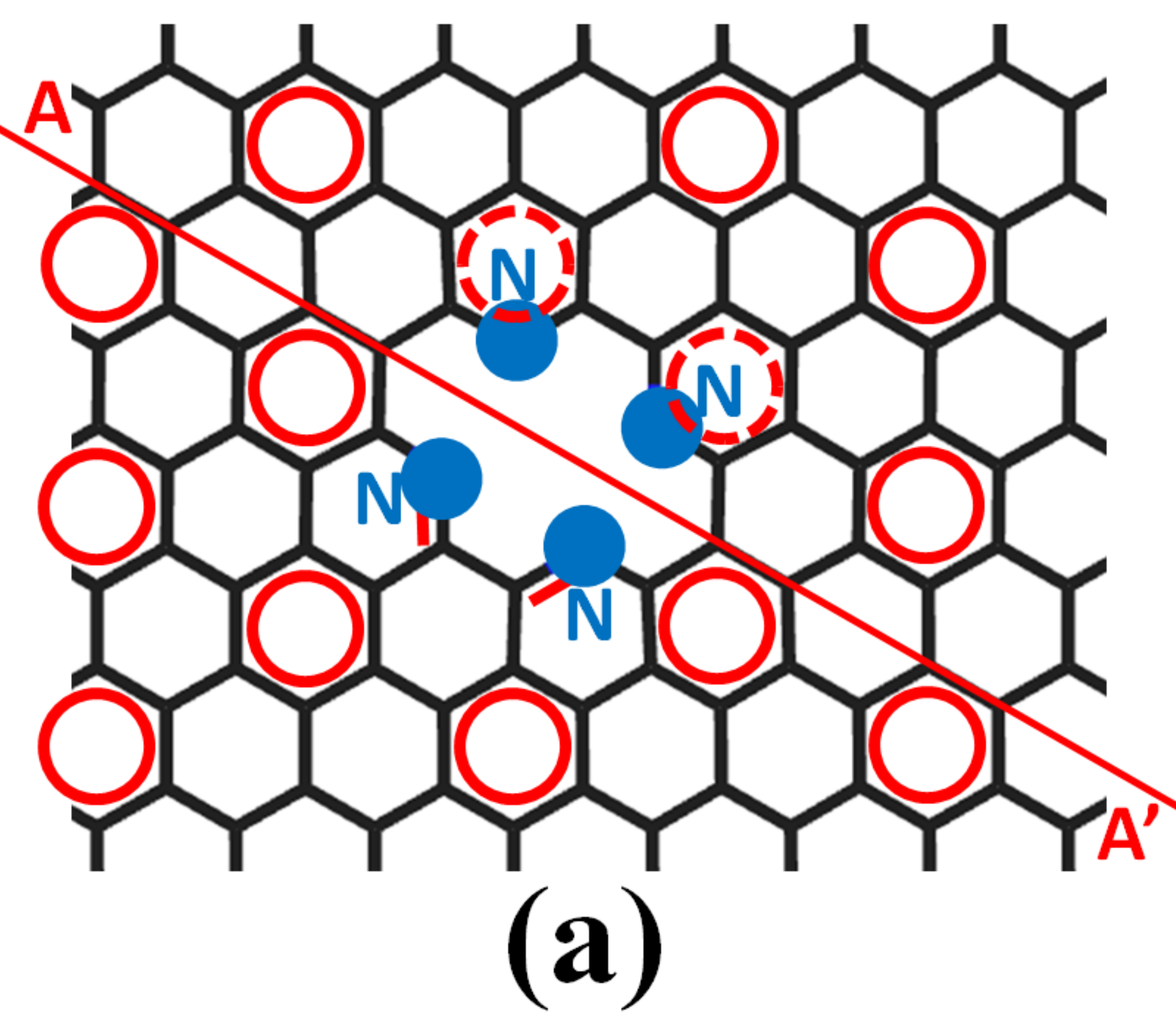}
\end{minipage}
\hspace*{.50cm}
\begin{minipage}[t]{5.0cm}
\includegraphics*[width=5.0cm]{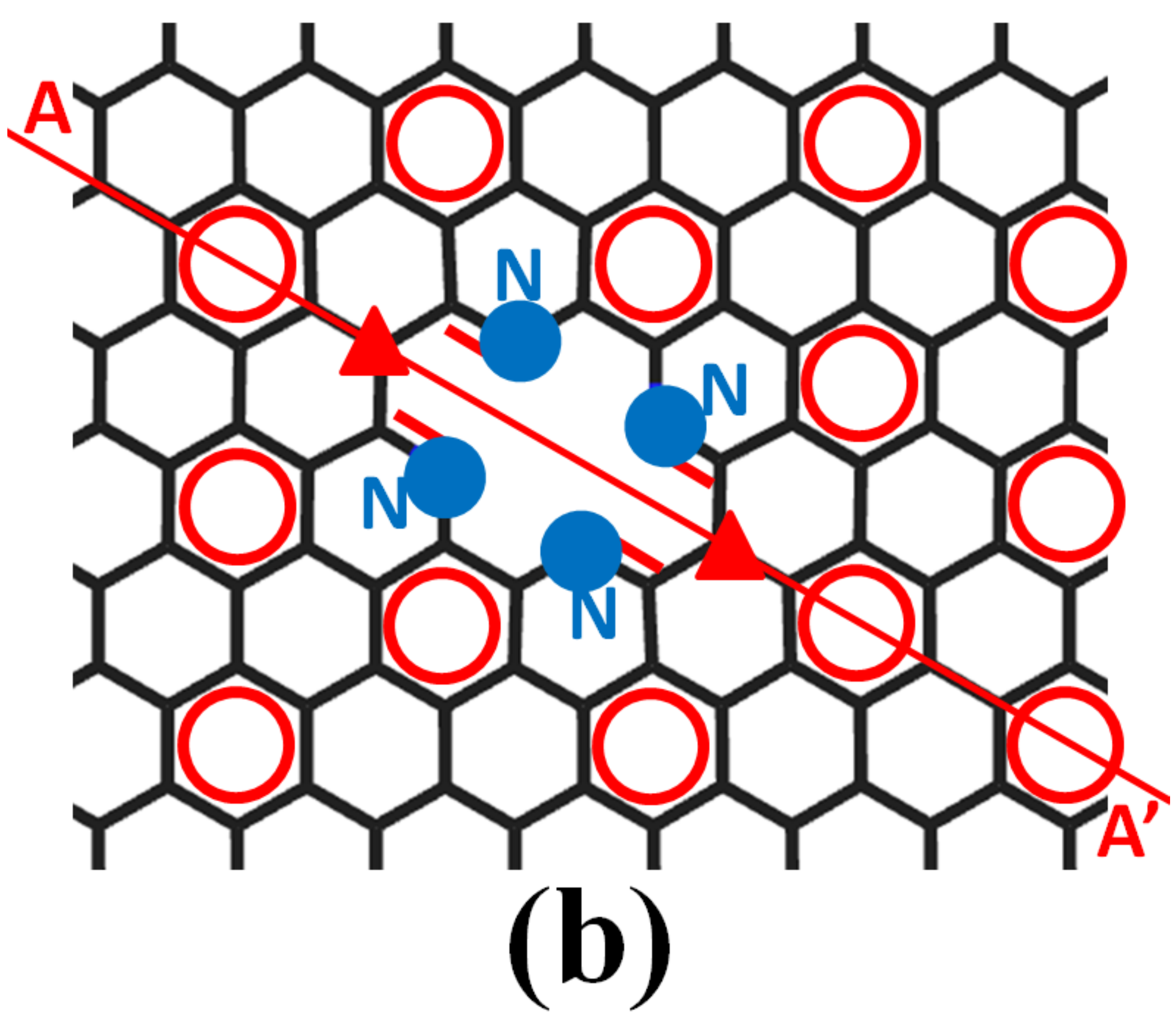}
\end{minipage}
\caption{\label{fig:12} (Color online) Two possible ways of arranging Clar sextets and double bonds around tetramized pyridinelike N at 5-8-5 DV. The solid line AA$^{\prime}$ shows the mirror symmetry axis. The dashed circle and triangle symbol stand for a pseudo-Clar sextet and a dangling $\pi$ orbital, respectively.}
\end{center}
\end{figure*}

Figure~\ref{fig:12} shows two ways of arranging the Clar sextets and double bonds around tetramized pyridinelike N at 5-8-5 DV.  The comparison of this case with the undoped DV clarifies the characteristic features of the tetramized pyridinelike N configuration.  In the present case, the N-N bond corresponding to the C2-C2$^{\prime}$  bond in Fig.~\ref{fig:9}(a) is broken, which makes the N$_\mathrm{C2}$-C1 distance shorter (1.325 \AA).  Therefore, the configuration of Fig.~\ref{fig:12}(a) is quite stable.  Moreover, the pattern in Fig.~\ref{fig:12}(a) has no dangling bonds. Therefore, this pattern is more stable than that in Fig.~\ref{fig:12}(b).

\subsection{Stone-Wales defect and N doping}
\label{sec:3D}
\begin{figure}[htbp!]
\begin{center}
\begin{minipage}[t]{10.0cm}
\includegraphics*[width=10.0cm]{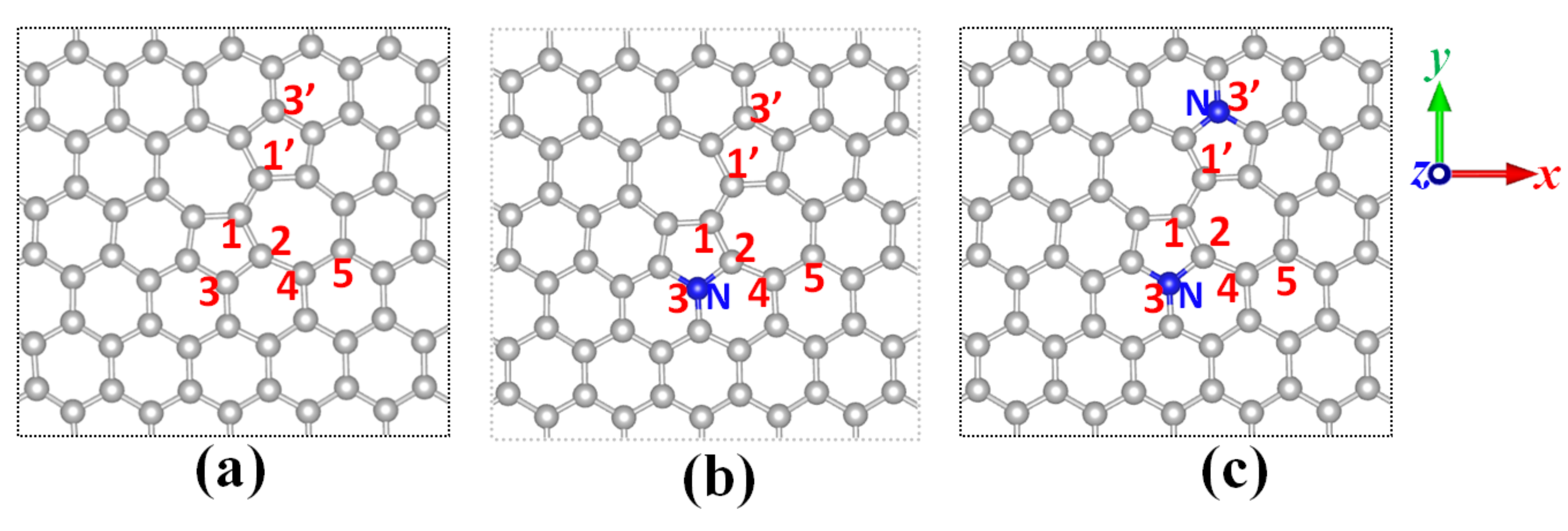}
\end{minipage}
\vspace*{0.2cm}
\begin{minipage}[t]{9.0cm}
\includegraphics*[width=9.0cm]{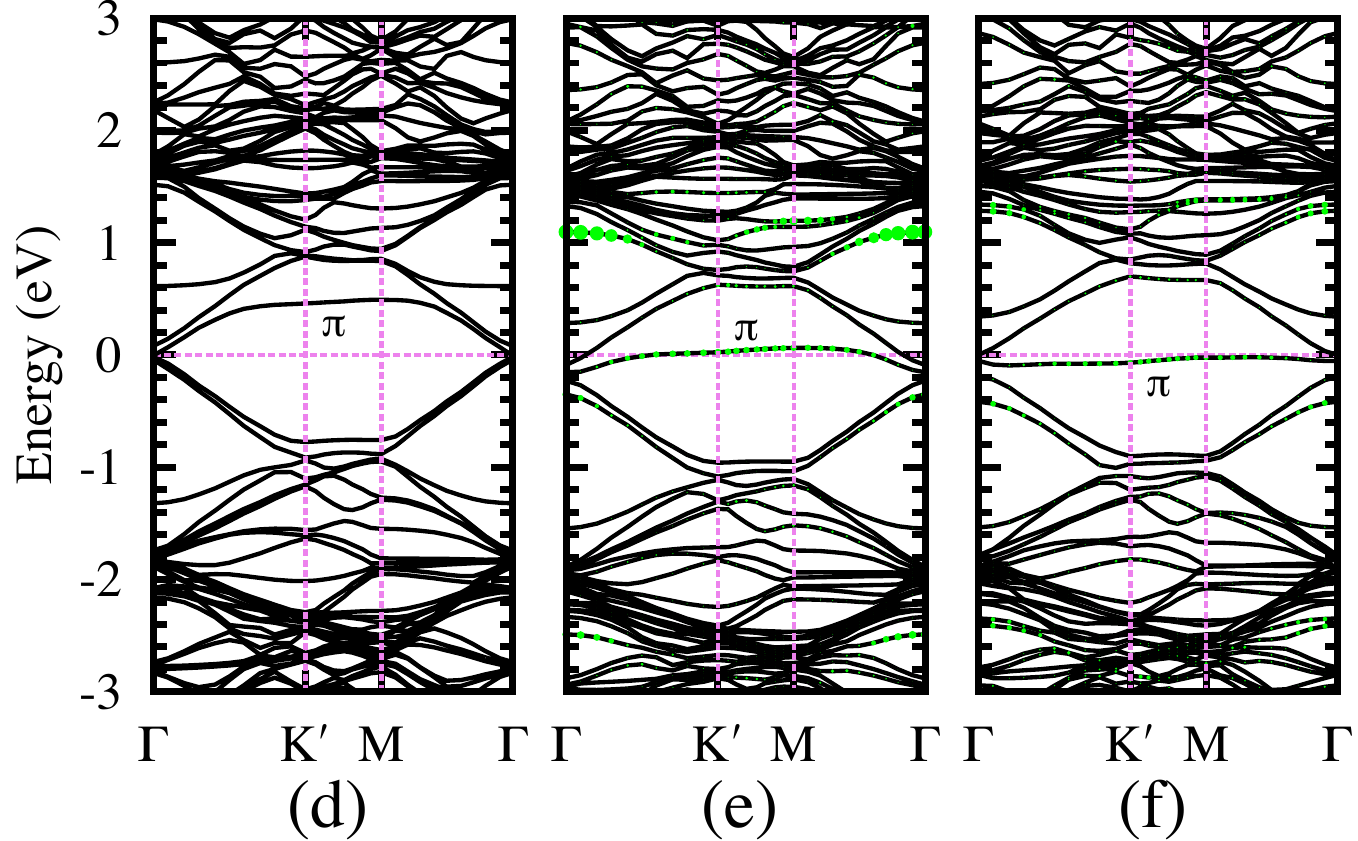}
\end{minipage}
\vspace*{.20cm}
\begin{minipage}[t]{8cm}
\includegraphics*[width=8.0cm]{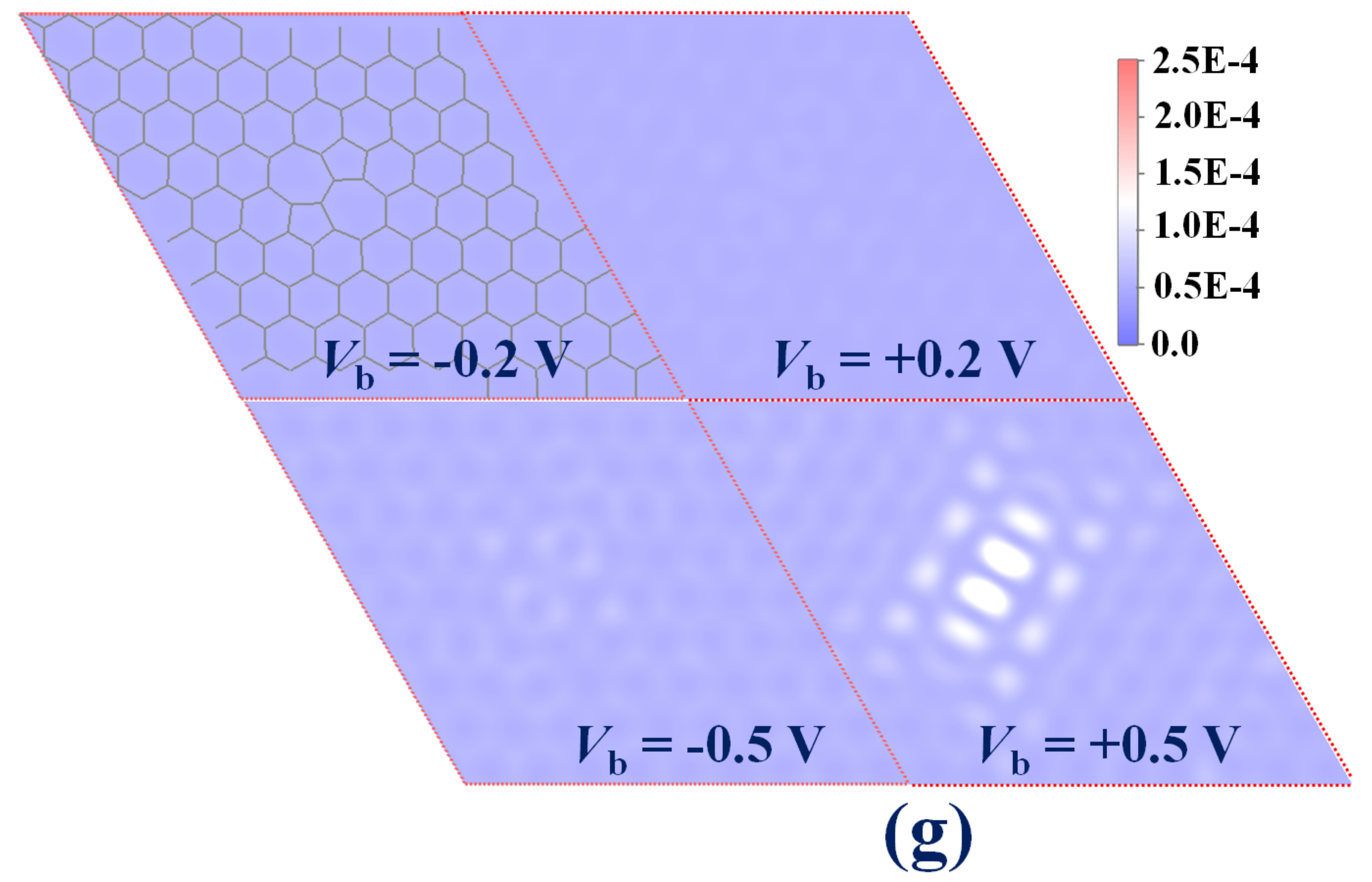}
\end{minipage}
\vspace*{.20cm}
\begin{minipage}[t]{8cm}
\includegraphics*[width=8cm]{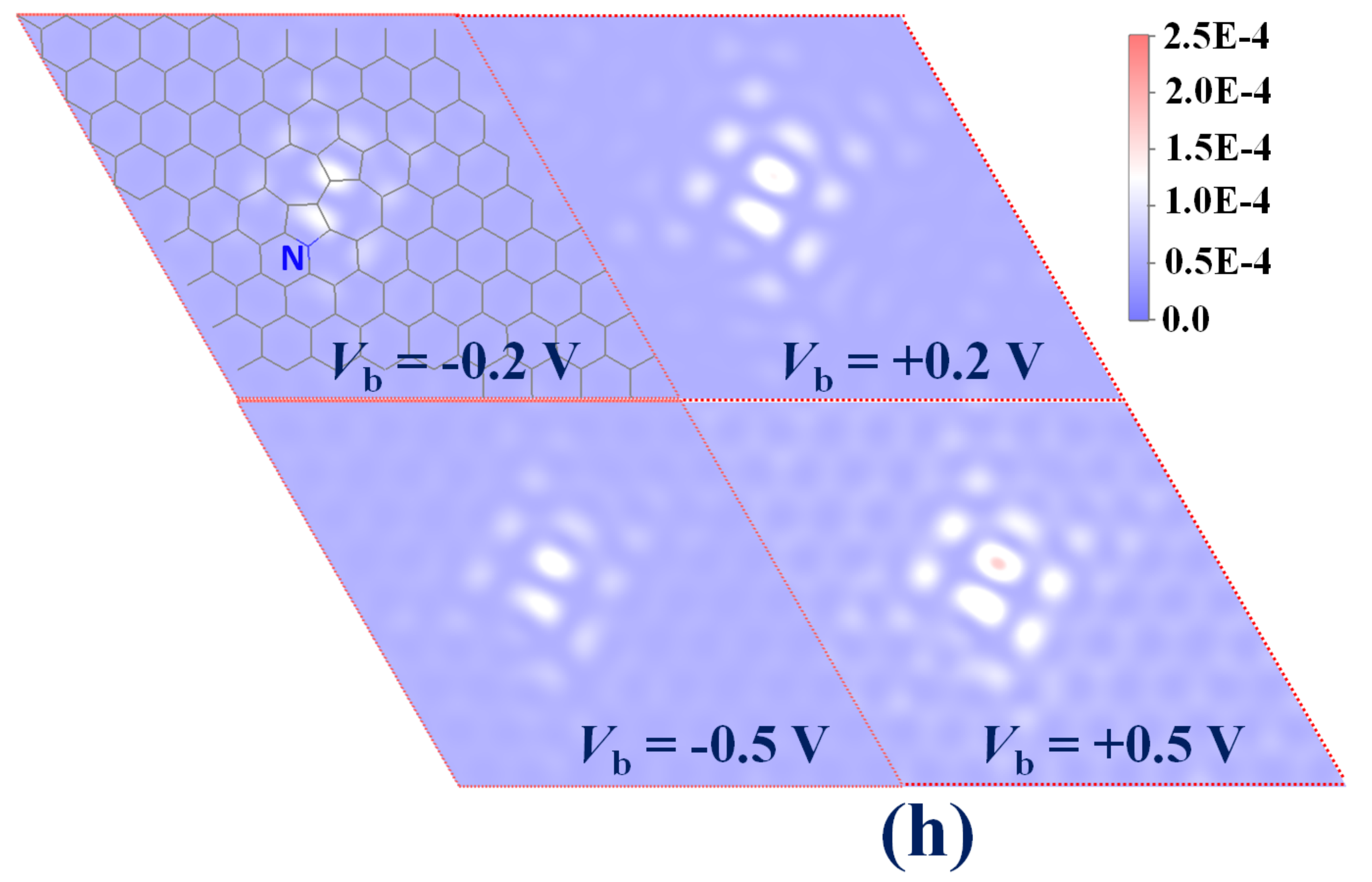}
\end{minipage}
\caption{\label{fig:13} (Color online) The atomic structures of (a) SW defect, (b) single m-graphitelike N atom (i.e., N$_\mathrm{C3}$) at SW defect, and (c) two  m-graphitelike N atoms (i.e., N$_\mathrm{C3}$+N$_\mathrm{C3^{\prime}}$) at SW defect. The band structures of graphene with SW defect in different cases: (d) before N doping, (e) N$_\mathrm{C3}$, and (f) N$_\mathrm{C3}$ + N$_\mathrm{C3^{\prime}}$. The zero of energy is set at  $E_{\mathrm{F}}$. Panels (e) and (f) show the fat-bands (green solid points) derived from the N $2p_z$ states. Simulated STM images under different bias voltages indicated in each panel and with a sample-tip distance of $d=2$ \AA: (g) SW defect and (h) N$_\mathrm{C3}$ at SW defect.}
\end{center}
\end{figure}

\begin{figure*}[htbp!]
\begin{center}
\begin{minipage}[t]{5.0cm}
\includegraphics*[width=5.0cm]{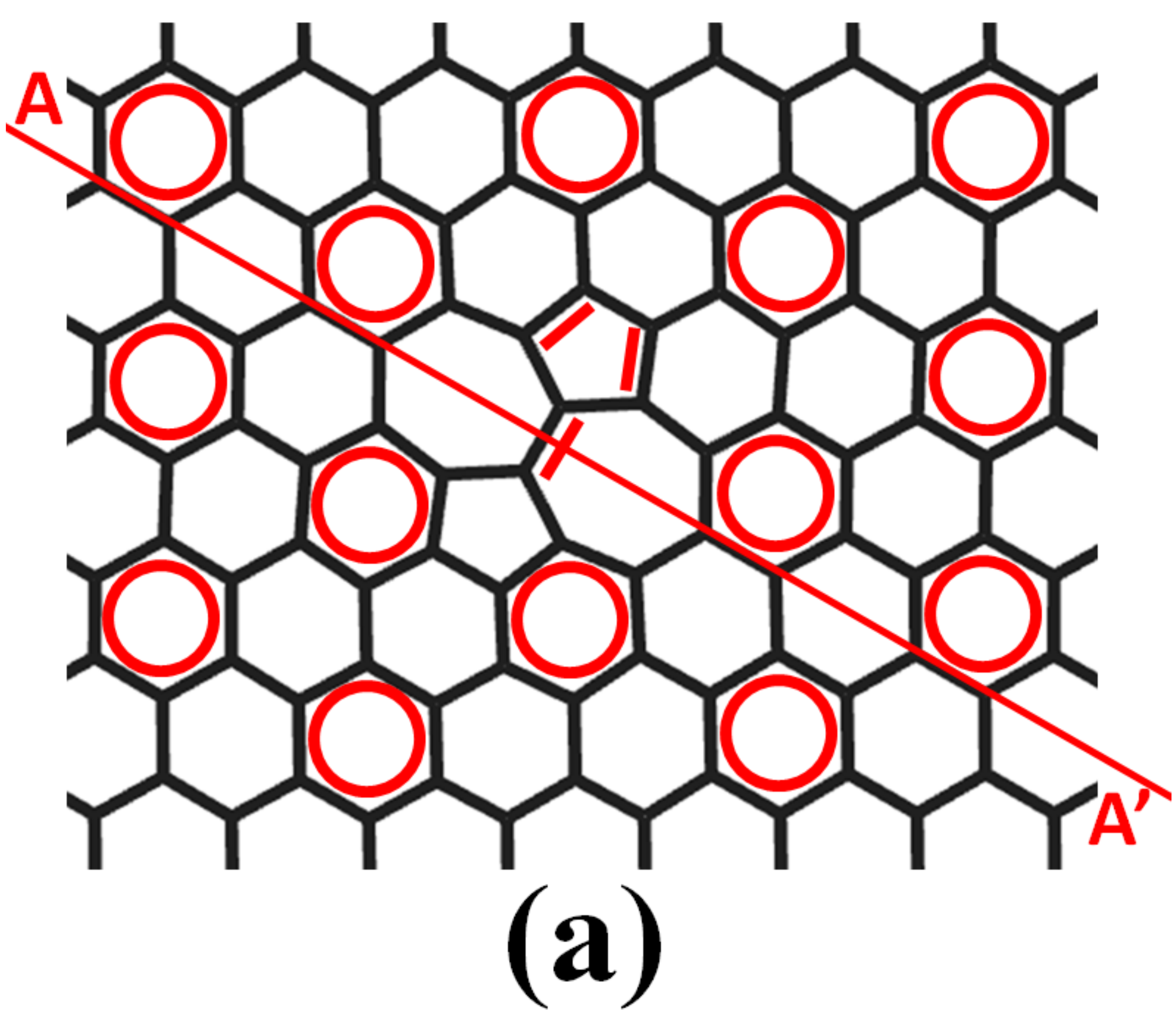}
\end{minipage}
\hspace*{0.5cm}
\begin{minipage}[t]{5.0cm}
\includegraphics*[width=5.0cm]{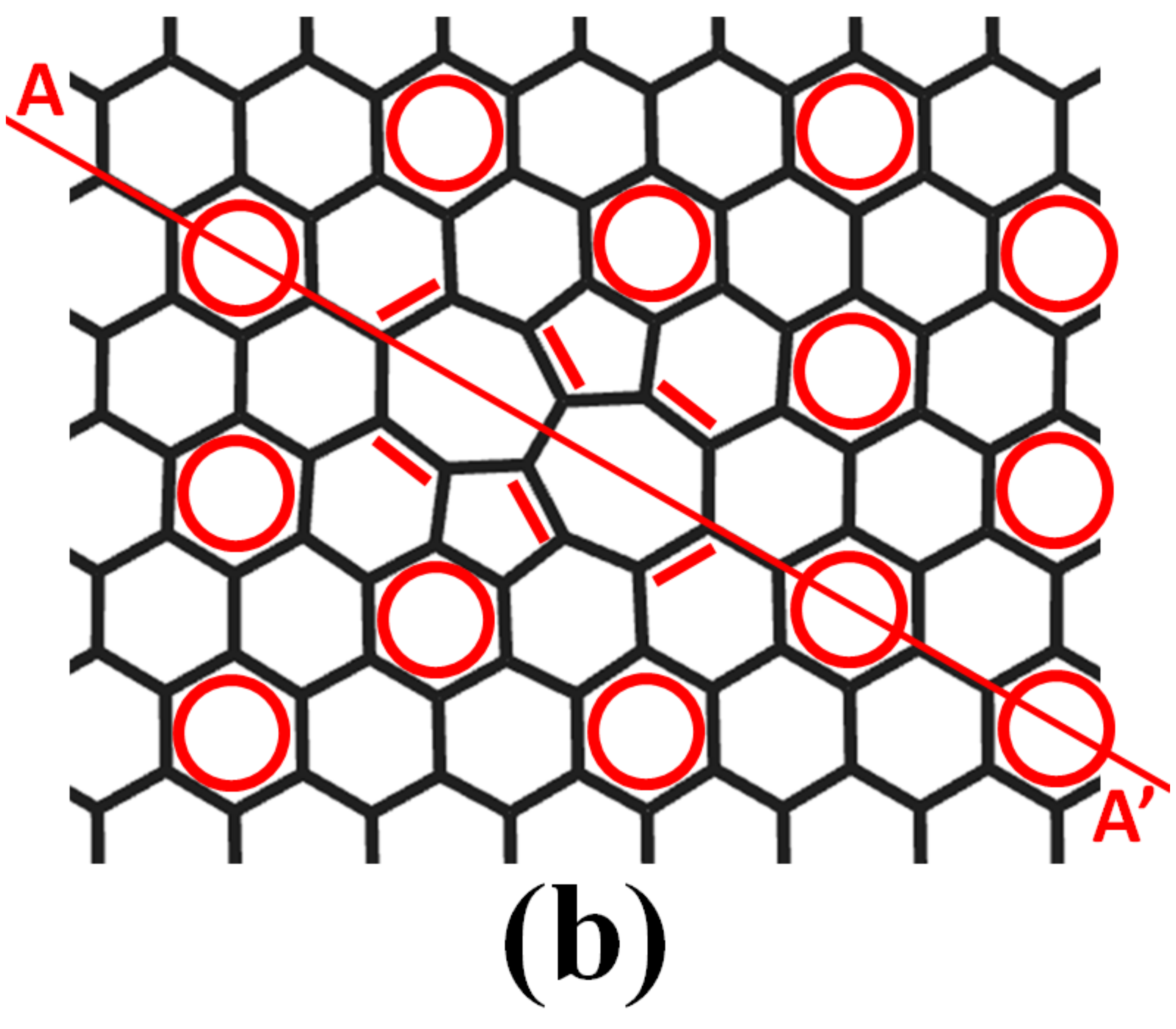}
\end{minipage}
\caption{\label{fig:14} (Color online) Two possible ways of arranging Clar sextets and double bonds around SW defect. The solid line AA$^{\prime}$ shows the mirror symmetry axis.}
\end{center}
\end{figure*}

The SW defect is formed by in-plane 90$^{\circ}$ rotation of the C1-C1$^{\prime}$ bond [see Fig.~\ref{fig:13}(a)]. This rotation not only changes the bonding direction of C1 to its nearest neighboring C2, but also results in the shortening of the C1-C1$^{\prime}$ bond by 0.10 \AA~and the elongation of C1-C2 bond by 0.04 \AA.  As seen from the band structure of graphene with the SW defect [Fig.~\ref{fig:6}(d)], the Fermi level coincides with the Dirac point and no additional carriers are introduced into graphene.~\cite{Dubois2011}
Nevertheless, the SW defect also induces a defect $\pi$ state which is located about 0.4 eV above $E_{\mathrm{F}}$ and hybridizes with one of the unoccupied bulk $\pi$ bands. The defect $\pi$ state is localized significantly at the C1 and C1$^{\prime}$ sites as seen from the simulated STM images for different bias voltages in Fig.~\ref{fig:13}(g) and more clearly in PDOS given in Fig.~S6 in the Supplemental Material~\cite{suppm}. Under the bias voltages of $\pm$ 0.2 and $-0.5$ V, the simulated STM images show only the background, indicating that no localized defect states appear below $E_\mathrm{F}+0.2$ eV. Under the applied bias voltage of $+0.5$ V, the bright spots in the simulated STM images form a pattern with rectangular shape and extend along the bond direction of the rotated C-C dimer. In particular, the rotated carbon atoms have strongest brightness in the simulated STM image. Two ways of Clar sextets and double bonds around SW defect are presented in Fig.~\ref{fig:14}. There are no dangling $\pi$ orbitals in both ways. This may correspond to no defect states below the Fermi level, as discussed above in the band calculations. Considering the C1-C1$^{\prime}$ bond is significantly shorten after rotation in SW defect, a double bond may exist at the C1-C1$^{\prime}$ dimer. Therefore the pattern in Fig~\ref{fig:14}(a) is more probable and it may correspond to the ground state of SW defect.

Upon N doping as in Fig.~\ref{fig:13}(b), the defect $\pi$ state comes down and partially occupied and small amount of electrons are doped into the bulk $\pi$ band [see Fig.~\ref{fig:13}(e)]. Therefore, a single m-graphitelike N at SW defect acts as an electron donor. From the simulated STM images shown in Fig.~\ref{fig:13}(h), we can see that the C1, C1$^{\prime}$ and C2 sites are brighter under the bias voltage of $\pm$0.2 V, suggesting these carbon sites may be more chemically reactive after N doping. The possibility of spin polarization of the defect $\pi$ state in this case will be discussed in Sec.~\ref{sec:3}.E. The defect band is pushed completely down in close proximity to $E_{\mathrm{F}}$ and the possibility of spin polarization disappears after doping two N atoms [see Figs.~\ref{fig:13}(c) and (f)].

\begin{figure*}[htbp!]
\begin{center}
\begin{minipage}[t]{5.0cm}
\includegraphics*[width=5.0cm]{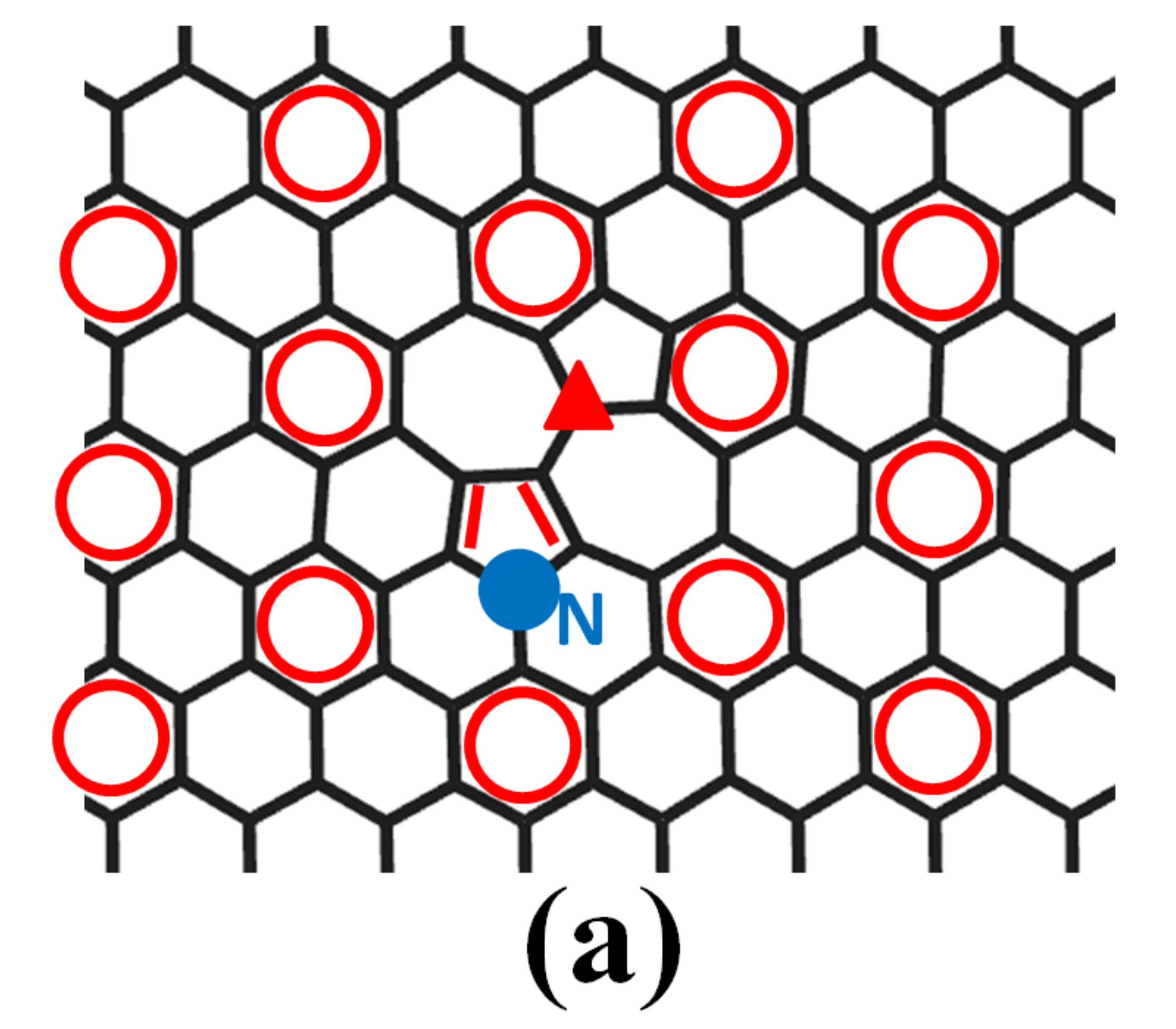}
\end{minipage}
\hspace*{0.3cm}
\begin{minipage}[t]{5.0cm}
\includegraphics*[width=5.0cm]{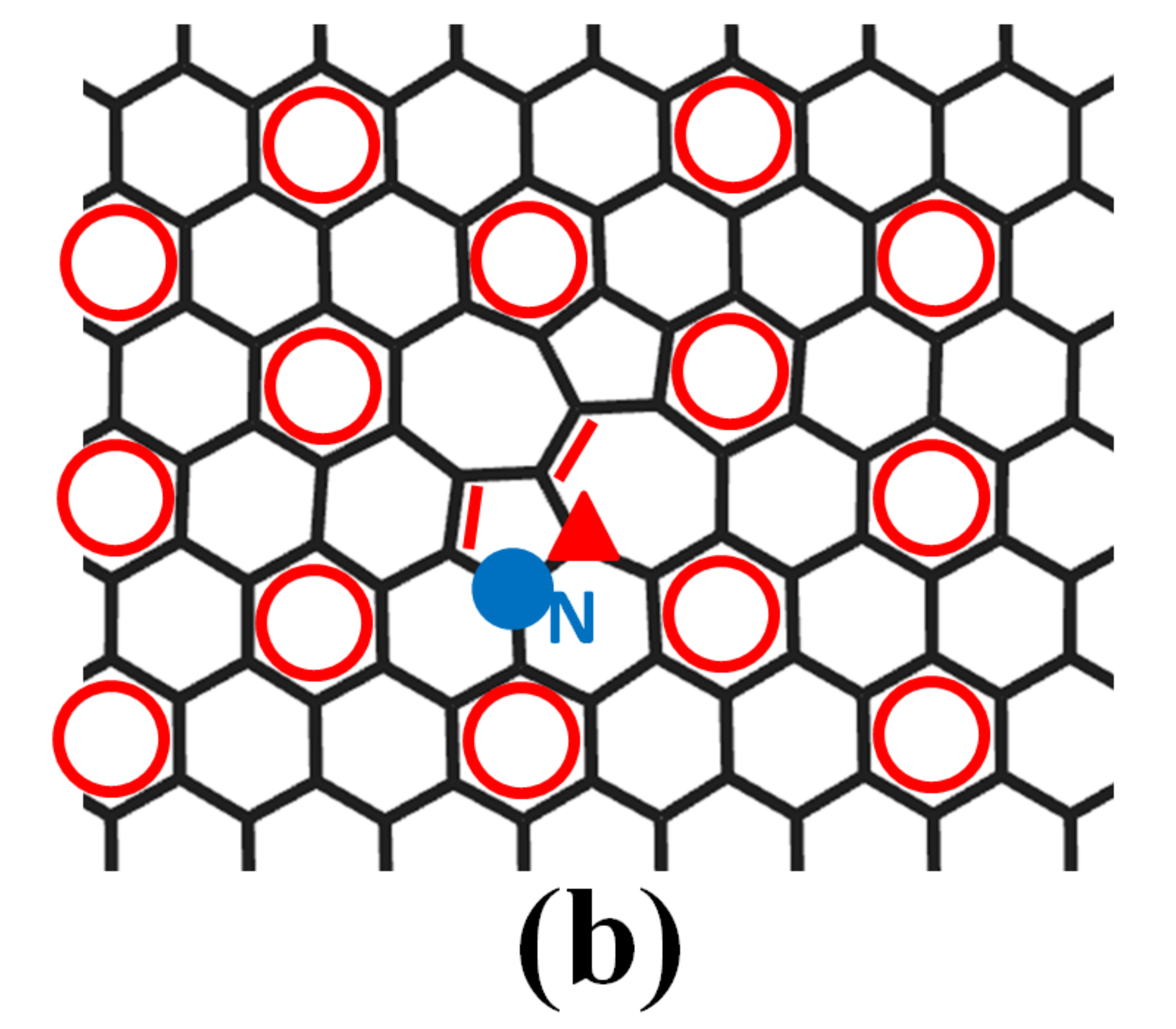}
\end{minipage}
\hspace*{0.3cm}
\begin{minipage}[t]{5.0cm}
\includegraphics*[width=5.0cm]{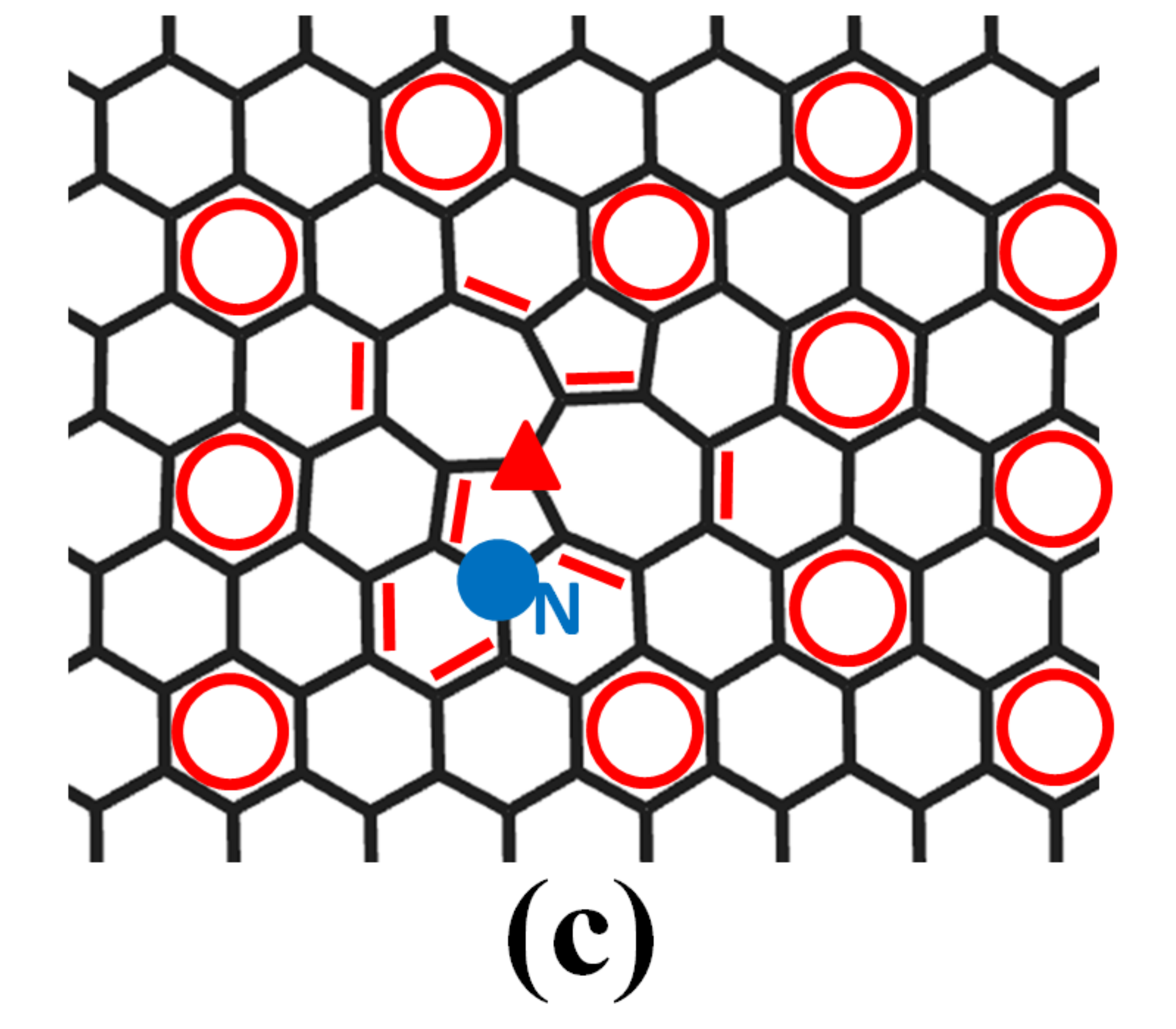}
\end{minipage}
\caption{\label{fig:15} (Color online) Three possible ways of arranging Clar sextets and double bonds around a m-graphitelike N at SW defect. The triangle symbol stands for a dangling $\pi$ orbital. }
\end{center}
\end{figure*}

From the STM images and the PDOS given in Fig.~S6 in the Supplemental Material~\cite{suppm}, we see that the DOS at $E_\mathrm{F}$ is large in the order of C1$^{\prime}$, C2 and C1 for a single m-graphitelike N at SW defect in Fig.~\ref{fig:13}(b).  As there are many possible ways of arranging Clar sextets and double bonds in this case, it is difficult to give a consistent picture of explaining the behaviors seen in PDOS and STM images using local bond picture.  Among many possible configurations, we only show three patterns where dangling bond exists at one of C1$^{\prime}$, C2 and C1 sites.  For Figs.~\ref{fig:15}(a) and (b), only one Clar sextet circle is missing, while three are missing in Fig.~\ref{fig:15}(c).  This aspect is consistent with weaker brightness and smaller PDOS at $E_\mathrm{F}$ at C1 site.  On the other hand, between (a) and (b) of Fig.~\ref{fig:15}, the configuration of (b) seems to be more probable because of the short bond length (1.34 \AA) of C1-C1$^{\prime}$, being inconsistent with the trend in STM image and PDOS.  At the present stage, we do not have a proper explanation for this inconsistency.  More subtle aspects may have to be taken into account.

\subsection{Magnetic properties of nitrogen doped at native point defects}
\label{sec:3E}
As briefly mentioned in Sec.~\ref{sec:3}B, Ref.~\onlinecite{Palacios2012} studied the concentration dependence of the spin polarization of defect $\pi$ state for MV and claimed that the $\pi$ state spin polarization would vanish in the dilute limit.  The MV case has subtle aspects. As shown in Figs.~\ref{fig:2}(a), \ref{fig:2}(c), and \ref{fig:2}(e), the dangling $\sigma$ state is definitely spin polarized to have 1.0 $\mu_\mathrm{B}$, which is maintained even in the dilute limit because of the strong localization of defect $\sigma$ state.  The $\sigma$ spin polarization produces the spin dependent exchange potential which is felt also by the defect $\pi$ state.  This means that even if the defect $\pi$ state is not spontaneously spin polarized, the $\pi$ state spin polarization will be induced as long as the $\pi$ state is not completely filled.  Looking at Fig.~3 of Ref.~\onlinecite{Palacios2012}, we note that as the unit cell size (defect concentration) increases (decreases), the Fermi level increases to the Dirac point.  On the other hand, the energy of the defect $\pi$ state also increases.  Therefore, a very careful and rigorous study is needed to check whether the $\pi$ state polarization will vanish or not in the dilute limit of MV.  Note that we are discussing quantities of O(1/$N$) with $N$ denoting the number of atoms in the supercell.  The concentration dependence of magnetic moment shown in Fig.~8 of Ref.~\onlinecite{Palacios2012} clearly shows that the $\pi$-state spin polarization decreases with decreasing the MV concentration.  Nevertheless, the magnetization value (including the $\sigma$ contribution) in the dilute limit in the figure may not be 1.0, although the authors fitted the calculated data with a function which takes 1.0 at the zero concentration limit.  A more rigorous treatment seems to be needed to reach a definite conclusion about the limiting value of the $\pi$-state spin polarization.

The subtlety of the defect $\pi$-state spin polarization at MV is caused by the existence of the defect $\sigma$ state spin polarization.  In this context, the H-MV case needs careful consideration, because the out-of-plane displacement of H and C1 in the spin-polarized state in Figs.~\ref{fig:2}(b), \ref{fig:2}(d), and \ref{fig:2}(f) produces $\sigma$-$\pi$ hybridization leading to $\sigma$ spin polarization at the C1 atom. As already mentioned in Sec.~\ref{sec:3}.B, the spin polarization totally vanishes both in $\sigma$ and $\pi$ states once H and C1 are confined within the atomic plane of graphene. Moreover, as the weight of $\sigma$ component in the defect states at the Fermi level is very small, the role of $\sigma$ state is different from that in the nonhydrogenated MV case. Only by increasing the unit cell size from $9\times 9$ to $12\times 12$, the ground state of H-MV is not spin polarized (see Table~\ref{tab:1}).

Other cases of possible $\pi$-state spin polarization are MV [Figs.~\ref{fig:5}(a) and (d)], 5-8-5 DV [Figs.~\ref{fig:11}(a) and (e); Figs.~\ref{fig:11}(b) and (f)] and SW defect [Figs.~\ref{fig:13}(b) and (e)] for all of which single N is doped at the defect. We first note that in all of these cases, there is no $\sigma$ spin polarization.  We checked the spin-polarized $\pi$ states for these cases and the results are summarized in Table~\ref{tab:1}.  It is found that a pyridinelike N at MV does not induce spin polarization. This is similar to the case of H-MV with H and C1 confined in the atomic plane of graphene. For m-graphitelike N atoms doped at 5-8-5 DV and SW defect, the $\pi$ defect level is spin polarized if we use the 9$\times$9 supercell, although the spin polarization energy is only of the order of meV.  However, the spin polarization is significantly reduced or totally suppressed by increasing the supercell size to 12$\times$12. Therefore, we can safely conclude that the $\pi$ state spin polarization will vanish in the dilute limit of 5-8-5 DV and SW concentrations.

\subsection{Simulated N 1\emph{s} core level spectra}
\label{sec:3G}
Finally, we discuss N 1\emph{s} core level spectra for different N bonding configurations near NPDs.

\subsubsection{N 1\emph{s} binding energy and its chemical shift}
\label{sec:3G1}
Table~\ref{tab:2} presents the chemical shifts of N 1\textit{s} core level binding energy for substitutional N dopants around NPDs with respect to graphitelike N in defect-free graphene.
The results obtained by the Kohn-Sham (KS) energy level of the ground state take into account the initial state effect. On the other hand, the ones from the total energy difference ($\Delta\mathrm{SCF}$) based on Eq.~(\ref{eq:4}) include the final state effect. Here our results indicate that although the two different estimates give about 20 eV difference in the N 1\textit{s} binding energy, its chemical shifts are determined mainly by the initial state effect, consistent with the trend of N dopants along edges of graphene nanoribbons~\cite{Wang2011prb,Wang2013} and nitrogen-containing molecules~\cite{Casanovas1996}. This is strikingly different from the situation in the C 1\textit{s} binding energies in graphene, where the final state effects govern the chemical shift.~\cite{Hou2011jpcc}  We do not understand quite well the reason of this difference.  Maybe the core-state binding energy of an impurity atom and that of a host atom may have qualitative difference in the final state effect. It is noted that the N dopants at the vertex sites of a pentagonal ring (i.e. m-graphitelike) in 5-8-5 DV and SW defect have quite small chemical shifts with respect to the graphitelike N. This small chemical shift is generally smaller than the peak width of the experimental spectra. The N 1\emph{s} core level of pyridinelike N at MV and 5-8-5 DV is about 3.8 eV shallower than the one of graphitelike N, which qualitatively agrees with the experimental results.~\cite{Niwa200993,Usachov2011} The chemical shift of pyridiniumlike N at MV is about $-$1.2 eV, suggesting that this type of pyridiniumlike N will contribute to the XPS peak generally assigned to pyrrorlike N.  We noticed that the chemical shift of pyridiniumlike N has significant dependence on its environment:
According to our recent work,~\cite{Wang2013} the corresponding values are $-$1.0 along the zigzag edge and $-$0.2 along the armchair edge.

\subsubsection{N \emph{K}-edge XAS spectra}
\label{sec:3G2}
\begin{figure}[htbp!]
\begin{center}
\includegraphics*[scale=0.8,angle=0]{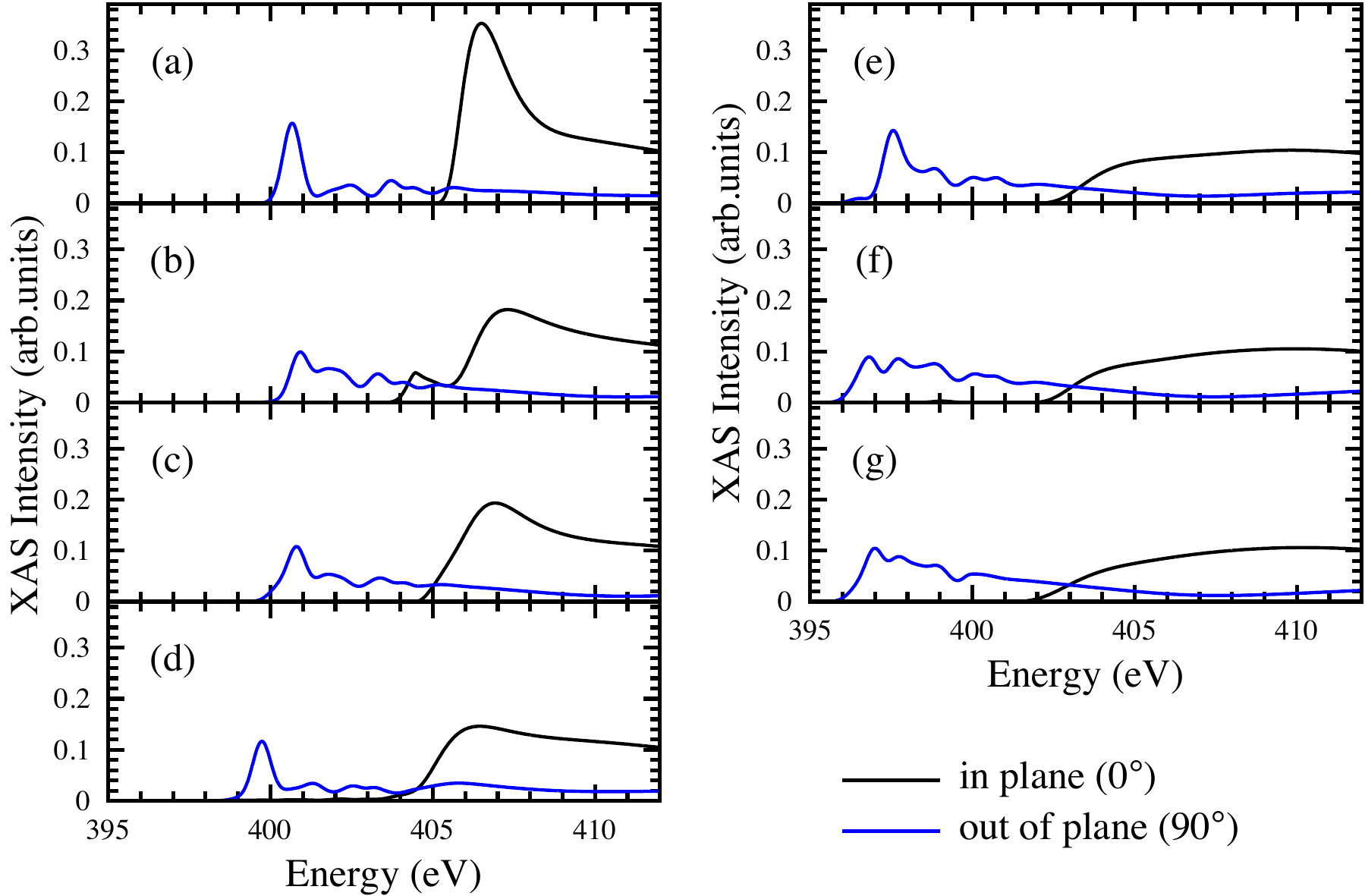}
\caption{\label{fig:16} (Color online) The computed N 1\textit{s}  XAS spectra of substitutional N atoms in different bonding configurations: (a) a graphitelike N in prefect graphene; (b) a m-graphitelike N at 5-8-5 DV [N$_\mathrm{C4}$ shown in Fig.~\ref{fig:11}(a)]; (c) a m-graphitelike N at SW defect [N$_\mathrm{C3}$ shown in Fig.~\ref{fig:13}(b)]; (d) a pyridiniumlike N at MV [N$_\mathrm{C1}$ shown in Fig.~\ref{fig:5}(c)]; (e) a single pyridinelike N at MV [N$_\mathrm{C1}$ shown in Fig.~\ref{fig:5}(a)]; (f) dimerized pyridinelike N at 5-8-5 DV [N$_\mathrm{C2}$  +N$_\mathrm{C2^{\prime}}$ shown in Fig.~\ref{fig:11}(c)]; and (g) tetramized pyridinelike N at 5-8-5 DV [shown in Fig.~\ref{fig:11}(d)]. }
\end{center}
\end{figure}

For simulating XAS spectra, we used a larger unit cell of $12\times12$ to reduce the interference between the excited states at neighboring cells.  Even with such a large unit cell, the interference effect cannot be totally removed as we point out in the following arguments.  The XAS spectra are shown for $\pi^{\ast}$ and $\sigma^{\ast}$ components separately in Fig.~\ref{fig:16}.

\paragraph{Graphitelike N} The simulated N \textit{K}-edge XAS spectrum for an isolated graphitelike N in graphene is presented in Fig.~\ref{fig:16}(a). Two main features are found in the theoretical spectrum, i.e., a sharp peak at 400.8 eV which is
due to the $\pi^{\ast}$ resonance and an intense peak at 407 eV which is contributed by the $\sigma^{\ast}$ resonance. The positions of the $\pi^{\ast}$ and $\sigma^{\ast}$ peaks in our calculations agree well with the experimental results.~\cite{Niwa200993,Zhao12011s,Usachov2011}  Small shoulders and a valley around 403 eV are the result of the interference effect mentioned above.

\paragraph{M-graphitelike N at pentagon} We examine the m-graphitelike N at two different types of NPDs, i.e., the 5-8-5 DV and SW defect and show their XAS spectra in Figs.~\ref{fig:16}(b) and \ref{fig:16}(c). The atomic structures for such N bonding configurations are presented in Figs.~\ref{fig:11}(a) and~\ref{fig:13}(b). The main features of two spectra look very similar, including the position and the profile. The sharp feature at 401 eV is attributed to N 1\textit{s}$\rightarrow \pi^{\ast}$ and the broad feature at 407 eV is associated with N 1\textit{s}$\rightarrow \sigma^{\ast}$. It is noticeable that the position and profile of the two main features of m-graphitelike N are also quite similar to those of graphitelike N, although both peaks are less sharp and the edge of $\sigma^{\ast}$ component is shifted to lower energy by about 1.0 eV for m-graphitelike N.

\paragraph{Pyridiniumlike N at MV} The simulated XAS spectrum for single pyridiniumlike N at MV is presented in Fig.~\ref{fig:16}(d).  This looks quite similar to the one for m-graphitelike N at SW defect [Fig.~\ref{fig:16}(c)] except lower energy shift  by about 1.0 eV.
We noticed that the present XAS spectrum is quite different from those along edges.~\cite{Wang2013}  In the latter there is a sharp peak of  N-H $\sigma^{\ast}$ origin between $\pi^{\ast}$ and $\sigma^{\ast}$ peaks.

\paragraph{Pyridinelike N at vacancies} We first consider three cases, i.e., single pyridinelike N at MV, two and four pyridinelike N at 5-8-5 DV. Their atomic structures can be referred to Figs.~\ref{fig:3}(a), \ref{fig:11}(c) and~\ref{fig:11}(d). The simulated XAS spectra are presented in Figs.~\ref{fig:16}(e), \ref{fig:16}(f), and \ref{fig:16}(g). For a single pyridinelike N at MV, there is a tiny peak at 396.5 eV which is associated with the transition to the unoccupied part of the defect $\pi$ state. A sharp peak appears at 397.5 eV in the simulated spectrum and it is attributed to the transition to the N component of the bulk $\pi^{\ast}$ state. However the feature associated with $\sigma^{\ast}$ states is quite broad. The XAS spectra for two and four pyridinelike N at 5-8-5 DV are almost the same, reflecting the similar band structures for them shown in Figs.~\ref{fig:11}(g) and \ref{fig:11}(h).  The broad low energy $\pi^{\ast}$ peak  comes from the transitions to the two flat $\pi^{\ast}$ bands.  However, the separation of the two flat bands will decrease as the defect concentration decreases similarly to the case of two flat bands for doped N in the defect-free graphene as discussed already. Other features are similar to those of single pyridinelike N at MV.
The position of the feature associated with N 1\textit{s}$\rightarrow \pi^{\ast}$ for pyridinelike N in the theoretical spectrum agrees well with the one reported in experiment.~\cite{Usachov2011, Saito2010}

\section{Conclusion}
\label{sec:4}
The electronic structures of N-doped graphene have been studied by performing the DFT calculations. We analyze the doping behavior of NPDs and substitutional N dopants in graphene. The origin and character of the localized states induced by NPDs are also carefully analyzed.
Our results show that MV acts as hole dopant and the hole induced by a MV can be compensated by a single pyridiniumlike N dopant (but not by a single pyridinelike N dopant) or by two N dopants. No carriers are induced by 5-8-5 DV and SW defect. The doping behavior of substitutional N is affected significantly by the defect-induced states. Single pyridinelike N at MV and single m-graphitelike N at 5-8-5 DV both exhibit hole dopants, while single m-graphitelike N at SW defect acts as electron donor. The dimerized and tetramized pyridinelike N at 5-8-5 DV do not induce any carriers.  The possibility of spin polarization of $\pi$ state by defects and N-doped defects has been carefully discussed.  Although some ambiguity still remains for MV for which the defect $\sigma$ state has stable spin polarization, the spin polarization of defect $\pi$ state without being accompanied with the $\sigma$ spin polarization will not exist in the dilute limit of defect concentration at least for the defects studied in the present work.  We have also discussed the characteristic features in the simulated STM images and the 1\textit{s} related XPS and XAS spectra for pyridinelike N at vacancies, pyridiniumlike N at MV, and m-graphitelike N at a five-membered ring of DV and SW defect. These will provide useful information for identifying the different N bonding configurations in N-graphene.  Moreover, the partial density of states at various sites around N-doped NPDs (presented mostly in Supplemental materials) and more directly the STM images for small bias voltages will also give us hints about the catalytically active carbon sites.  These results obtained by the standard first-principles electronic structure calculations were explained by the local bond picture, particularly in connection with STM images for some cases studied here.

\section*{Acknowledgement}
This work was performed under  Project No.10000829-0 and Project No.10000832-0 at the New Energy and Industrial Technology Development Organization (NEDO).
The computation was performed using the supercomputing facilities in the Center for Information Science in JAIST. Parts of the computations were done on TSUBAME Grid Cluster at the Global Scientific Information and Computing Center of the Tokyo Institute of Technology.

\bibliography{reference1}

\clearpage
\newpage

\begin{table}
\caption{\label{tab:1} Total magnetic moment ($M_\mathrm{t}$, $\mu_\mathrm{B}$ per cell) and the spin polarization energy ($E_\mathrm{SP}$, in eV) of doped nitrogen near the NPDs in the $9\times9$ and $12\times12$ supercells of graphene. Here we present the results for the pyridinelike N at MV [N$_\mathrm{C1}$ shown in Fig.~\ref{fig:5}(a)] and 5-8-5 DV [N$_\mathrm{C2}$ shown in Fig.~\ref{fig:11}(b)], and m-graphitelike N at 5-8-5 DV [N$_\mathrm{C4}$ shown in Fig.~\ref{fig:11}(a)] and SW defect [N$_\mathrm{C3}$ shown in Fig.~\ref{fig:13}(b)]. The results for H-MV are also shown for comparison. $E_\mathrm{SP}$ is defined as the total energy difference between the spin-polarized state and the non-spin-polarized state. }
\begin{ruledtabular}
\begin{tabular}{ccc cc c}
   Configuration  & \multicolumn{2}{c}{pyridinelike} &\multicolumn{2}{c}{m-graphitelike}  &     \\
       \cline{2-3}\cline{4-5}
     N in Figure         &  N$_\mathrm{C1}$, \ref{fig:5}(a)    & N$_\mathrm{C2}$, \ref{fig:11}(b) & N$_\mathrm{C4}$,~\ref{fig:11}(a) & N$_\mathrm{C3}$,~\ref{fig:13}(b)&      \\
   Defect            &    MV     &  5-8-5 DV&   5-8-5 DV & SW defect & H-MV \\
   \hline
  $M_\mathrm{t}$, $9\times9$  &  0.00     &  0.21        &  0.50    &    0.53  & 0.55      \\
  $E_\mathrm{SP}$, $9\times9$    & 0.000   & -0.001         &  -0.002   &   -0.003  &  -0.005   \\
 $M_\mathrm{t}$, $12\times12$  &  0.00  &   0.00          &  0.00     &   0.08  & 0.00  \\
$E_\mathrm{SP}$,$12\times12$   & 0.000  &   0.000         & 0.000    &  -0.001  & 0.000   \\

\end{tabular}
\end{ruledtabular}
\end{table}

\begin{table}
\caption{\label{tab:2} N 1\emph{s} core level binding energy [$E_\mathrm{b}$(N 1\emph{s}), in eV] with respect to the Fermi level for graphitelike N in defect-free graphene (N$_\mathrm{C}$), m-graphitelike N at 5-8-5 DV and SW defect [N$_\mathrm{C4}$ and N$_\mathrm{C3}$ shown in Figs.~\ref{fig:11}(a) and~\ref{fig:13}(b), respectively], pyridiniumlike N at MV [N$_\mathrm{C1}$-H shown in Fig.~\ref{fig:5}(c)], and pyridinelike N at MV and 5-8-5 DV [N$_\mathrm{C1}$ and N$_\mathrm{C2}$ shown in Figs.~\ref{fig:5}(a),~\ref{fig:11}(c) and~\ref{fig:11}(d), respectively] estimated by the initial Kohn-Sham energy level (KS)
and the total energy difference ($\Delta$SCF). The values in parentheses are the chemical shifts with respect to the graphitelike N. }
\begin{ruledtabular}
\begin{tabular}{ccc cc ccc}
            &  \multicolumn{7}{c}{$E_\mathrm{b}$(N 1\emph{s})}          \\
            \cline{2-8}
   Configuration  & graphitelike  &\multicolumn{2}{c}{m-graphitelike} & pyridiniumlike & \multicolumn{3}{c}{pyridinelike}\\
       \cline{3-4}    \cline{6-8}
     N in Figure         &  N$_\mathrm{C}$    & N$_\mathrm{C4}$, \ref{fig:11}(a) & N$_\mathrm{C3}$,~\ref{fig:13}(b) & N$_\mathrm{C1}$-H,~\ref{fig:5}(c)& N$_\mathrm{C1}$,~\ref{fig:5}(a)  & N$_\mathrm{C2}$,~\ref{fig:11}(c) &N$_\mathrm{C2}$,~\ref{fig:11}(d)  \\
   Defect            &    defect-free     &    5-8-5 DV & SW defect & MV&  MV & 5-8-5 DV &  5-8-5 DV\\
   \hline
   KS            &  379.78     &  378.98    &  379.17   &  378.07  & 375.73 & 375.82 &375.83  \\
                 &  (0.00)     &  (-0.80)   & (-0.61)   & (-1.71)  & (-4.05) &  (-3.96) &(-3.95)\\
                    $\Delta$SCF   & 400.16 &  399.95 &       399.93 &    398.93    &  396.40  & 396.39  & 396.41 \\
                 & (0.00)      & (-0.21)    & (-0.23)  & (-1.23)  & (-3.76) & (-3.77) & (-3.75) \\

\end{tabular}
\end{ruledtabular}
\end{table}

\clearpage
\newpage

\end{document}